\documentclass[fleqn,usenatbib]{mnras}
\usepackage{newtxtext,newtxmath}
\usepackage{url}
\usepackage[T1]{fontenc}
\usepackage{ae,aecompl}
\usepackage{graphicx}
\usepackage{amsfonts,subfigure}                                                           
\usepackage{bm}%allows bold fonts for Greek letters               
\usepackage{times}         
\usepackage[usenames,dvipsnames]{xcolor}
\usepackage{soul}
\usepackage{comment}
\usepackage{upgreek}
\usepackage{tablefootnote}
\usepackage{threeparttable}

\def\symbolfootnote[#1]#2{\begingroup%
def\thefootnote{\fnsymbol{footnote}}\footnote[#1]{#2}\endgroup}

\newcommand{\um}{\mbox{\textmu m}}
\newcommand{\configure}{\textsc{configure}}
\newcommand{\scheduler}{\textsc{scheduler}}
\newcommand{\HI}{\textsc{H\,i}}
\newcommand{\Msun}{\mbox{$\mathrm{M}_{\odot}$}}
\newcommand{\Teff}{\mbox{$T_{\mathrm{eff}}$}}

\title[The WEAVE Survey]{The wide-field, multiplexed, spectroscopic facility WEAVE: Survey design, overview, and simulated implementation}

\author[S. Jin et al.]{
\parbox{\textwidth}{
\Large
Shoko Jin$^{1,2,3,4\star}$,
Scott C.~Trager$^{2\diamond}$,
Gavin B.~Dalton$^{1,3\ast}$,
J.~Alfonso L.~Aguerri$^{5,6}$,
J.~E.~Drew$^{7,8}$,
Jes\'us Falc\'on-Barroso$^{5,6}$,
Boris T.~G\"ansicke$^{9}$,
Vanessa Hill$^{10}$,
Angela Iovino$^{11}$,
Matthew M.~Pieri$^{12}$
% ...
}
\vspace{0.4cm}
\\
\parbox{\textwidth}{
The authors' affiliations are shown in Appendix~\ref{sec:affiliations}.}
}

\date{}
\date{Accepted XXX. Received YYY; in original form ZZZ}

\pubyear{2022}

\begin{document}
\label{firstpage}
\pagerange{\pageref{firstpage}--\pageref{lastpage}}
\maketitle

\clearpage

\begin{abstract}

WEAVE, the new wide-field, massively multiplexed spectroscopic survey facility for the William Herschel Telescope, saw first light in late 2022. WEAVE comprises a new 2-degree field-of-view prime-focus corrector system, a nearly 1000-multiplex fibre positioner, 20 individually deployable `mini' integral field units (IFUs), and a single large IFU. These fibre systems feed a dual-beam spectrograph covering the wavelength range 366$-$959\,nm at $R\sim5000$, or two shorter ranges at $R\sim20\,000$. After summarizing the design and implementation of WEAVE and its data systems, we present the organization, science drivers, and design of a five- to seven-year programme of eight individual surveys to: (i) study our Galaxy's origins by completing Gaia's phase-space information, providing metallicities to its limiting magnitude for $\sim$3 million stars and detailed abundances for $\sim1.5$ million brighter field and open-cluster stars; (ii) survey $\sim0.4$ million Galactic-plane OBA stars, young stellar objects, and nearby gas to understand the evolution of young stars and their environments; (iii) perform an extensive spectral survey of white dwarfs; (iv) survey $\sim400$ neutral-hydrogen-selected galaxies with the IFUs; 
(v) study properties and kinematics of stellar populations and ionized gas in $z<0.5$ cluster galaxies; (vi) survey stellar populations and kinematics in $\sim25\,000$ field galaxies at $0.3\lesssim z \lesssim 0.7$; (vii) study the cosmic evolution of accretion and star formation using $>1$ million spectra of LOFAR-selected radio sources; (viii) trace structures using intergalactic/circumgalactic gas at $z>2$. Finally, we describe the WEAVE Operational Rehearsals using the WEAVE Simulator.

\end{abstract}

\begin{keywords}
surveys -- instrumentation: spectrographs -- Galaxy: general -- stars: general -- galaxies: general -- cosmology: observations
\end{keywords}

\section{Introduction}
\label{sec:Introduction}

The last decade has seen the opening of new windows on the Universe, made possible by the development of new instrumentation. Two of these new windows are ultra-high-precision astrometry, enabled by the {\it Gaia} mission \citep{GaiaMission2016} and providing us with radically new views on the structure, dynamics, and evolution of our own Milky Way system, and low-frequency radio astronomy, enabled by LOFAR \citep{vanhaarlem13}. The latter is already revealing millions of active galactic nuclei (AGNs) and star-forming galaxies across cosmic time, and their interactions with their local circumgalactic and intergalactic media \citep[e.g.][]{shimwell19,williams19,duncan19,2021A&A...648A...1T,2021A&A...648A...2S,2021A&A...648A...3K,2021A&A...648A...4D}, while the former has provided unprecedented insights into -- or constraints on -- subject matters such as star formation in the Galactic disc \citep[e.g.][]{2020NatAs...4..965R}, the dynamics of the Milky Way disc \citep[e.g.][]{Antoja2018,2019MNRAS.482.1417B,2019MNRAS.485.3134L}, the mass of the Milky Way \citep[e.g.][]{2019A&A...621A..56P,2019MNRAS.485.3296W,2020MNRAS.494.4291C}, the dynamics of satellite galaxies \citep[e.g.][]{2022A&A...657A..54B}, the effect of the Large Magellanic Cloud on the dynamics of the Milky Way and stellar streams, \citep[e.g.][]{2021MNRAS.501.2279V}, stellar streams in the Galactic halo \citep[e.g.][]{Malhan2018}, a major accretion event \citep[{\it Gaia} Enceladus Sausage,  e.g.][]{Belokurov2018,Helmi2018}, the chemo-dynamics of stellar streams \citep{2019MNRAS.482.3426M} and determination of stellar ages in the disc \citep{2019MNRAS.486.1167B} through combining {\it Gaia} data with those from complementary surveys, and the Galactic disc as seen from its component stellar clusters \citep{2020A&A...640A...1C}.\looseness-1

At the start of the last decade, the European astronomical community, through the European Commission-funded ASTRONET programme, laid out a set of recommendations for solving pressing astrophysical problems in the following decade and the infrastructure required for those solutions. A clear recommendation from both the ASTRONET Wide-Field Spectroscopy Working Group\footnote{\url{https://www.astronet-eu.org/sites/default/files/d26-vsdef-2.pdf}} and the report on Europe's 2--4\,m telescopes by the European Telescope Strategic Review Committee\footnote{\url{https://www.astronet-eu.org/sites/default/files/plaquettet2_4m-final-2.pdf}} was for an intermediate-resolution ($R\sim5000$), wide-field ($\gtrsim 1^\circ$ diameter) multi-object spectrograph with a multiplex of $>500$ (and preferably $>1000$) to follow up the {\it Gaia} mission and to allow for precision-cosmology experiments; an extension to even higher resolution ($R\gtrsim20\,000$) with the same instrument was deemed desirable, as this extension would allow for chemo-dynamical labelling of structures in the Milky Way.\looseness-1

Roughly simultaneously, in 2010 January, the Isaac Newton Group of telescopes (ING) held a workshop on `Science with the William Herschel Telescope 2010--2020',\footnote{\url{http://www.ing.iac.es/conferences/wht201020/}} where ideas for wide-field spectroscopy with the William Herschel Telescope\footnote{\url{http://www.ing.iac.es/astronomy/telescopes/wht/}} (WHT) were discussed. Previous, on-going, and then-planned surveys (e.g.\ SDSS: \citealt{2000AJ....120.1579Y}; SEGUE-1: \citealt{2009AJ....137.4377Y}; SEGUE-2: \citealt{2011AJ....142...72E}; APOGEE-1: \citealt{2017AJ....154...28B}; APOGEE-2: \citealt{2011AJ....142...72E}; RAVE: \citealt{2006AJ....132.1645S}; LAMOST: \citealt{2012RAA....12.1197C,2015RAA....15.1095L}; Gaia-ESO: \citealt{2012Msngr.147...25G,2022A&A...666A.121R,2022A&A...666A.120G}; GALAH: \citealt{2015MNRAS.449.2604D}) have demonstrated the immense power of wide-field spectroscopy for answering important questions about the formation and evolution of galaxies and their stars and the structure of the Universe, leading to the conclusion -- fully supported by the ASTRONET recommendations -- that such a survey instrument was vital for the ING's long-term vision. During follow-up discussions, a small team developed a set of desirable requirements for an instrument capable of both providing the missing aspects of the (then) upcoming survey instruments {\it Gaia}, LOFAR, and Apertif \citep{2008AIPC.1035..265V,2022A&A...658A.146V}, and science goals extending beyond those three instruments: these are listed in Table~\ref{table:requirements}. These requirements and a well-developed design for such an instrument were presented to the community in 2015, while members of the community presented their own aspirations for multi-object spectroscopy for the coming decade at the same conference \citep{2016ASPC..507.....S}, resulting in a much broader science case covering cosmology, galaxy evolution, and stellar evolution on scales from the local Solar neighbourhood to redshifts in excess of 4.\looseness-1

\begin{table}
\centering
\caption{Desirable requirements for follow-up of the {\it Gaia} mission \citep{GaiaMission2016}, and LOFAR \citep{vanhaarlem13} and Apertif \citep{2008AIPC.1035..265V} surveys.}
\label{table:requirements}
\begin{threeparttable}
\begin{tabular}{ll}
\hline
Instrument & Desirable requirements \\ \hline
{\it Gaia}\tnote{\textit{a,b}}
& $R=5000$ for radial velocities at $17\leq V\leq20\,\mathrm{mag}$ \\
& $R=20\,000$ for stellar abundances at $12\leq V\leq17\,\mathrm{mag}$ \\
& $10^7$ stars over $10^4$ contiguous square degrees \\
LOFAR\tnote{\textit{c,d}}
& Continuous wavelength coverage over $370$--$980$ nm \\
& $V\leq21.5\,\mathrm{mag}$ at $S/N=5$ in continuum for redshifts \\
& $10^7$ galaxies over $10^4$ square degrees \\
Apertif\tnote{\textit{e,f}}
& Large IFU and mini-IFUs for spatially resolved spectra of\\
& gas-rich galaxies \\
& $10^4$ galaxies over $10^4$ continuous square degrees \\ \hline
\end{tabular}
\begin{tablenotes}
\item[\textit{a}] \url{https://www.esa.int/Science_Exploration/Space_Science/Gaia}
\item[\textit{b}] \url{https://www.cosmos.esa.int/web/gaia/data-release-3}
\item[\textit{c}] \url{https://science.astron.nl/telescopes/lofar/}
\item[\textit{d}] \url{https://lofar-surveys.org/dr2_release.html}
\item[\textit{e}] \url{https://science.astron.nl/telescopes/wsrt-apertif/}
\item[\textit{f}] \url{https://tinyurl.com/2p8d64av}
\end{tablenotes}
\end{threeparttable}
\end{table}

The WHT Enhanced Area Velocity Explorer \citep[{WEAVE\footnote{\url{https://ingconfluence.ing.iac.es/confluence//display/WEAV}},}][]{2012SPIE.8446E..0PD,2016SPIE.9908E..1GD} is the result of these discussions: a next-generation wide-field spectroscopic survey facility for the $4.2\,\mathrm{m}$ William Herschel Telescope at the Observatorio del Roque de los Muchachos on La Palma, Spain. It provides the spectroscopic follow-up required for full scientific exploitation of the {\it Gaia}, LOFAR and Apertif surveys (for references, see Table~\ref{table:requirements}) in the Northern Hemisphere and, furthermore, to study advanced phases of stellar evolution, galaxy evolution in different environments over the last 5--8 Gyr, and the changing scale of the Universe and therefore its most basic parameters since near its beginning. The facility comprises a new $2^\circ$ field-of-view prime-focus corrector system with a 1000-multiplex fibre positioner, 20 individually deployable `mini' integral field units (mIFUs), and a single `large' integral field unit (LIFU). The fibres feed a dual-beam spectrograph that will provide full coverage of most of the visible spectrum in a single exposure at a spectral resolution of $\sim5000$, or a more limited wavelength coverage in both arms at a resolution of $\sim20\,000$. The timely arrival of WEAVE at the precise moment when {\it Gaia} and LOFAR are producing their groundbreaking results guarantees its utility to the community in the coming half-decade and beyond.\looseness-1

The design of WEAVE was driven by the fact that neither {\it Gaia} nor LOFAR provide all of the information required to fully exploit the data they produce. {\it Gaia} produces radial velocities mostly for sources with magnitude $G<14.5$ at present \citep[Gaia DR3;][]{2023A&A...674A...5K} and $G\sim 16-17$\footnote{This $G$ magnitude range corresponds to the $G_{\mathrm{RVS}} \leq 16$ limit expected \citep{2023A&A...674A...6S}.} by the end of the {\it Gaia} mission. In both cases, these limiting magnitudes are where the {\it Gaia} completeness is reasonably high, while measurements are still available in a tail of fainter stars. These limits are also far brighter than the astrometric magnitude limit of $G<20-21$ in both cases. Similarly, although the exquisite low-frequency continuum sensitivity of the LOFAR Surveys Key Science Project \citep{rottgering11} observations in the LOFAR Two-metre Sky Survey \citep[LoTSS; e.g.][]{shimwell17,shimwell19,2022A&A...659A...1S,2021A&A...648A...1T} enables the detection of extremely faint radio sources, the redshift (i.e. distance) information that is necessary to distinguish between their apparent and intrinsic properties must be obtained via other means. Extensive optical spectroscopy over a large contiguous wavelength range is the ideal way of achieving this.\looseness-1

Furthermore, other pathfinder telescopes for the Square Kilometer Array,\footnote{\url{http:///www.skatelescope.org}} such as the Apertif focal-plane array system \citep{2008AIPC.1035..265V, 2022A&A...658A.146V} on the Westerbork Synthesis Radio Telescope, will reveal the neutral hydrogen sky through wide-field surveys not previously possible. These surveys are also incomplete, as connecting the \HI\ sky -- inherently three-dimensional (positions plus radial velocity are simultaneously observed) -- to the optical sky requires three-dimensional optical data. Obtaining such data is now possible through large-field and/or multiple integral field units \citep[e.g.][]{2006PASP..118..129K,2015AJ....149...77D}. An instrument with both kinds of IFU capabilities, capable of targeting large and multiple small targets, is necessary for the full exploitation of upcoming \HI\  surveys.\looseness-1

The following eight independent surveys will be carried out with WEAVE over a period of (at least) five years, producing more than 30 million spectra of nearly 10 million objects:
\begin{enumerate}
\item a survey of the Milky Way galaxy, providing radial velocities and stellar abundances for stars too faint for these quantities to be measured by {\it Gaia} (WEAVE Galactic Archaeology; Section~\ref{subsec:GA});\looseness-1
\item a survey characterizing young and massive stellar populations and the interstellar medium, and thus probing star formation and evolution, across the Galactic Plane (Stellar, Circumstellar and Interstellar Physics; Section~\ref{subsec:SCIP});
\item a survey studying the death of stars and constraining the local star-formation history of the Galaxy through its white dwarf population (WEAVE White Dwarfs; Section~\ref{subsec:WDs});\looseness-1
\item a survey of the stellar and gaseous kinematics and physical properties of gas-rich galaxies, providing a necessary optical complement to Apertif's neutral hydrogen surveys of the local Universe (WEAVE-Apertif; Section~\ref{subsec:WA});\looseness-1
\item a survey probing the evolution of galaxies as a function of environment, from the cores of rich clusters to their lively environs, going from their smallest members out to the field at cosmological distances (WEAVE Galaxy Clusters; Section~\ref{subsec:WC});\looseness-1
\item a survey providing the first detailed view of the stellar population properties of galaxies at $z=0.3$--$0.7$ as a function of galaxy mass and environment,
yielding star formation histories, stellar ages, stellar and gas metallicities, dust attenuation, gas kinematics, and stellar velocity dispersions (Stellar Populations at intermediate redshifts Survey; Section~\ref{subsec:StePS});\looseness-1
\item a survey probing galaxy evolution over cosmic time, providing much-desired redshifts and galaxy properties of LOFAR's radio sources (WEAVE-LOFAR; Section~\ref{subsec:WL});\looseness-1
\item a survey of large-scale structure using quasar absorption lines as a cosmic ruler to probe the expansion of the Universe, which also extends the study of gaseous environments to larger scales and earlier epochs (WEAVE-QSO; Section~\ref{subsec:WQ}).\looseness-1
\end{enumerate}
The WEAVE Survey will provide data that will help answer the questions: How did our Galaxy form and the stars within it evolve? How were other galaxies assembled? What are dark matter and dark energy? On these topics, WEAVE will be complementary to the surveys carried out by 4MOST \citep{2019Msngr.175....3D}, which has a similar design, including spectral range and resolution, and will operate in the Southern Hemisphere on a similarly sized telescope, DESI \citep{2016arXiv161100036D,2016arXiv161100037D}, which will focus largely on baryonic acoustic oscillations of galaxies, SDSS-V \citep{2017arXiv171103234K}, which will have complementary, lower resolution optical spectra and, additionally, infrared spectroscopy,
and PFS \citep{2018SPIE10702E..1CT}, which will cover a smaller field on a larger telescope that is not fully dedicated to survey operations.\looseness-1

This paper first presents the design of the WEAVE instrument -- in terms of both hardware and software -- as relevant to WEAVE Survey preparation and execution (Section~\ref{sec:instr}), followed by the structure of the WEAVE work and workforce including an overview of the key cross-team working groups involved in and responsible for the preparation and execution of the Survey (Section~\ref{sec:Consortia}). We then present overviews of each of the eight component surveys that comprise the WEAVE Survey (Section~\ref{sec:Survey}), with each subsection summarizing the particular survey's science case and survey plan, as well as providing the relevant introduction and background to the field. This is followed by a description of the WEAVE Simulator (Section~\ref{sec:sim}) and a summary of the simulated implementation of the Survey through `Operational Rehearsals' (Section~\ref{sec:OpRs}). Section~\ref{sec:summary} concludes the paper.\looseness-1

\section{The WEAVE Facility}
\label{sec:instr}

The WHT is an alt-azimuth telescope with a Cassegrain focus and two Nasmyth foci \citep[for a description of the original design features of the WHT, see][]{1985VA.....28..531B}. WEAVE's predecessor on the WHT was the prime-focus instrument Autofib-2 \citep[AF2;][]{1994SPIE.2198..125P}, which had 150 fibres deployable over a 1-degree-diameter field of view. WEAVE has nearly 1000 fibres deployable over a field of view four times that of AF2 and, with four times the number of resolution elements of AF2 and three to four times its throughput, dramatically increases the resolution and multiplex power of the telescope.

Full details of the WEAVE design and performance will be presented in a separate paper (Dalton et al., in preparation). Here, we summarize only the key parameters of the WEAVE facility as they relate to the detailed design of the component surveys making up the total WEAVE Survey \citep[cf.][]{2016SPIE.9908E..1GD}.\looseness-1

\subsection{A new top-end for the WHT}
\begin{figure}
	\includegraphics[width=\columnwidth]{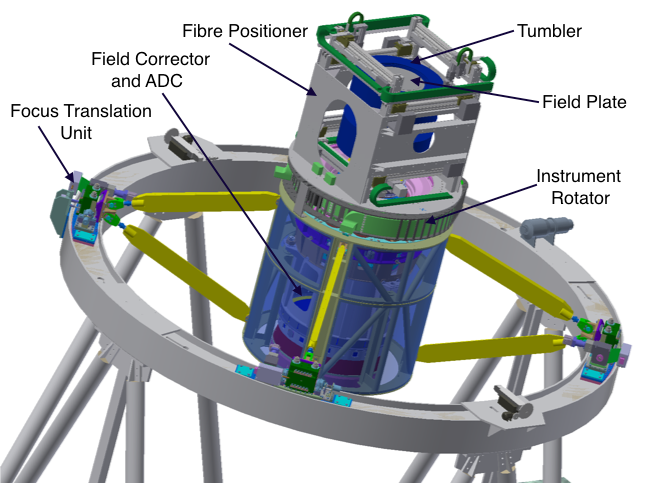}
    \caption{A computer-aided diagram (CAD) representation of the WEAVE top-end assembly, incorporating the prime-focus (field) corrector system, instrument rotator and fibre positioner. The four mounting units on the outer ring of the telescope structure provide for focus and tilt correction of the whole system.}
    \label{fig:TopEnd}
\end{figure}

As part of the new top-end for the WHT (Fig.~\ref{fig:TopEnd}), the WEAVE prime-focus corrector system delivers a 2-degree-diameter field of view, with a flat focal plane at $f/2.78$ with full correction for atmospheric dispersion from 370--1000$\,\mathrm{nm}$ to $65^\circ$ zenith distance \citep{2014SPIE.9147E..73A}. The telecentricity of the input beam will deviate by no more than $4^\circ$ at the edge of the field. For a point-source image in the absence of seeing, the corrector delivers 80 per cent encircled energy within 0.71\,arcsec, corresponding to a full width at half maximum (FWHM) of 0.41\,arcsec. The effect of differential image distortion with changing zenith distance is 0.13\,arcsec across the field between $0^\circ$ and $65^\circ$ zenith distance. \looseness-1

\subsection{The WEAVE fibres and fibre positioner}
\label{subsec:FIB+POS}

\begin{figure}
	\includegraphics[width=\columnwidth]{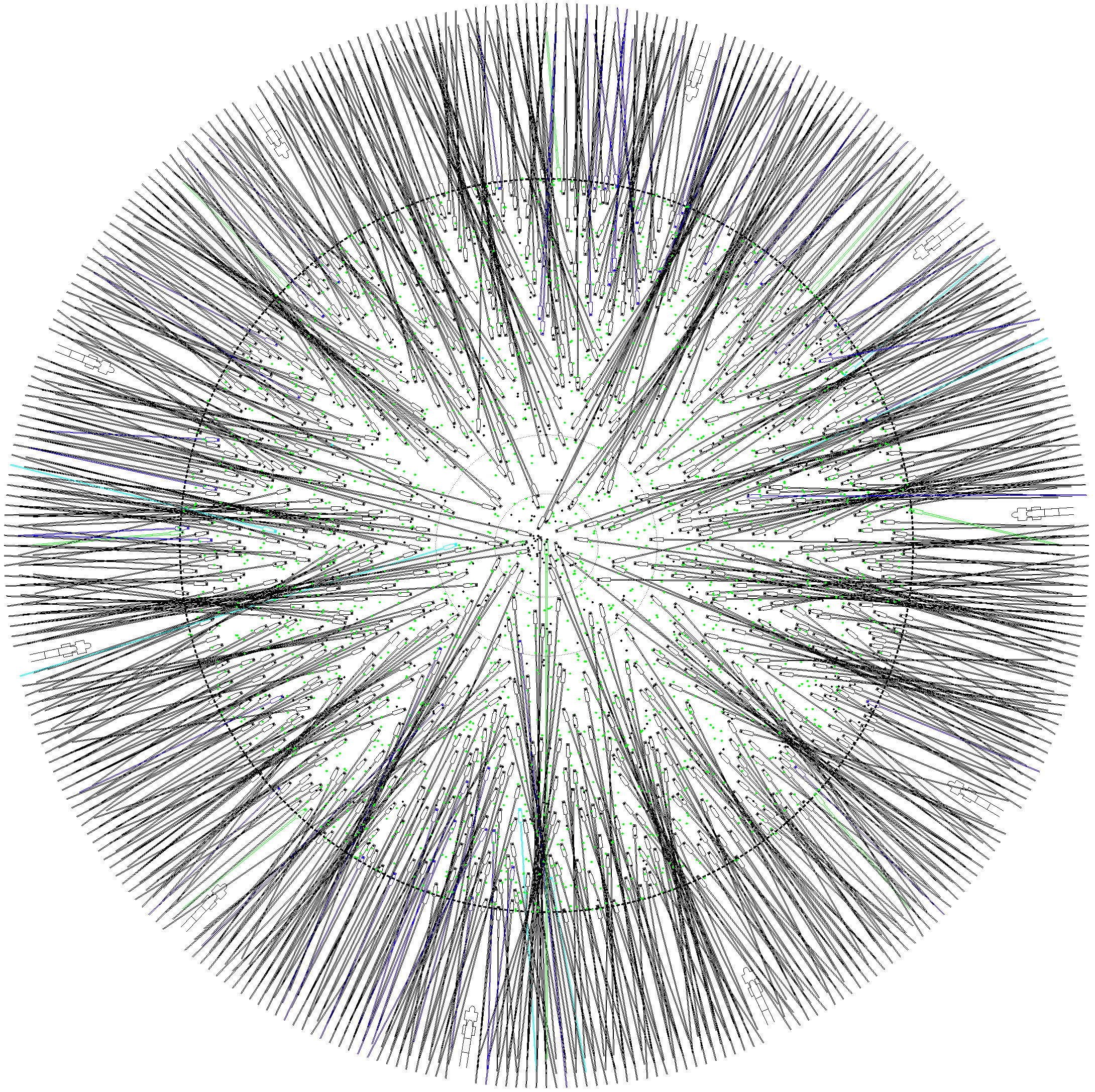}
    \caption{Configured field for OB \#3182 from OpR3b (see Section~\ref{sec:OpRs}) as an example of a representative WEAVE-Survey MOS field. The black, dashed, outermost circle indicates WEAVE's 2-degree-diameter field of view. MOS fibres and targets are colour-coded as follows -- black: science (821); cyan: calibration stars (19); blue: blank sky (100); green: guide stars (8). The two faint inner circles denote the maximum extent of the outer two tiers of MOS fibres. As this particular field is a configuration on plate B, the `park' locations of the 10 pairs of mIFU bundles can also be seen along the periphery. Note that the mIFUs cannot be used concurrently with the MOS fibres.\looseness-1}
    \label{fig:configure}
\end{figure}

\begin{table*}
\caption{Key parameters of the four focal-plane modes of WEAVE. MOS-B and mIFU modes both use plate B, but cannot be employed simultaneously.}\label{table:FOC_modes}
\begin{tabular}{lllll}
\hline
Focal-plane mode & MOS fibres (MOS-A) & MOS fibres (MOS-B) & mini-IFUs (mIFU) & Large IFU (LIFU)\\\hline
Tumbler position & 0\degr & 180\degr & 180\degr & 90\degr\\
Diameter of individual fibres & 1.3\,arcsec\;(85\,\um) & 1.3\,arcsec\;(85\,\um) & 1.3\,arcsec\; (85\,\um) & 2.6\,arcsec\;(170\,\um)\\
Multiplex & 960 fibres & 940 fibres & 20 IFUs & 1 IFU\\
Diameter of field over which & 2\degr & 2\degr & 2\degr & on axis\\
deployable & & & & \\
Minimum separation on sky & $\sim$60\,arcsec & $\sim$60\,arcsec & $\sim$60\,arcsec & --\\
Fibres per IFU & -- & -- & 37 & 547 \\
IFU field of view & -- & -- & 11\,$\times$\,12\,arcsec${^2}$ & 90\,$\times$\,78\,arcsec${^2}$ \\
IFU filling factor & -- & -- & 0.50 & 0.55 \\
Fibres for auto-guiding & 8\,$\times$\,3\,arcsec${^2}$ coherent bundles & 8\,$\times$\,3\,arcsec${^2}$ coherent bundles & 8\,$\times$\,3\,arcsec${^2}$ coherent bundles & Separate camera\\
& & & &  (4\,$\times$\,3.7\,arcmin${^2}$ field of view)\\ 
Fibres for sky subtraction & $\sim$5--10$\%$ of science fibres & $\sim$5--10$\%$ of science fibres & one of the mIFUs & 8 peripheral bundles \\
&&&&of 7 fibres each\\
Configuration time & $\sim55$ min & $\sim55$ min & <20 min & $\sim 1$ min\\
\hline
\end{tabular}
\end{table*}

Behind the prime-focus corrector system, fibres are deployed in the focal plane using a buffered positioner system developed from the 2dF concept \citep{2002MNRAS.333..279L}: the positioner can deploy up to 1008 fibre buttons on each of two field plates using a pair of Cartesian robots sharing a common $x$-axis. Each science-fibre button is $5\,\mathrm{mm}\times2\,\mathrm{mm}\times7\,\mathrm{mm}$ in size and terminates in a $1.5\,\mathrm{mm}\times1.5\,\mathrm{mm}$ prism with a concave upper surface to expand the telescope beam to $f/3.2$. Each multi-object spectrograph (MOS) science fibre carries a $85\,\um$ core fibre that subtends 1.3\,arcsec on the sky. Eight buttons on each field hold coherent fibre imaging bundles to be used for acquisition and guiding. On the first plate (`plate A'), each of the eight guide-fibre bundles is packaged so that each bundle replaces six potential science fibres, leaving 960 science fibres available.
On the second plate (`plate B'), 10 groups of six science fibres are replaced by pairs of deployable mini integral field units (mIFUs). On plate B, the guide-fibre bundles each replace only a single science fibre to leave 940 science fibres available. The two MOS plates are at either end of the positioner, such that when one plate is being used for the current observation, the other can be set up for the upcoming observation. A MOS field can be configured in under one hour, thus setting the length of a typical WEAVE observation. Once the current observation has finished, the positioner tumbles to reverse the roles of the two plates \citep{2014SPIE.9147E..34L}. This results in a marked improvement in observing efficiency as compared with AF2, as there will, in principle, be no loss in observing time due to field configurations when switching between the two MOS plates for observations with the MOS or mIFU modes with WEAVE. Table~\ref{table:FOC_modes} summarizes the key parameters of the four focal-plane modes of WEAVE. \looseness-1  

The 20 mIFUs are packaged similarly to the MOS fibres, but with slightly larger buttons ($7\,\mathrm{mm}\times5\,\mathrm{mm}\times7\,\mathrm{mm}$ in size), each holding a hexagonal close-packed array of 37 fibres with the same size and aperture as the MOS fibres. The outer diameter of the fibre buffer layer is $120\,\um$, so each mIFU covers roughly 11\,$\times$\,12\,arcsec$^2$ on the sky with a filling factor of $\approx0.50$. The buffered arrangement of the two field plates also affords an intermediate location (i.e.\ at 90$^{\circ}$ to the MOS plates), which accommodates a single, large integral field unit (LIFU) containing a hexagonal array of 547 fibres with $170\,\um$ cores 2.6\,arcsec, subtending 90\,$\times$\,78\,arcsec$^2$ on the sky with a filling factor of $\approx0.54$, along with eight peripheral bundles of seven fibres, arranged in a ring of radius 150\,arcsec from the centre of the array, to be used for sky subtraction. The LIFU package includes a discrete guiding camera mounted within the same physical structure as the LIFU head.\looseness-1

Each MOS observation is prepared in advance using a purpose-built software tool \citep[{\configure}:][see Fig.~\ref{fig:configure}]{2014SPIE.9152E..0PT}, which builds on the simulated annealing approach of \cite{2006MNRAS.371.1537M} with some elements of the original `Oxford' algorithm \citep{2001MNRAS.328.1039C} to facilitate an optimal, on-the-fly assignment of fibres for sky-background determination without impacting the results of the annealing process. The output of \configure\ is an XML file containing all of the information required to specify an observation, including observing constraints, spectrograph configuration, and calibration lamp sequences. Observations for the mIFU and LIFU modes follow a similar but slightly simpler approach. Using {\configure}, the mIFUs in a given field can be manually allocated to targets such that the central fibre of each mIFU is located at a desired target position. Final sky coordinates for the remaining 36 fibres are output by {\configure}, based on the sky position angle of the mIFU field resulting from this allocation. Each mIFU-field setup  requires at least one of the mIFUs to be dedicated for calibration purposes. The central seven fibres are identified with the primary calibration target (i.e.\ a white dwarf), and the remaining 30 fibres are labelled as sky positions.
LIFU observations are prepared using a similar tool, with the output XML file providing information on guide-star positions and the precise on-sky position for each fibre within the LIFU.\looseness-1

\subsection{The WEAVE spectrograph}
\label{subsec:SPE}

\begin{figure}
	\includegraphics[width=\columnwidth]{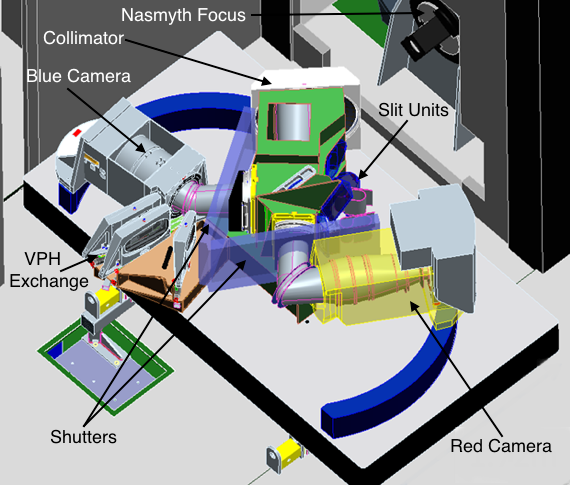}
    \caption{A CAD representation of the WEAVE spectrograph installed in the GHRIL instrument room on the WHT Nasmyth platform. The spectrograph cameras are pictured in a configuration unused in practice, with the blue camera at the LR position and the red camera at the HR position. As a scale reference, the optical bench is approximately 3-m long.}
    \label{fig:Spectrograph}
\end{figure}

All four sets of science fibres pass through a cable-wrap around the prime-focus instrument rotator, along the vanes of the prime-focus support spider and down the telescope structure, to where they pass over the elevation axis and into the GHRIL instrument room\footnote{\url{https://www.ing.iac.es//PR/inst.php?tel=wht&inst=GHRIL}} on the Nasmyth platform that houses the WEAVE spectrograph (see Fig.~\ref{fig:Spectrograph}).

The spectrograph \citep{2014SPIE.9147E..6HR} provides full wavelength coverage over the range 366--959$\,\mathrm{nm}$ at a nominal resolving power of $R\sim5000$ (the low-resolution, `LR', mode), or a pair of restricted wavelength ranges at $R\sim20\,000$ (the high-resolution, `HR', mode; note that these resolving powers are halved when using the LIFU with its two-times-larger fibres). This is achieved by a dual-beam design with an off-axis catadioptric\footnote{A catadioptric optical system combines both reflection and refraction to reduce aberration effects.} $f/3.1$ collimator, which delivers a 190$\,$mm collimated beam. The dichroic is located before the collimator correcting lenses so that these can be optimized for transmission in each arm. Each arm feeds an $f/1.8$ transmissive camera consisting of 8 lenses (one asphere on the first lens), designed such that the lens prescriptions are common to both cameras, and only the coatings and lens spacings differ between the two. The useful focal plane of each camera is 8192 (spectral) $\times$ 6144 (spatial) pixels, populated by a pair of e2V CCD 231-C6 CCDs, with the physical size of the pixels being 15~$\um$. This oversize of the detector avoids use of a custom format, and also allows the gap in the spectral direction to be chosen in each camera to fall in a region of low interest. At the same time, placing this gap close to the centre of the image plane makes the control of Littrow ghosts easy by using VPH gratings with slightly slanted fringes \citep{2007PASP..119.1069B,WEAVE.VPHGs}, such that the image of the ghost is formed in the gap. Each arm of the spectrograph contains a rolling shutter mechanism and a grating exchange mechanism to allow the choice between two resolution modes, with the associated change of beam angles accommodated by articulation of the two cameras. The available modes and associated nominal spectral coverage are listed in Table~\ref{table:SPE_modes}.\looseness-1

\begin{table}
\centering
\caption{Available resolution modes (LR = low resolution, HR = high resolution) and nominal useful spectral coverage for the WEAVE spectrograph in the MOS/mIFU modes, including the wavelength gap between the chips.}
\label{table:SPE_modes}
\begin{threeparttable} 
  \begin{tabular}{lrccc}
    \hline
    Spectrograph & \multicolumn{1}{|c|}{$R$\tnote{\textit{a}}} & $\lambda_\mathrm{low} $(\AA)  & $\lambda_\mathrm{high}$ (\AA) & gap in $\lambda$\\
    mode & & & &  coverage (\AA)\\\hline
    LR Blue & 5\,000 & 3660 & 6060 & 5491--5539 \\
    LR Red & 5\,000 & 5790 & 9590 & 7590--7669 \\
    HR Blue-1\tnote{\textit{b}} & 20\,000 & 4040 & 4650 & 4525--4536 \\
    HR Blue-2\tnote{\textit{c}} & 20\,000 & 4730 & 5450 & 5302--5315 \\
    HR Red & 20\,000 & 5950 & 6850 & 6412--6431 \\ \hline
\end{tabular}
\begin{tablenotes}
\item[\textit{a}] Due to the larger fibre cores of the LIFU, the resolution in this mode is halved with respect to the MOS/mIFU cases.
\item[\textit{b}] `Blue' HR grating.
\item[\textit{c}] `Green' HR grating. Note that the two HR Blue modes cannot be used simultaneously.
\end{tablenotes}
\end{threeparttable}
\end{table}

Direct illumination of each camera's focal plane is achieved by a series of LEDs that are mounted on the spectrograph's shutter mechanisms. These are used to provide a direct measure of the pixel-to-pixel response variations of the detector, and a measure of the non-linearity present in the detector read-out electronics.\looseness-1

\subsection{Calibration unit}
\label{sec:calibs}
The WHT has a set of projector lamps mounted at one of the broken-Cassegrain foci. These can be used to direct light to the primary focus via the tertiary mirror. Investigations during the development of WEAVE have shown that the illumination is adequate -- but not perfect -- out to the edge of the WEAVE field of view \citep{2016SPIE.9908E..8RD}. Observations of a quartz-halogen lamp are used to determine the traces of the individual fibre spectra on the detector mosaic for each camera, and to provide a rough indication of the illumination variations for each fibre. Observations of a ThArCr lamp are used to determine a wavelength solution for each fibre. With this lamp, we find that it is possible to obtain a sufficient density of lines in both low- and high-resolution modes.\looseness-1

Mapping of the wings of the fibre profiles on the detector is achieved by configuring three fields for each plate, each sampling every third fibre along the slit and observing high signal-to-noise arc and flat-field lamp exposures. As the fibres in these configurations can all be placed on a circle at around a 20-arcmin radius, these observations can also be used to determine fibre-to-fibre intrinsic transmission variations.\looseness-1

Flux calibration is achieved on-sky using calibrator stars within each WEAVE field (Section~\ref{sec:fluxcal}).\looseness-1

\subsection{Observatory Control System (OCS)}
\label{subsec:OCS}
WEAVE introduces a new top-level OCS to the WHT that permits queue-based observations \citep{2018SPIE10704E..2AP}. The XML files generated by \configure\ are converted to Observation Blocks (OBs), which are stored in a database and can be scheduled in a queue for the operator, based on the prevailing conditions and target priorities. At the heart of the OCS is a sequencer task that coordinates the actions of the telescope, spectrograph, positioner, and calibration system using a CORBA-based\footnote{\url{https://www.corba.org}} architecture and a centralised noticeboard. In this context, the positioner has two functions: one for the observation of the current field and performing any actions required of the focal-plane imager, and one for the positioning robots concerning the parallel setting-up of the next observation.\looseness-1

The `WEAVE observation queue scheduler' \citep[\scheduler:][]{2018SPIE10704E..0WF} is bespoke software written for use by the WEAVE operator to make informed decisions about which OBs in the database are most suitable for observation, at least one hour ahead of time, given the typical observation length (which in turn is driven by the configuration time of a MOS field). Requirements set by the OBs on instrument configuration (i.e.\ observing mode) and observing conditions (e.g.\ sky brightness, seeing, etc.) are combined with current and predicted weather conditions and current instrument configuration to allocate scores to OBs in the OB database. For example, poor seeing conditions might trigger a LIFU observation, during which no configuration is possible on either field plate. If this occurs during the night, the most efficient subsequent observation, once the seeing conditions start improving, is a mIFU field using plate B, as configuring a mIFU field takes less time (under 20 min) compared with a MOS field (just under one hour), during which plate A can be set up for a MOS observation. If such a LIFU observation takes place at the start of the night, both field plates will already have been set up with suitable fields before the beginning of the night. Note that each MOS field configuration is valid only for a specific length of time, depending on its elevation, due to the distortion introduced by the effects of differential refraction.

\subsection{WEAVE Data Processing and Archiving}
\label{subsec:SPA}
From the outset, it was envisaged that WEAVE would require a dedicated data-processing pipeline to ensure the best outcomes from the facility. Given the scale and complexity of this task, the activities of this Science Processing and Analysis (SPA) system have been divided into three distinct sub-systems: the Core Processing System (CPS) for basic data reduction and calibration, the Advanced Processing System (APS) for producing derived data products, and the WEAVE Archive System (WAS) for data archiving.\looseness-1

All WEAVE data frames, including data for individual open-time programmes, are retrieved from the observatory in near real-time by the CPS \citep{2014SPIE.9152E..0RW} and processed to remove instrumental signatures, calibrate wavelength and flux scales (see Section~\ref{sec:fluxcal}), and subtract sky background. A subset of the CPS pipeline will run at the telescope to provide a quick-look analysis tool \citep{2019hsax.conf..608P} for the observers for the purposes of quality control during observations.
For each OB, the output from the CPS consists of a multi-spectrum image for each science frame in each arm (with sky and variance components) and a stacked multi-spectrum image for each arm. These data are passed to the WAS, together with the raw frames and processed calibration data.\looseness-1

The APS picks up the processed science frames and produces joined spectra in both the low and high-resolution modes, before providing an analysis of each spectrum using a variety of methods \citep[e.g.\ \textsc{ferre}, \texttt{redrock},\footnote{\url{https://github.com/desihub/redrock}} \textsc{ppxf}, and \textsc{gandalf}:][the last two of which are wrapped in \textsc{gist}: \citealt{2019ascl.soft07025B,2019A&A...628A.117B}]{2006ApJ...636..804A,2010PASP..122..248B,2012ascl.soft10002C, 2017ascl.soft08012S}. For extragalactic sources, the analysis yields redshift, galaxy classification, velocity dispersion, spectral indices, and equivalent widths for a list of standard emission lines. For each stellar source, the analysis produces velocity, temperature, surface gravity, and abundance measures for a variety of elements. For IFU spectra, the APS produces maps of velocity, velocity dispersion, and abundances. The APS also allows for additional analysis software (`Contributed Software') to be developed by the science teams and incorporated into the processing chain. All data products produced by the APS are stored in FITS tables and sent to the WAS.\looseness-1

The WAS is the official archive for WEAVE data products \citep{2016SPIE.9913E..1XG}. It is based on Apache Solr\footnote{\url{http://lucene.apache.org/solr/}} and provides interactive front-end access to WEAVE data and data products for WEAVE data users. The WAS not only stores the results from WEAVE's two main pipelines, namely the CPS and the APS, but also stores Contributed Data Products (CDPs), which are specific data products produced by WEAVE Science Team members in addition to the APS products. To preserve and ensure the high quality of data available to WAS users, CDPs are agreed in advance by the Science Team with the Science Executive (see Section~\ref{sec:Consortia}). As well as search and download options, the WAS user interface also offers plotting tools for the visualisation of spectra and other data products. The WAS manages both internal and public data releases. Internal data releases are expected on a regular basis, at least once a semester, while public data releases will be yearly following the first public data release two years after the start of Survey operations. Data from the public data releases will also be published to the Virtual Observatory\footnote{\url{https://ivoa.net/}} using the Table Access Protocol.\footnote{\url{https://www.ivoa.net/documents/TAP/20190927/REC-TAP-1.1.html}}\looseness-1

\section{WEAVE's Organizational Structure}
\label{sec:Consortia}
\begin{figure}
    \centering
    \includegraphics[width=\columnwidth]{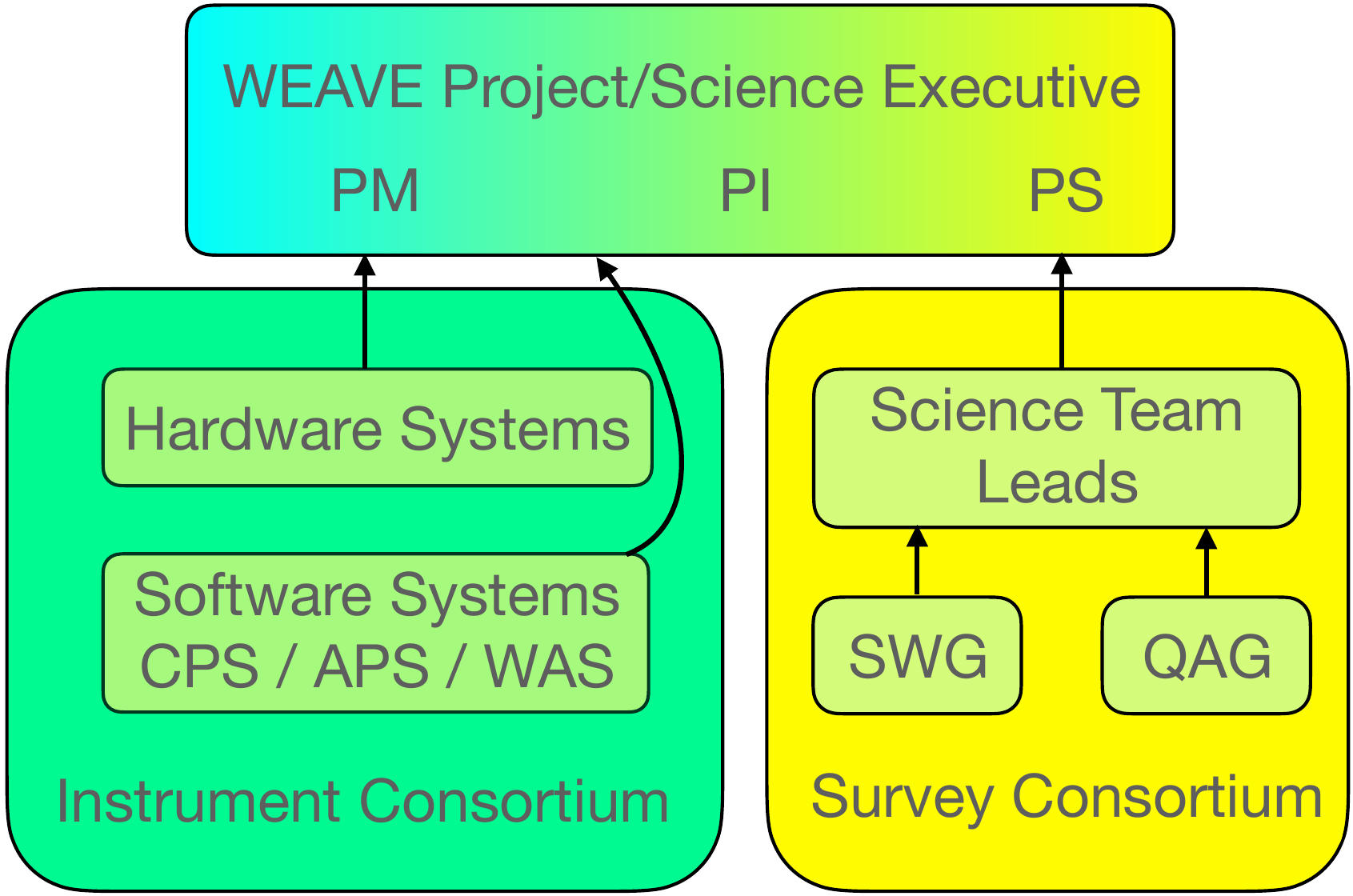}
    \caption{A simplified version of the WEAVE Project Management Team structure (the `WEAVE organogram'). Arrows point in the direction of reporting. PM: Project Manager; PI: Principal Investigator; PS: Project Scientist; CPS: Core Processing System; APS: Advanced Processing System; WAS: WEAVE Archive System; SWG: Survey Working Group; QAG: Quality Assurance Group.}
    \label{fig:organogram}
\end{figure}

The work and workforce within WEAVE are divided into two entities: the WEAVE Instrument Consortium and the WEAVE Survey Consortium. The former, also commonly referred to as `the WEAVE Project', is responsible for the delivery of the WEAVE facility to the ING. The latter,
also commonly referred to as `the WEAVE Science Team', is responsible for preparing and executing a five-year survey. A simplified view of the organizational structure is shown in Fig.~\ref{fig:organogram}.

The WEAVE Project is responsible for the delivery of the complete instrument along with data-processing software and associated hardware for the data-processing pipelines, a data archive system (along with a user interface), and other useful tools such as a quick-look GUI, an operational repository, OB templates, and a file-submission platform (see Sections~\ref{sec:instr} and \ref{sec:OpRs}). The WEAVE Project is overseen by a Project Board with representatives from the funding parties, and managed through the Project Executive.\footnote{\url{https://ingconfluence.ing.iac.es/confluence/display/WEAV/Project+Executive}} The latter comprises the Project Manager and Principal Investigators (PIs) from each of the communities in WEAVE-funding countries. These PIs also currently compose the Science Executive, which oversees the Survey Consortium.\looseness-1

The WEAVE Survey Consortium -- also known as the WEAVE Science Team -- is composed of eight individual science teams, preparing to execute the eight respective surveys. Each team is led by a Science Team Lead (STL) and each survey has its own science case and survey plan. The WEAVE Survey Consortium currently consists of over 500 astronomers\footnote{\url{https://ingconfluence.ing.iac.es/confluence/display/WEAV/Science+Team}} spread over 12 countries, where membership is contingent on satisfying eligibility requirements, e.g.\ any astronomer working in an ING country may request to join the Survey Consortium. An individual joins the Survey Consortium by becoming a member of one or more of the eight science teams. WEAVE membership entitles its members to proprietary WEAVE-data access during Survey operations, while for certain science teams, this also includes ancillary data provided through a Memorandum of Understanding (MoU) with another survey. The component surveys comprising the global WEAVE Survey are described further in Section~\ref{sec:Survey}.

\subsection{Survey Working Group (SWG)}
\label{subsec:SWG}

During Survey operations, the main responsibility of the SWG is to provide catalogues of targets and configured fields in the form of XML files using \configure\ (see Section~\ref{subsec:FIB+POS}) on a trimesterly basis, i.e.\ every three months. The SWG is composed of representatives from the WEAVE Science Team, each of whom reports to the relevant Science Team Lead, and is currently led by the Project Scientist and Deputy Project Scientist. The SWG has been fundamental in helping to define the structure and details of key data-flow processes, testing many steps of the Survey preparation process from the creation of input target catalogues and testing of the \configure\ software for creating configured fields to trialling the trimesterly XML submission process via the WEAVE Automated Submission Platform (WASP) in the Operational Rehearsals (see Section~\ref{sec:OpRs}).\looseness-1

\subsection{On-Island Survey Management Team (OISMT)}
\label{subsec:OISMT}
The OISMT at the ING is responsible for many aspects of the running of the WEAVE facility, ranging from the commissioning of the WEAVE instrument to ensuring the day-to-day running of WEAVE in the contexts of both the WEAVE Survey and PI programmes\footnote{WEAVE will be available for community PI programmes in the same manner as for instruments on other ING telescopes, via the usual ING call for proposals. Further details can be found at \url{https://www.ing.iac.es/astronomy/observing/INGinfo_home.html}} once the instrument is on sky. With support from the OISMT, a dedicated WEAVE instrument specialist will be responsible for managing the execution of nightly OBs, aided by the observation queue created by the \scheduler\ based on current and predicted weather conditions and the observing requirements of OBs in the OB database, held at the OCS. Regular communication between the OISMT and the SWG is key for ensuring the smooth and optimized scheduling of OBs from the WEAVE Survey and PI programmes combined, and this interface has been tested extensively during the most recent Operational Rehearsal (see Section~\ref{sec:OpRs}).\looseness-1

\subsection{Quality Assurance Group (QAG)}
\label{subsec:QAG}

Although various checks are performed on the spectra being processed by the CPS and APS before reaching the WAS, some unexpected results requiring closer inspection are anticipated, especially at the start of Survey operations. To allow such spectra to be caught and investigated quickly, the QAG of each science team will perform regular checks on data taken for their survey to ensure that the spectra are of sufficient quality.\looseness-1

The work of the QAG will also be instrumental in decisions to be made on timescales of several months to a year. Although each of the eight science teams comprising the global WEAVE Survey has an internally and externally reviewed five-year survey plan based on its individual science case prior to start of Survey operations, the detailed planning of which fields/targets are submitted for observations (and when) is a real-time decision-making operation that will continue to take place on a trimesterly cycle. The QAG will report its findings of their quality-assessment work to the STL and SWG member(s) of their team on a regular basis. This, in turn, will form the basis of decisions made by the STL and SWG on any necessary strategic planning for upcoming submissions on how to proceed with their survey. Reports from the QAG will also be combined with individual survey completion rates and other considerations such as right-ascension pressure for top-level Survey-planning decisions to be made by the Science Executive.\looseness-1

The QAG workflow and its tasks have been tested intensively during Operational Rehearsals involving the WEAVE Science Team (see Section~\ref{sec:OpRs}).\looseness-1

\subsection{IFU Working Group (IWG)}
\label{subsec:IWG}

The IWG deals with planning observational and software aspects of WEAVE's integral-field modes. The IWG consists of members of the software systems, SWG, QAG, and ad-hoc members of the science teams as required, and discusses issues ranging from observation preparation to dithering strategies, data reduction, and analysis techniques. Given the complexity of designing IFU observations within the context of the WEAVE observation preparation system, the IWG has been instrumental in the development and testing of the `IFU workflow' software for preparation of IFU observations, which in turn has also significantly contributed to the parallel development of the `MOS workflow' software for preparation of the multi-object-spectroscopy observations. During Survey operations, the work carried out by the IWG will serve as additional input to the SWG and the QAG of science teams that include IFU observations.\looseness-1

\section{The WEAVE Survey}
\label{sec:Survey}

\begin{table*}
\caption{WEAVE survey parameters, listed in order of survey appearance in the text. Numbers provided are best estimates at the time of writing for a seven-year WEAVE Survey.}
\label{table:WEAVE_surveys}
\begin{threeparttable}
\begin{tabular}{llrrrllr}
\hline
WEAVE survey & main targets & \multicolumn{1}{c}{number of} & \multicolumn{1}{c}{area} & Survey & observing & resolution & \multicolumn{1}{c}{redshift} \\
(short-hand) &  & \multicolumn{1}{c}{objects} & \multicolumn{1}{c}{(deg$^{2}$)} & fraction\tnote{\textit{a}} & mode(s) & modes(s) &  \\
\hline 
Galactic Archaeology & MSTO\tnote{\textit{b}}\, stars, RGB\tnote{\textit{b}}\, stars,& $\sim1.6\times10^6$  & 8750 & 0.168 & MOS & LR & 0 \\
\hspace{0.2cm} (GA-LRhighlat) &  \quad BHB\tnote{\textit{b}}\, stars, RR Lyrae &  &  &  & & \\
Galactic Archaeology & Red Clump stars,& $\sim1.1\times10^6$ & 1370 & 0.110 & MOS & LR &  0 \\
\hspace{0.2cm} (GA-LRdisc) & \quad RGB stars  &  &  &  & & \\ 
Galactic Archaeology & Main sequence \& & $\sim1.6\times10^6$ & 5650 & 0.309 & MOS & HR &  0 \\
\hspace{0.2cm} (GA-HR) & \quad RGB stars  &  &  &  & & \\
Galactic Archaeology &  Stars in open clusters & $\sim1\times10^5$ & 375 & 0.029 & MOS & HR & 0 \\
\hspace{0.2cm} (GA-OC) & \quad and star forming regions &  &  &  & & \\
Stellar, Circumstellar and & OBA stars, ionized nebulae, & $\sim4\times10^5 $ & 1230 & 0.069 & MOS, LIFU & LR, HR & 0 \\
\hspace{0.2cm} Interstellar Physics (SCIP) & \quad young stars, compact objects &  &  &  & & \\
White Dwarfs &  white dwarfs & $\gtrsim5\times10^4$ & $\gtrsim10\,000$ & $\sim0.012$ & MOS, mIFU\tnote{\textit{c}}, & LR, HR & 0 \\
&  &  &  &  & \quad LIFU\tnote{\textit{c}} & &\\
WEAVE-Apertif & \HI-detected, mostly late-type & 400 (LR), & 0.2\tnote{\textit{d}} & 0.061 & LIFU &  LR, HR & $< 0.04$ \\
& \quad galaxies & 100 (HR) & & & & &\\
WEAVE Galaxy Clusters & galaxies in dense environments & $\sim2\times10^5$ & 1350 & 0.064 & MOS, mIFU, LIFU & LR & $< 0.5$ \\
Stellar Populations at intermediate & field galaxies  &$\sim2.5\times10^4$  & 25 & 0.026 & MOS & LR & 0.3--0.7 \\
\hspace{0.2cm} redshifts Survey (StePS) &  &  &  &  &  & &\\
WEAVE-LOFAR & 150\,MHz sources & $\sim7\times10^5$ & 8950 & 0.109 & MOS, mIFU, LIFU & LR & $<6.9$ \\
WEAVE-QSO  & bright, $r<21.5$: $z>2.2$;  & $\sim4\times10^5$ & 8950 & 0.056 & MOS & LR, HR & $>2.2$ \\
& $21.5<r<23.5$: $2.5<z<3$\tnote{\textit{e}} & &  &  &  & &\\
\hline
\end{tabular}
\begin{tablenotes}
\item[\textit{a}]`Survey fraction' denotes the fraction of total WEAVE-Survey fibres hours currently planned for use for a given WEAVE survey. 
\item[\textit{b}]MSTO: Main Sequence Turnoff; RGB: Red Giant Branch; BHB: Blue Horizontal Branch.
\item[\textit{c}]White dwarfs used purely for IFU-mode calibrations.
\item[\textit{d}]Area coverage calculated using the field of view of the LIFU, not the field of view of WEAVE.
\item[\textit{e}]Applies to WQ-Wide (when J-PAS target selection is available). WQ-HighDens will target $r<23.5$ for $z>2.2$ over $\sim$420\,deg$^2$.
\end{tablenotes}
\end{threeparttable}
\end{table*}

The WEAVE Survey is composed of eight individual surveys, each with its own dedicated science team, whose science goals span a wide range of topics covering various fields of stellar, Galactic, and extragalactic astronomy. The surveys will together use approximately 1150 nights over the course of five years of WHT time. Their science goals play crucial roles in complementing major space- and ground-based programmes in the current and coming decade, including {\it Gaia}, LOFAR, and Apertif, by providing a dedicated wide-field optical spectroscopic instrument in the Northern Hemisphere. WEAVE will also be accessible to the wider astronomical community through open calls outside of the allocated survey time.

In this section, an overview of each of the planned surveys of the global WEAVE Survey is provided, starting from surveys focusing on the nearby Universe and extending towards higher redshifts (see also Table\,\ref{table:WEAVE_surveys}).\looseness-1

\subsection{The WEAVE Galactic Archaeology (GA) Survey}
\label{subsec:GA}

How did our Galaxy and its components assemble and evolve over time? This question is key to the general understanding of galaxy formation over cosmic times, as the Milky Way is the system for which we can hope to constrain in most detail the physical processes that play a role in the formation and evolution of galaxies. These processes leave behind specific signatures that are encoded in the location, kinematics, and chemistry of stars \citep[e.g.][]{2002ARA&A..40..487F}. The ultimate goal of Galactic Archaeology as a field of study is to obtain a comprehensive census of the positions, orbits, ages, and chemical compositions of stars in all major stellar structures of our own Galaxy to enable a complete reconstruction of its formation and subsequent evolution.\looseness-1

Many questions remain open in understanding how the complex structure of the Milky Way was assembled, with its stellar populations occupying a bulge, a thin and thick disc, and a halo. Whether the different stellar populations have truly different origins, what the relative importance is of internal ({\em in-situ} star formation, secular evolution of the disc, etc.) and external (accretion events, gas accretion through filaments, etc.) processes in forming and shaping the Galaxy, and how universal these processes are in the evolution of galaxies in general, are questions to which Galactic Archaeology as a field of astronomical study tries to find answers. The most outstanding open questions currently include: What is the assembly history of the Milky Way mass? What is the role of the structure currently known as the Milky Way thick disc at the earliest times, and what is its relationship with the Milky Way bulge and halo? What are the relative fractions and properties of the Galactic stellar halo, formed within the Milky Way ({\em in-situ}) and accreted from `building blocks' of satellite systems? What is the general shape of the potential of the Milky Way's halo out to large distances, and how lumpy is this potential? Is the Galactic disc duality (that refers to the existence of a thin and thick disc) real, and what is the cause of this duality if so? What causes the [$\alpha$/Fe] bimodality in the Galactic disc? What is the Galactic disc potential in detail, and how important are deviations from axisymmetry and stationarity? What are the origins of the chemical elements in the Galaxy?\looseness-1

The European Space Agency mission {\it Gaia} \citep[][]{GaiaMission2016} was launched in 2013 December. During its nominal five-year mission (and up to 10 years pending approvals of mission extensions), {\it Gaia} is already producing -- and will continue to produce --  the most accurate astrometric data (positions, parallaxes and hence geometrical distances, and proper motions) ever produced in the optical domain for more than a billion stars in the Milky Way. In addition, {\it Gaia} is providing very accurate photometry in three bandpasses \citep[][]{Rielloetal2021}, complemented since its DR3 by shallower photometry in a fourth bandpass \citep[$G_{\mathrm{RVS}}$;][]{2023A&A...674A...6S}, as well as spectrophotometric information from the Gaia BP/RP spectrophotometer \citep[][]{2023A&A...674A...2D}.

However, for a number of objectives, {\it Gaia} will not be sufficient by itself, and ground-based spectroscopic information provided by large-scale surveys such as WEAVE are mandatory. {\it Gaia}'s on-board Radial Velocity Spectrograph (RVS), with a resolution of $\sim11\,500$ over the wavelength range 845--872\,nm \citep{2018A&A...616A...5C,2023A&A...674A...5K}, will not reach the same depth as the rest of the {\it Gaia} instruments,\footnote{{\it Gaia} will deliver astrometry down to $G\sim21$, while {\it Gaia} RVS will deliver radial velocities only down to $G\sim16$--$17$.} leaving the vast majority of the survey without the third (radial) velocity dimension. Furthermore, the RVS's limited wavelength coverage and resolution will not allow for detailed studies of a large variety of elemental abundances, with reliable detailed elemental abundances only for sources brighter than $G\approx 12$\footnote{This limiting magnitude was obtained by extracting from the DR3 Astrophysical Parameters catalogue in the {\it Gaia} archive all stars with at least a [Ca/Fe] abundance measurement with a corresponding flag of 0, which shows a strong drop in number count for $G>12.5$.} \citep{
%recioetalGDR3Gspspec22
2023A&A...674A..29R}.

\begin{figure}
	\includegraphics[width=\columnwidth]{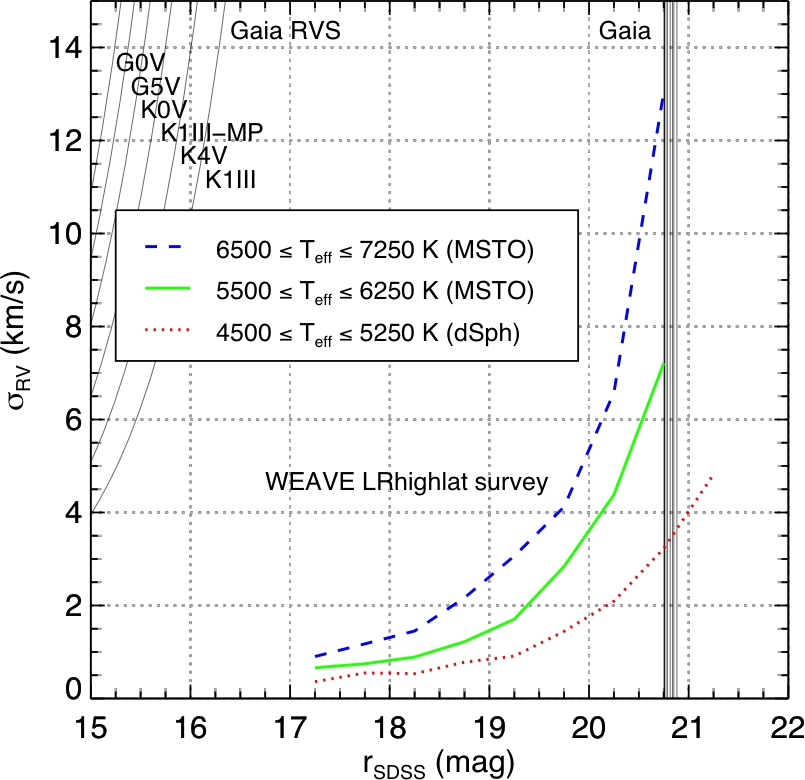}
    \caption{Expected radial velocity accuracies for WEAVE in its $R=5000$ mode (thick solid coloured lines) for different stellar effective temperature ranges, based on the WEAVE `Operational Rehearsal' (OpR) simulations (specifically OpR2.5 -- see also Section~\ref{sec:OpRs}; final improvements to the APS pipeline based on OpR3 data have confirmed the results presented here, albeit on a smaller sample), together with the post-launch predictions for {\it Gaia} at the end of the nominal five-year mission (thin solid coloured lines) for different stellar types. Dashed coloured lines represent {\it Gaia} end-of-mission photometric limits for astrometry of different spectral types.
    }
    \label{fig:eVlos}
\end{figure}

The WEAVE Galactic Archaeology survey is tailored to complement {\it Gaia} in two primary ways:
\begin{itemize}
\item by providing measurements based on low-resolution spectra of radial velocities (expected precision $\pm 1$--$2\,\mathrm{km\,s^{-1}}$), effective temperature, surface gravity, and metallicity (expected precision $\pm0.2\,\mathrm{dex}$) for $\sim 1.8$--2.6 million targets in the faint part of the {\it Gaia} catalogues with magnitudes in the range $16<G<20.7$, which are too faint for the RVS (see Fig.~\ref{fig:eVlos});
\item by yielding accurate measurements based on high-resolution spectra of abundance ratios covering the main nucleosynthetic channels (light, $\alpha$-, Fe-peak, and $s$- and $r$-process neutron-capture elements, to better than $\pm0.05$--$0.2\,\mathrm{dex}$), depending on the element and type of star, and radial velocities (to better than $\sim \pm 0.5\, \mathrm{km\,s^{-1}}$) for $\sim 1.1$--1.6 million stars in the magnitude range $12<G<16$, for which the RVS will again not be able to provide the relevant data.
\end{itemize}

Studying the origin and evolution of the main different stellar components of the Milky Way (halo, thin and thick discs) calls for massive spectroscopic surveys to provide line-of-sight velocities, distances, and chemical abundances of several million stars, and WEAVE is a cornerstone for this endeavour, as one of the first -- and for its high-resolution capability, the only -- high-multiplex wide-field optical multi-object spectrograph in the Northern Hemisphere. These spectroscopic surveys complement {\it Gaia}'s superb astrometry while benefiting from {\it Gaia} for optimal target selection. A WEAVE survey starting in the coming months will be particularly timely, benefiting from {\it Gaia} Data Release 3 in 2022 June \citep[DR3: ][]{2023A&A...674A...1G}, and producing its final science in the same time-frame as the {\it Gaia} end-of-mission catalogues (the {\it Gaia} legacy release, planned for 2030). {\it Gaia}'s third data release \citep[DR3, released as the EDR3 on 2020 December 4 and DR3 on 2022 June 13:][]{2023A&A...674A...1G} are being used for WEAVE target selection, providing a homogeneous source for astrometry \citep[positions, parallaxes, proper motions --][]{2021A&A...649A...2L} and broad-band photometry \citep{2021A&A...649A...3R}, to enhance the capabilities of WEAVE to build large source catalogues tailored to studying specific aspects of the Milky Way's stellar populations, with well-controlled and reproducible target selection criteria.\looseness-1

\begin{figure*}
\includegraphics[width=\hsize]{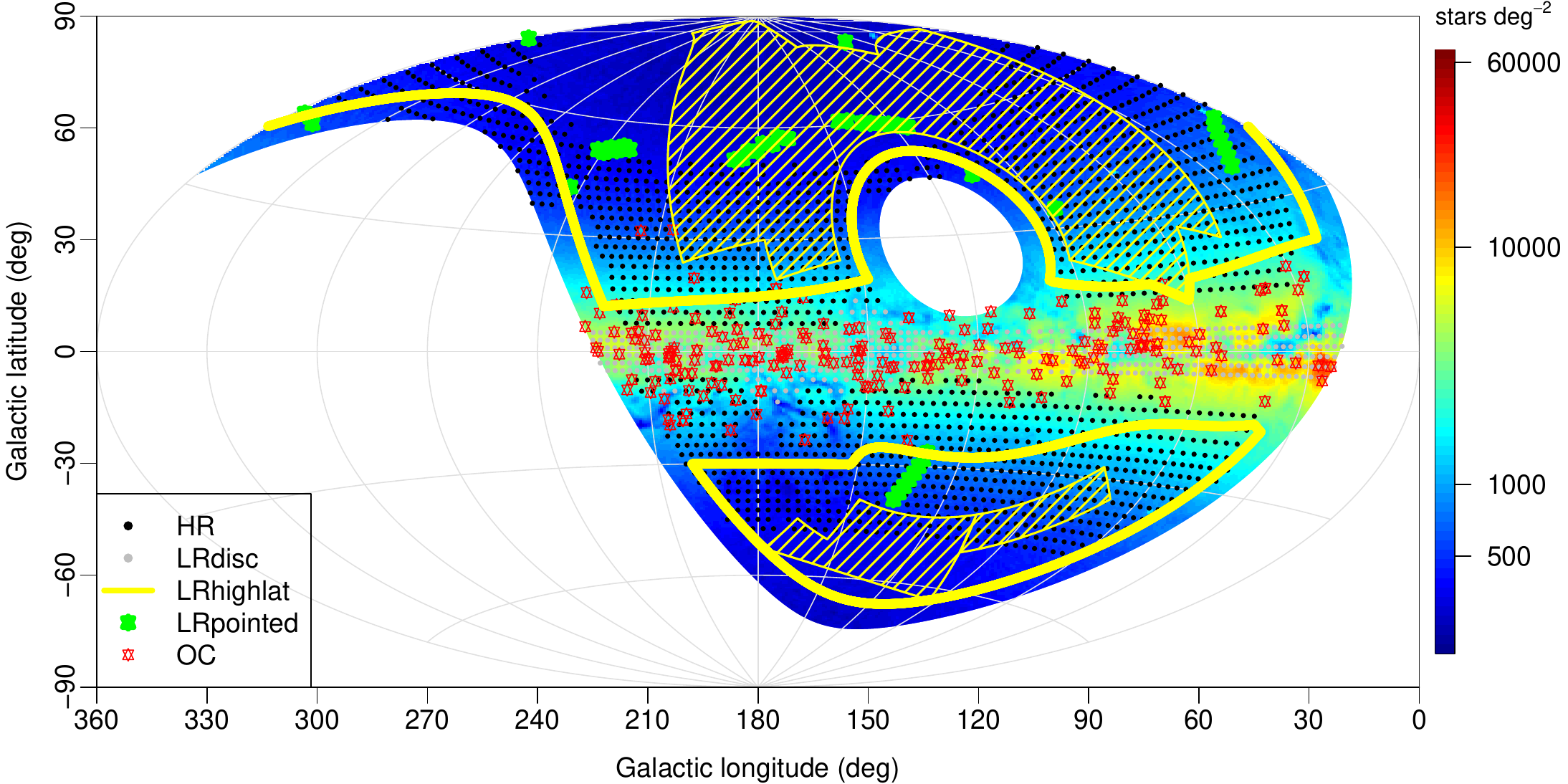}
\caption{Footprint of the WEAVE GA surveys on the sky in Galactic coordinates $(\ell,b)$ in an Aitoff projection overlaid on the {\it Gaia} DR3 \citep[][]{2023A&A...674A...1G} density contours of stars with parallax measures, $G<16$, and $-10^\circ<\mathrm{\delta}<80^\circ$ (in stars\,deg$^{-2}$). 
HR chemo-dynamical survey: black dots; 
Open Clusters survey: red stars;
disc-dynamics LR survey: dark green dots; 
high-latitude LR survey (shared with the WEAVE-LOFAR and WEAVE-QSO surveys) yellow outline, where the hashed yellow outline region shows the provisional area to be surveyed at $\sim 100$ per cent coverage factor, while the remaining area for this survey will be surveyed with a $\sim 30$ per cent coverage factor.  Known stellar streams and dwarf spheroidal galaxies covered within the pointed part of the high-latitude LR survey: light green stars. The available declination range arises from the impact of differential atmospheric refraction on targets near the edge of the field on the typical 1-hour WEAVE-Survey OB. Near-polar targets like NGC 188 require short exposures outside of the normal WEAVE-Survey OBs.}
\label{fig:GAall_lb}
\end{figure*}

\begin{figure}
\includegraphics[width=\columnwidth]{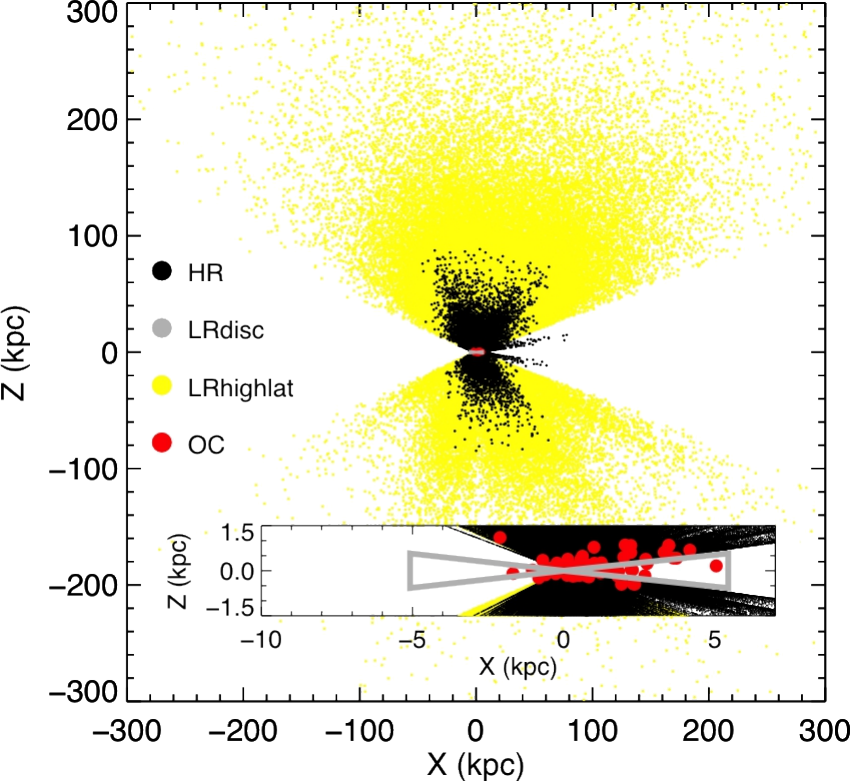}
\caption{Expected coverage of the WEAVE GA sub-surveys in the $(X,Z)$ plane (Cartesian coordinates centred on the Sun's position in the Galaxy) with the Galactic Centre located at $(-8.5,0)$\,kpc. This is based on the target-selection-scheme footprints of each of the sub-surveys, applied to simulations performed with GUMS \citep[{\it Gaia} Universe Model Snapshot;][]{RobinGUMS2012} and Galaxia \citep[][]{SharmaGalaxia2011}, taking into account the expected uncertainties on the {\it Gaia} astrometric data used in the target selection, estimated using the PyGaia toolkit (for {\it Gaia} performances, see: \url{https://www.cosmos.esa.int/web/gaia/science-performance}).}
\label{fig:GAall_XZ}
\end{figure}

The WEAVE Galactic Archaeology surveys fall into four independent categories listed below, according to the spectral resolution of the observations (LR or HR) and the stellar populations or Galactic locations targeted, as illustrated in Figs~\ref{fig:GAall_lb} and \ref{fig:GAall_XZ}. These complementary surveys, together with the SCIP and White Dwarfs surveys (see Sections~\ref{subsec:SCIP} and \ref{subsec:WDs}, respectively), nearly completely sample the Milky Way's stellar populations accessible from the Northern Hemisphere.\looseness-1

\begin{itemize}
\item \textbf{The high-latitude LR survey (`LR-highlat')}: This low-resolution wide-area survey of 6000--8500\,deg$^2$ at high Galactic latitude will observe $1.10\times10^6$ (five-year survey) to $1.55\times10^6$ (seven-year survey) stars in the magnitude range $15<r_\mathrm{SDSS}<20$--$21$\footnote{Magnitudes with subscript `SDSS' refer to Sloan Digital Sky Survey magnitudes \citep[see][]{2017ApJS..233...25A}.} to complement {\it Gaia} with line-of-sight velocities and metallicities for stars too faint for the RVS, and thereby yield the best-yet constraints on the halo potential's lumpiness and total mass to large distances \citep[$\sim$100\,kpc, e.g.][]{2020MNRAS.494.4291C}, address the formation of the Galactic halo, and characterize its progenitors (including substructures such as streams and dwarf galaxies). The search for traces of accretion events has recently been exemplified by {\it Gaia}'s finding of a significant merger, known as {\it Gaia} Enceladus or the {\it Gaia} Sausage, that has been shown to make up a significant fraction of the inner Galactic halo \citep[e.g.][]{Belokurov2018, Helmi2018, Haywood2018,2019MNRAS.488.1235M,Gallartetal2019,2020ApJ...901...48N,2022ApJ...926..107M}. A number of new cold streams have also been discovered thanks to {\it Gaia} astrometry \citep[e.g.][]{Malhan2018,2021ApJ...914..123I}, which anticipates the wealth of discoveries awaiting us with the combination of {\it Gaia} and WEAVE. While the main target of this survey is the Galactic halo, the survey will also probe the thick disc of the Milky Way, complementing the HR survey (see section below).
To achieve these goals, the LR-highlat wide-area survey will target the following: colour- and magnitude-selected main-sequence turnoff (MSTO) stars; red-giant-branch (RGB) stars selected by colour, magnitude, and {\it Gaia} astrometric data (parallaxes and proper motions, to reject local red main-sequence stars from the selection); extremely metal-poor star candidates selected from narrow-band photometry \citep[e.g.\ Pristine:][]{Starkenburg2017}, broad-band photometry \citep[e.g.\ SDSSS;][]{2018ApJS..235...42A} or {\it Gaia}'s on-board spectrophotometry \citep[{\it Gaia} Data Release 3 and onwards; see][]{2023A&A...674A...6S}; blue-horizontal-branch (BHB) stars and blue stragglers \citep[using, for example, a combination of colours including the $u$ band, e.g.][]{Deason2012}; RR Lyrae stars \citep[e.g.][]{Sesar2017,GaiaDR2Variables,2021MNRAS.502.5686I,2023A&A...674A..18C}; hyper-velocity stars \citep[e.g.][]{2018ApJ...866...39B,2019MNRAS.487.4025C,2022MNRAS.512.2350E,2022MNRAS.515..767M}. This survey is foreseen to share fibres on the sky with the WEAVE-LOFAR and WEAVE-QSO surveys (see Sections~\ref{subsec:WL} and \ref{subsec:WQ}, respectively), and hence share the same footprint. In addition to this wide-area component, the LR-highlat survey will drill a few deeper (down to at least $r_\mathrm{SDSS}=21$) 4-hour pointings (and even up to 7 hours in some specific cases) around specific objects of interest, namely dwarf spheroidal galaxies and known stellar streams, where the multi-epoch data will allow for the detection of binary stars to refine the line-of-sight velocity dispersion profiles of these objects and to understand better the dark matter distribution \citep[e.g.][]{2019A&A...621A..56P,2020MNRAS.494.4291C}. This pointed survey is expected to cover a total of $\sim200\,\mathrm{deg}^2$ on the sky.\looseness-1

\item \textbf{The disc-dynamics LR survey (`LR-disc')}: This low-resolution, high radial-velocity accuracy survey will observe $\sim500$ lines of sight through the Galactic disc at very low latitudes to probe fundamental Galactic disc dynamics processes in a volume inaccessible to {\it Gaia} alone. The survey will observe $7\,\times 10^5$ (five-year survey) to $1.1\times10^6$ (seven-year survey) stars selected mainly to be red-clump giants with Pan-STARRS1 \citep{2016arXiv161205560C} $r$-band magnitudes in the range $15<r_\mathrm{PanSTARRS}<19,$\footnote{Magnitudes with subscript `PanSTARRS' refer to PanSTARRS PS1 magnitudes \citep{2012ApJ...750...99T}.} spanning evenly the distance range $\sim$2--8\,kpc from the Sun, or from $\sim2$--15\,kpc in Galactocentric radius. Towards the outer disc, all RGB stars will be allowed into the selection function. The fundamental goal of this survey is to understand the phase-space distribution of stars of the Galactic disc, in particular those effects associated with mergers and/or resonances of non-axisymmetric perturbations (bar and spiral arms), both in the inner and outer disc \citep[see also][]{Famaeyetal2016}. This survey is likely to provide unique insights into the effects of various internal and external perturbers on the secular evolution of the Galactic disc at large scales, in a similar fashion to the local non-stationarity found locally in {\it Gaia} DR2 \citep[e.g.][]{GaiaDR2Katz2018, Antoja2018,2019MNRAS.482.1417B,Bland-Hawthornetal2021,2019MNRAS.490.4740B,2020NatAs...4..965R}. Thanks to the Northern location of WEAVE, this survey will also provide a unique view of the Galactic Anticentre \citep[cf.][]{2021A&A...649A...8G}. \looseness-1

\item \textbf{The HR chemo-dynamical survey (`HR')}: This high-resolution survey, covering $\sim 4000$--5600\,deg$^2$ at intermediate and high Galactic latitudes, will target stars with $12<G<16$ to complement Gaia's phase-space information with full chemical information, thereby opening the full 15+ dimensional space of the chemo-dynamical picture of the three major Galactic populations: the thin and thick discs and the halo. The thick disc will be sampled to large distances, allowing us to study its connection to the thin disc and the inner halo, together with its age-density and age-kinematics relations through mono-abundance and mono-age groups \citep[e.g.][]{2017MNRAS.471.3057M,2020MNRAS.492.3631M}. Ancient accretions that have now dissolved in the halo and discs will be detected chemo-dynamically through the wealth of abundance information for different nucleosynthetic channels that will be provided by the HR spectra \citep[eg.][]{2019MNRAS.482.3426M,DiMatteoetal2019}. In the volume within which {\it Gaia} parallaxes are most accurate (i.e.\ accurate to better than 10 per cent), high-quality stellar parameters (effective temperature, gravity, metallicity, and [$\alpha$/Fe]) for $>7\times10^5$ (five-year survey) to $1.1\times10^6$ (seven-year survey) MSTO and sub-giant stars will also allow us to constrain stellar ages to exquisite precision \citep{2019MNRAS.486.1167B,2023A&A...669A.104K} within a sphere of radius $\sim2$\,kpc. The target selection function, based on absolute magnitudes and their associated uncertainties from {\it Gaia} DR3 (or from later releases), will include all stars with absolute magnitudes $M_{G}<4.5$, thus also selecting intrinsically bright targets such as RGB stars that will probe the Milky Way's thin and thick discs and halo to distances of $\sim10$\,kpc and beyond. Targets of special interest will be given higher priority to ensure proper sampling. These include very and extremely metal-poor stars with prior information on metallicity \citep[e.g.\ Pristine:][]{Starkenburg2017}, RR Lyrae \citep[see e.g.][]{2023A&A...674A..18C} and Cepheid \citep[both in this HR survey, as well as in the LRdisc survey above; see e.g.][]{2023A&A...674A..17R} variable stars, known exoplanet hosts, bright members of known streams \citep[e.g.][]{Martinetal2022}, and stars with high tangential velocities to increase the halo fraction in the sample \citep[from {\it Gaia} DR3; see e.g.][]{GaiaDR2Babusiaux2018,Koppelmanetal2021}.\looseness-1

\item \textbf{The Open Clusters survey (`OC')}: This is a survey (mostly in high resolution) of a sizeable sample ($\sim120$) of old and young open clusters (OCs) and tiling three cluster-formation regions, aimed at obtaining homogeneous information on radial velocities and chemical information. This will complement Gaia's superb distances and proper motions of individual stars in known and newly discovered OCs \citep[e.g.][and references therein]{2020A&A...635A..45C,2022A&A...661A.118C} up to 5\,kpc from the Sun, with a precision in distance of 1 per cent up to a distance of 1.5\,kpc and 10 per cent for more distant targets, leading to an accurate and reliable definition of membership even for the most distant objects. This survey will constrain our understanding of the formation and disruption processes of OCs \citep[][and references therein]{2019A&A...627A.119C,2022Univ....8..111C}, the chemical evolution of the Milky Way disc using OCs as probes \citep{2019A&A...623A..80C,2020A&A...640A...1C,2022A&A...668A..16S}, and provide insights into star formation and early stellar evolution \citep{2019A&A...627A.173V,2021A&A...651A..84M}.\looseness-1 
\end{itemize}

A careful validation and accurate calibration of the derived atmospheric parameters, measured abundance ratios, and radial velocities will ensure that the data from the aforementioned four GA sub-surveys will be on the same scale. The WEAVE GA sub-surveys will therefore include dedicated observations for this purpose, including the observation of globular clusters, well-studied open clusters, field giant stars with stellar parameters derived from asteroseismic data from the CoRoT \citep[e.g.][]{2017A&A...597A..30A} or Kepler \citep[e.g.][]{2010Sci...327..977B,2010A&A...522A...1K,2018ApJS..236...42Y} missions, and field stars \citep[{\it Gaia} benchmark stars;][]{2015A&A...582A..49H,2016A&A...592A..70H}, as well as stars in previous high-resolution spectroscopy studies \citep[e.g.][]{2016A&A...591A.118S}. This will allow us to combine the results of the GA sub-surveys both among themselves and with several other stellar Galactic surveys including Gaia-ESO \citep{2012Msngr.147...25G}, GALAH \citep{2021MNRAS.506..150B}, APOGEE \citep{2017AJ....154...94M}, 4MOST \citep{2019Msngr.175....3D}, and others; see e.g. \cite{refId0} for a comparison of metallicity determinations for FGK stars across public data releases from several such spectroscopic surveys.\looseness-1

In addition to dedicated calibration fields, it is highly desirable that a significant overlap is built between the high and low spectral resolution (HR and LR, respectively) GA surveys in order to ensure homogeneity of the stellar parameters and individual chemical abundances that are derived between these two instrumental modes. While calibration stars (cluster and reference field stars observed specifically with both modes) provide the absolute anchor to these quantities through comparisons with independent estimates, it is desirable to build an overlap of stars surveyed in both LR and HR that would enable us to better map the systematics between stellar parameters and abundances derived from these two types of WEAVE spectra.
Furthermore, it is expected that it will not be possible to measure the abundances of some elements in the restricted wavelength range of the HR mode; among these are, for example, nitrogen (and carbon in the green+red HR mode). A programme to utilize sub-optimal and/or relatively under-subscribed observing conditions has thus been devised to observe a fraction of the GA-HR survey observing blocks in LR mode, to be executed in weather conditions when the main surveys cannot be executed. Thanks to the limited depth of the GA-HR survey (12$\leq$$G$$\leq$16), this programme is expected to be executed in bright time and modest seeing conditions (seeing$<$2~arcsec), thus complementing the suite of programmes with observations suitable for the relatively under-subscribed bright time.
This GA programme will allow us: (i) to map systematics between the LR and HR analysis of WEAVE spectra over a large part of the Hertzsprung-Russell diagram, and reflecting the target selection of the GA surveys, with sufficiently large statistics to allow this mapping to be worked out with machine-learning methods \citep[e.g.][]{nandakumaretal2022}; and (ii) to potentially build a significant sample in which the HR suite of elemental abundances is complemented by the few elements that are only accessible (or better measured) in LR.\looseness-1

\subsection{The WEAVE Stellar, Circumstellar and Interstellar Physics (SCIP) Survey}
\label{subsec:SCIP}

\begin{figure}
	\includegraphics[width=\columnwidth]{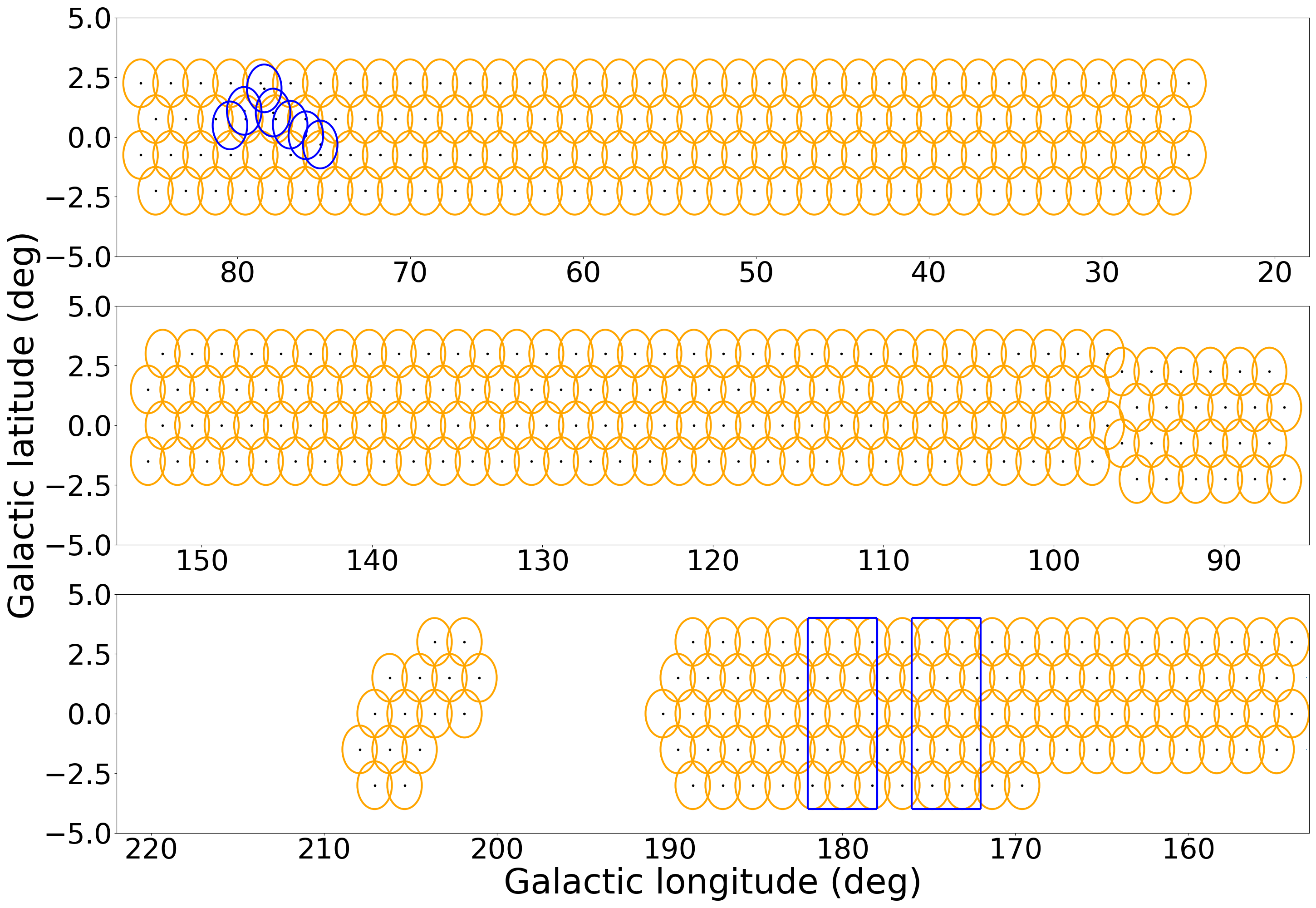}
     \caption{The SCIP survey footprint, mapped in Galactic coordinates, starting from $\ell \simeq 25^{\circ}$ (top right) and ending finally at $\ell \simeq 210^{\circ}$ (bottom left), with an island of pointings that encloses the Rosette Nebula, NGC 2264 and the Monoceros supernova remnant. The pattern of LR pointings, covering $\sim$1200\,deg$^2$, is traced by the orange circles representing the WEAVE field of view. In the second Galactic quadrant ($95^{\circ} < \ell < 170^{\circ}$), the footprint is shifted towards positive $b$ to better track the warp of the Galactic disc.  The regions to be studied intensively using the blue and red HR gratings are outlined in blue: these are in Cygnus (top panel, seven tailored pointings) and the Galactic Anticentre (using the same field centres as the LR survey).\looseness-1
    }
    \label{fig:SCIP}
\end{figure}

\begin{figure}
	\includegraphics[width=\columnwidth]{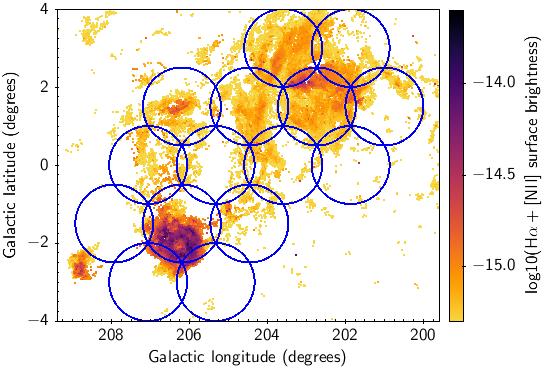}
    \caption{The distribution of potential WEAVE diffuse-ISM targets for the Rosette/NGC 2264/Monoceros supernova remnant region, derived from IGAPS/IPHAS H$\alpha$+[NII] narrow-band imagery \citep[see \url{www.igapsimages.org} and][]{2021A&A...655A..49G}. The planned SCIP LR fields are superimposed as blue circles. The candidate fibre positions, before final downsizing for configuration, are plotted on a colour scale according to narrow-band surface brightness (in $\mathrm{erg}\,\mathrm{cm}^{-2}\,\mathrm{s}^{-1}\,\mathrm{arcsec}^{-2}$). The brightest diffuse structure here is the Rosette Nebula, at the lower left.  In the upper right is the NGC 2264 region, with the $\sim$220-arcmin-diameter Monoceros supernova remnant in between.}\looseness-1
\label{fig:diffuse-ism}
\end{figure}

\begin{figure}
\begin{center}
	\includegraphics[width=\columnwidth]{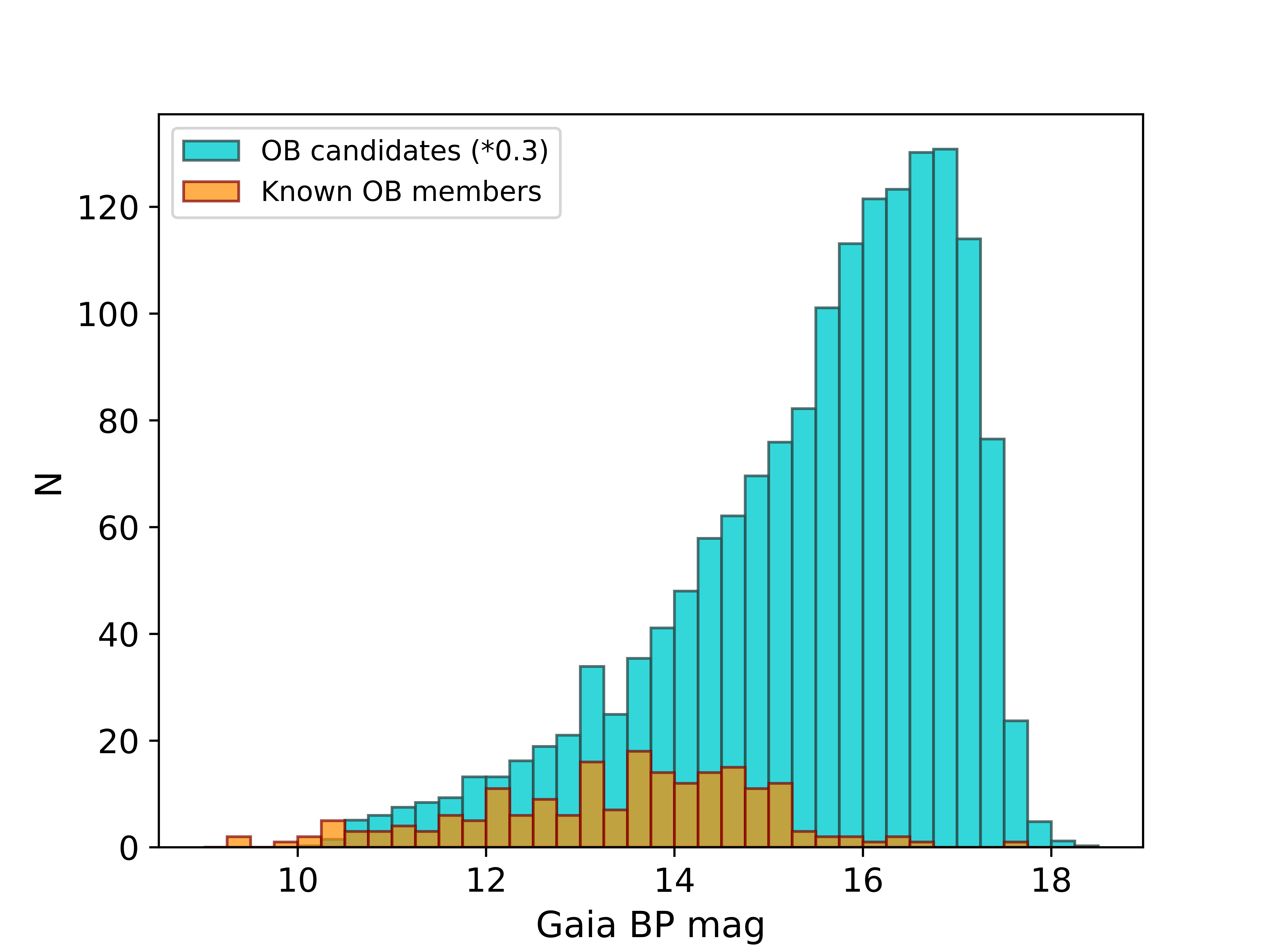}
    \caption{The discovery space for the high-resolution Cygnus component of the SCIP survey, expressed in {\it Gaia} BP magnitudes. Known OB stars are shown in orange, while those stars available for confirmation with WEAVE are shown in cyan after scaling down by a factor of 0.3. These fall within the seven fields planned for the survey in the Galactic longitude range $75^{\circ} < \ell < 81^{\circ}$ (see Fig.~\ref{fig:SCIP}).
    }
\label{fig:hrcyg_cand}
\end{center}
\end{figure}

Advancements in characterizing the stellar and interstellar processes currently at work in the Milky Way are helping us to better understand the physics that shape the appearances, content, and evolutionary histories of galaxies in the wider and younger Universe. Only locally can we easily access individual examples of the least well-described (and often short-lived) phases in stellar evolution down to sub-solar luminosity, and study the relationship between stars and the interstellar medium (ISM) at maximum angular resolution. Topics of current significance include: environmental factors in the formation of stars \citep{2016ApJ...828...48V, 2021A&A...650A.157G}; mapping the ISM in terms of its  extinction and its ionisation \citep{2018A&A...613A...9M, 2022A&A...658A.166D, 2021MNRAS.504.1039R}; filling in the Galactic massive-star Hertzsprung–Russell diagram \citep{2020A&A...642A.168B, 2020A&A...638A.157H}; how young stellar objects disperse into the field \citep{2019AJ....158..122K, 2020NewAR..9001549W}; the approach to stellar end states via cooling white dwarfs, X-ray, and other interacting binaries \citep{2020MNRAS.494.3799P, 2021MNRAS.504.2420I}.\looseness-1

SCIP will use the products of the IPHAS \citep{2005MNRAS.362..753D,2014MNRAS.444.3230B,2020A&A...638A..18M} and UVEX \citep{2009MNRAS.399..323G,2020A&A...638A..18M} surveys executed on the Isaac Newton Telescope and (to a lesser extent) VPHAS$+$ \citep{2014MNRAS.440.2036D} at the VLT Survey Telescope, for much of its target selection. Between them, these surveys map the whole Galactic plane at $\sim$1~arcsec resolution within $|b|<5\degr$ and in five optical bands.
These photometric surveys have captured more than half a billion point sources in the Milky Way's main mass component. Well-tuned samples of a range of object classes are being extracted from them for WEAVE, mostly within the magnitude ranges $11<i<18$ and $13<B<20$.  SCIP will pursue a number of scientifically distinct goals in Galactic astronomy, permitting full use of the available fibres in every pointing. The SCIP survey footprint is shown in Fig.~\ref{fig:SCIP}.\looseness-1  

The SCIP survey targets the high-mass and young/old extremes of stellar evolution along with the ISM on a large scale, capturing both the emissive, ionized nebulous component, and the ISM seen in absorption against background starlight. Since over half of the planned targets are relatively luminous OBA stars, the total space volume accessed will be very large, creating the opportunity to map Galactic disc structure out to heliocentric distances of $\sim10\,\mathrm{kpc}$. The $\sim200\,000$ BA stars to be surveyed, falling within the SCIP footprint, will provide a comprehensive account of young stellar kinematics in the Galactic disc on combining them with {\it Gaia} proper motions. The feasibility of this via fibre spectroscopy has been tested by \cite{harris2018}, where it was shown that stellar parameters of sufficient quality can be derived from the far-red calcium-triplet region of the spectrum alone. WEAVE in its $R\sim5000$ mode captures this region well, and is accessible for Galactic plane extinctions up to at least $A_V \sim 6$.
The $\sim20\,000$ O and B stars to be targeted will be most abundant in the first Galactic quadrant ($\ell < 90^{\circ}$). These spectra will build a more complete picture of the Galactic massive-star Hertzsprung--Russell diagram, and inform data-starved stellar evolution models. Parameter determination will proceed via the methods presented by \cite{SimonDiaz10}, \cite{2018A&A...613A..65H}, \cite{Carneiro19}, and others.\looseness-1

Pan-optical multi-object spectroscopy of stellar sight-lines at $R\sim5000$ provides many types of information, including: radial velocities to precisions of a few $\mathrm{km\,s^{-1}}$; reliable stellar effective temperatures and surface gravities; overall metallicity and first evidence of chemical peculiarity; detection of the more obvious spectroscopic binaries; interstellar absorption features; a suite of nebular/circumstellar emission line diagnostics; mass transfer/loss signatures; and markers for magnetic activity. In WEAVE's HR mode at $R = 20\,000$, observations in the Cygnus region and the Galactic Anticentre will push even further, from the measurement of projected stellar rotation speeds down to lower limits, to the determination of individual heavy-element abundances. Greater precision of radial velocity measurements will better expose the velocity dispersions of the local environments. Spectroscopic binaries will be detected via careful programmes of repeat observations.\looseness-1

The SCIP programme will be split between a LR survey, taking roughly two-thirds of the SCIP survey time, and two HR components that will equally split the remainder in bright time.\looseness-1

\begin{itemize}
\item {\bf The LR survey}: Around 1200\,deg$^2$ of the northern Galactic Plane will be observed. The largest target group is comprised of the higher-mass stars: this includes OBA stars and their evolved counterparts (including red supergiants), emission-line objects, and Cepheid variables.
The aim is to densely sample most Galactic longitudes in the thin disc accessible from La Palma. A novel feature of SCIP's target selection is the use of H$\alpha$ imagery from IPHAS to identify, in addition, a web of H$\alpha$-bright diffuse-ISM positions spanning \ion{H}{ii} regions, supernova remnants, and other nebulae \citep[][and Fig.~\ref{fig:diffuse-ism}]{2021A&A...655A..49G}.  These will complement the O- and B-star targets, with spectroscopy of the associated ionized ISM. The targeted range is from surface brightnesses of $6\times10^{-16}\,\mathrm{erg\,cm\,^{-2}\,s^{-1}\,arcsec^{-2}}$ up to $\sim 10^{-13}\,\mathrm{erg\,cm\,^{-2}\,s^{-1}\,arcsec^{-2}}$, with full abundance analysis achievable above 3--$5\times 10^{-15}\,\mathrm{erg\,cm\,^{-2}\,s^{-1}\,arcsec^{-2}}$. Radial velocities will be obtained for all targets.
A third focus is on the unbiased exploration of the incidence of young stellar objects in order to understand how they disperse across the Galactic field. The target selection draws significantly from infrared and optical photometric surveys and {\it Gaia} astrometry, yielding candidates of young stellar objects outside of the traditionally explored cores of OB associations and open clusters. The majority of targets will be selected using the method of \cite{2023MNRAS.521..354W}; for HAeBe stars, see\cite{2020A&A...638A..21V}. WEAVE spectroscopy will enable definitive classifications, as well as measures of radial velocities and accretion rates. A minority of fibres per pointing will also be allocated to compact binaries and white dwarfs (see also Section~\ref{subsec:WDs}).\looseness-1

\item {\bf The HR surveys}: These also place high priority on O, B, and A stars. In the Cygnus region, the aim is to obtain spectroscopy of prominent OB-associations \citep[for a recent review, see][]{2022arXiv220310007W} to build a full picture of kinematics and elemental abundances across the region, extending the studies first initiated by \cite{Berlanas18}. As shown in Fig.~\ref{fig:hrcyg_cand}, the discovery potential is very large.  In the Anticentre fields, the emphasis shifts to observations of late-B and A stars, where the science aims concern both the stars themselves and describing Galactic disc kinematics as imprinted on this young, dynamically cold population. Measuring constraints on binarity via repeat exposures is an important aspect in both HR programme components. Most stars in the Universe form in binaries or higher-order multiples \citep{2022IAUS..366...83S}. In their review, \cite{duchene2013} cite studies that indicate that $\sim$50 per cent of B and A stars have binary companions \citep[see also][]{2017ApJS..230...15M,2022A&A...658A..69B}. In the case of O stars, it has been estimated that over half will experience binary mass exchange in their lifetimes \citep{sana2013,2014ApJS..215...15S}.\looseness-1
\end{itemize}

The normal SCIP survey planning for MOS observations assumes an upper limit to the seeing of 1.2\,arcsec, appropriate for point sources observed using fibres of diameter 1.3\,arcsec. This constraint is unnecessary for extended sources, such as Galactic planetary nebulae (PNe), where abundance patterns as a function of position are an important goal \citep{2022MNRAS.510.5444G}. In poor seeing conditions ($\sim$2\,arcsec or more) that are unlikely to be usable for primary science objectives of the WEAVE surveys, SCIP will exploit the WEAVE (m/L)IFU capability to provide complete spatial chemo-dynamical maps of a significant sample of PNe. The key science aim is to probe the physical processes governing the late-stage evolution of intermediate-mass stars, and to determine how the abundances and dynamics of larger PNe are governed by their central stars (single or binary). PNe targets have been picked out that suit the 1.3--1.5-arcmin field of view of the LIFU sampled by 2.6\,arcsec fibres. These are large, bright nebulae where, in some, the central stars are known to be close binaries; such objects have markedly different chemical properties to other PNe \citep{2018MNRAS.480.4589W}. Our list includes well-known objects that have yet to be mapped comprehensively (e.g. the Owl and Cat’s-Eye nebulae). There will also be instances of several smaller nebulae in the field of view, for which the mIFU mode could also be usefully utilized.\looseness-1

\subsection{The WEAVE White Dwarfs survey}
\label{subsec:WDs}

\begin{figure}
	\includegraphics[width=\columnwidth]{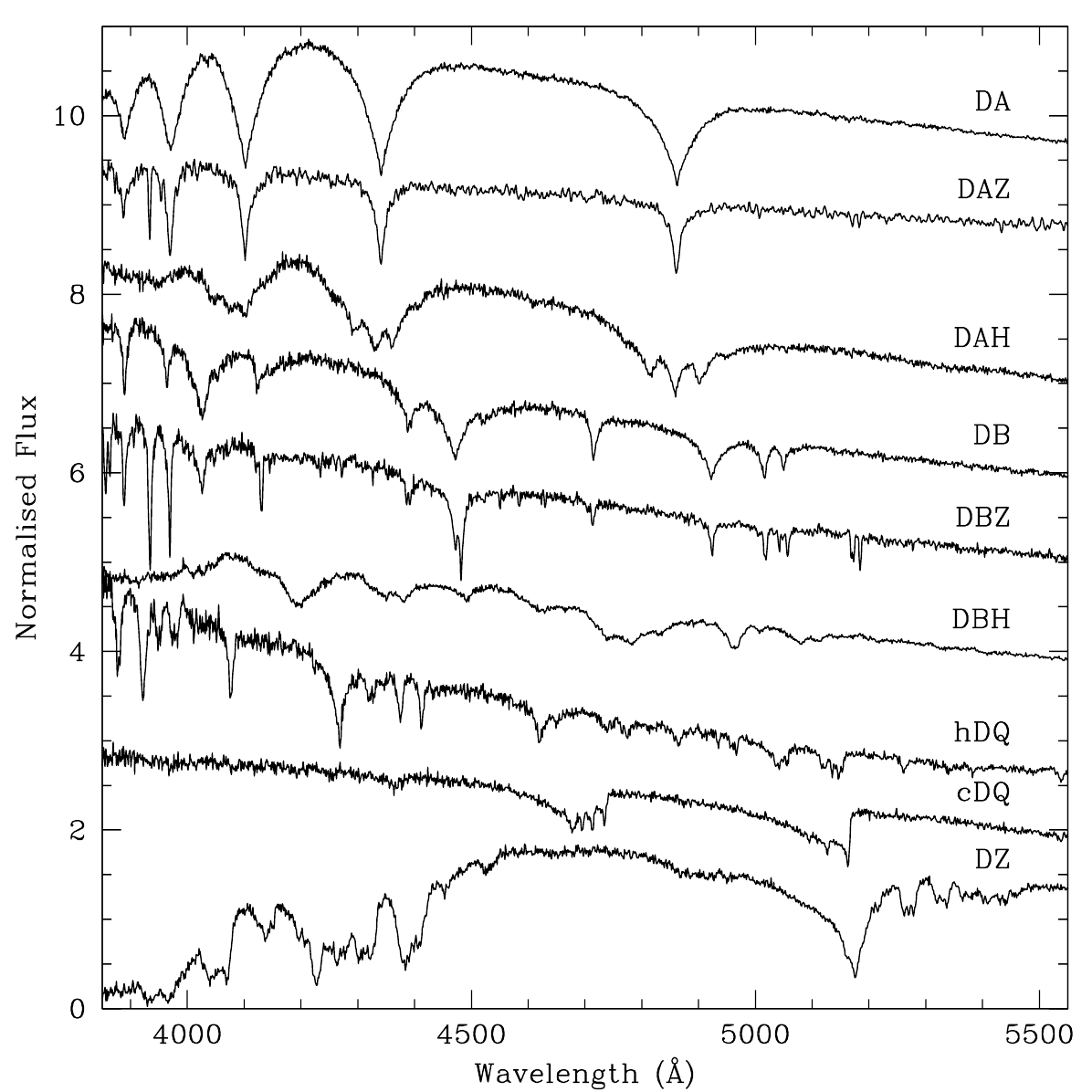}
    \caption{Examples showing a variety of white dwarf spectra in the range 3850-5550\,\AA. White dwarfs have large surface gravities ($\log g\simeq8$), which result in the chemical stratification of their constituents. Most white dwarfs have atmospheres dominated by hydrogen or helium, and their spectra contain only Balmer (DA) or He (DB) lines.  However, $\simeq20$ per cent of white dwarfs exhibit spectroscopic peculiarities. Accretion of disrupted planetesimals results in photospheric contamination by metals (DAZ, DBZ, DZ), providing a unique window into the bulk abundances of exoplanetary bodies. Cool white dwarfs with deep convective envelopes may dredge up carbon and oxygen from their cores (cDQ), allowing sensitive tests of stellar evolution. A small number of hotter white dwarfs with carbon-dominated atmospheres (hDQ) are thought to be products of white dwarf mergers, possibly descending from R\,Corona Borealis stars. Finally, up to 10 per cent show magnetic fields, across all atmospheric compositions (e.g.\ DAH, DBH), and serve as laboratories for atomic physics under extreme conditions. The WEAVE Survey will target $\simeq$100\,000 white dwarfs selected from {\it Gaia} DR3 (Fig.\,\ref{fig:wdhrd}).}
    \label{fig:wdspectra}
\end{figure}

\begin{figure}
	\centerline{\includegraphics[width=\columnwidth]{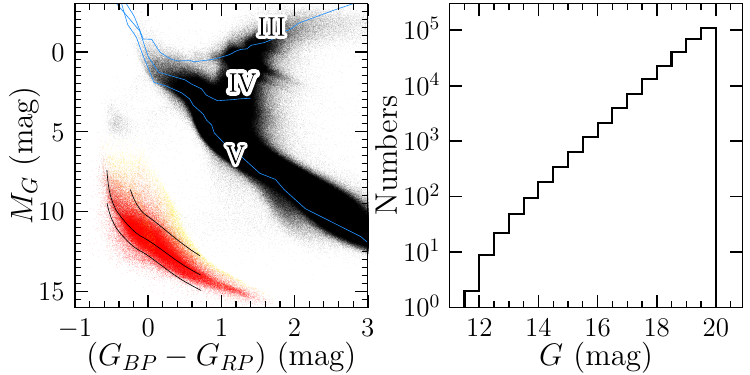}}
    \caption{Hertzsprung-Russell diagram (left) and magnitude distribution (right) of the 104\,198 high-confidence white dwarf candidate targets (red, $\mathrm{Dec}>-10^{\circ}$ and $G\le20$), selected from {\it Gaia} DR2 \citep{gentile-fusilloetal19-1}. Synthetic photometry was computed for main-sequence stars, sub-giants, and giants (the blue lines labelled V, IV, and III, respectively) using the spectral library of \citet{pickles98-1}, and for white dwarfs with masses of 0.2, 0.6, and 1.0\,M$_\odot$ (black lines, top to bottom) using the cooling sequences of \citet{holberg+bergeron06-1} and model spectra from \citet{koester10-1}.}
    \label{fig:wdhrd}
\end{figure}

All stars born with masses $\lesssim8$--$10\,\Msun$ eventually become white dwarfs \citep{ritossaetal99-1, dohertyetal17-1}: Earth-sized electron-degenerate stellar embers. The short main-sequence lifetime of stars with $M\gtrsim1.5\,\Msun$ implies that the majority of A/F-type stars formed throughout the history of the Galaxy are now white dwarfs. As such, white dwarfs play a central role across a variety of areas in astrophysics. Homogeneous samples of white dwarfs with accurate physical parameters are essential for constraining and calibrating stellar evolution theory, including mass loss on the asymptotic giant branch (intimately linked to the initial-to-final mass relation, e.g.\ \citealt{williamsetal09-1, cummingsetal18-1}), internal rotation profiles and loss of angular momentum \citep{hermesetal17-1}, and fundamental nuclear reaction rates \citep{kunzetal02-1}, with important implications for stellar population synthesis and galaxy evolution theory \citep{marastonetal98-1, kaliraietal14-1}. Because of their well-constrained cooling ages, white dwarfs provide an insight into the age of the Galactic disc \citep{wingetetal87-1, oswaltetal96-1}, open clusters \citep{garcia-berroetal10-1}, and globular clusters \citep{hansenetal07-1}, and can even trace variations in the Galactic star-formation rate \citep{tremblayetal14-1}.\looseness-1

White dwarfs will be observed as flux calibrators in WEAVE Survey observations, and the resulting spectra will provide a rich database for white-dwarf-enabled science.\looseness-1

\subsubsection{White dwarfs as flux calibrators}
\label{sec:fluxcal}

White dwarfs are routinely used as spectro-photometric flux standards \citep{bohlinetal01-1, bohlinetal14-1, moehleretal14-1}, as their spectra are extremely simple (i.e.\ pure hydrogen atmospheres in most cases) and can easily be modelled to high precision (1--2 per cent across all wavelengths) with only two free parameters: effective temperature and surface gravity. Moreover, in a magnitude-limited sample, the majority of white dwarfs have effective temperatures $\gtrsim8000$\,K, i.e.\ they are blue objects with significant amounts of flux at the shortest wavelengths.\looseness-1

The surface density and magnitude distribution of the {\it Gaia} white-dwarf population \cite[][]{2012MNRAS.426.1767P} is well-matched to the field of view of WEAVE and the aperture of the WHT. In the MOS mode, 10--15 white dwarfs are expected to be observed per WEAVE pointing for calibration purposes in the LR configuration. The number of suitable white dwarfs will likely be around 3--5 for the HR MOS configuration due to the magnitude limit for this mode, dictated by the observational constraints set by the science targets in the field, being brighter than for a LR observation (see Section~\ref{subsec:WDs}). In the mIFU and LIFU modes, constraints set by the number and placement of the IFU(s) in the field mean that no more than one white dwarf is expected to be observed per pointing; bootstrapping from other exposures or other information (such as known magnitudes) are required for these observations.\looseness-1

\subsubsection{White dwarf science with WEAVE}

A significant fraction of local white dwarfs are members of binaries \citep{toonenetal17-1}, providing a benchmark population for investigating complex interactions \citep[e.g.][]{zorotovicetal10-1}, including the progenitors of type Ia supernovae. Degenerate stars with precise cooling ages in wide binaries can also be used to calibrate main-sequence ages \citep{rebassa-mansergasetal16-1}. The host stars of virtually all planetary systems -- including the Sun -- will evolve into white dwarfs, and a sizable fraction of white dwarfs host the remnants of planetary systems \citep{farihietal09-1, koesteretal14-1}. White dwarfs that accrete tidally disrupted planetesimals display photospheric trace metals, which provides a unique opportunity to measure the bulk composition of extra-solar planets \citep{zuckermanetal07-1, gaensickeetal12-1, xuetal17-1}. Finally, white dwarfs are laboratories of extreme physics that are unachievable on Earth, including atomic and molecular physics in the presence of strong magnetic fields \citep{guan06-1} and high-density plasmas \citep{kowalski06-1}.\looseness-1

The extremely large surface gravities of white dwarfs, $\log g\simeq8$, result in chemical stratification and atmospheric compositions dominated by hydrogen and/or helium (e.g.\ \citealt{eisensteinetal06-1, giammicheleetal12-1}).  About 20 per cent of white dwarfs display  spectroscopic peculiarities, including metal pollution from accreted planetary debris or from dredge-up of core material, and Zeeman-splitting in magnetic fields of up to $10^{9}\,\mathrm{MG}$; see Fig.~\ref{fig:wdspectra} for examples. Spectroscopy spanning the full optical range is therefore critically important for the study of these degenerate stellar remnants.\looseness-1

Because of their small size, white dwarfs are intrinsically faint, and difficult to distinguish from more distant main-sequence stars with similar colours. Consequently, most known white dwarfs that were identified as ultraviolet-excess objects \citep{greenetal86-1,liebertetal05-1,kleinmanetal13-1} are young ($\lesssim1$\,Gyr) and hot ($\Teff\gtrsim10,000$\,K), and are utterly unrepresentative of the Galactic white dwarf population as a whole. {\it Gaia} DR2 finally provided the information necessary to break the degeneracy between nearby white dwarfs and background main-sequence stars: accurate parallaxes. \citet{gentile-fusilloetal19-1} assembled the first unbiased all-sky magnitude-limited ($G\simeq20$) sample of 260\,000 white dwarf candidates using {\it Gaia} DR2, followed by an updated catalogue using {\it Gaia} EDR3 data in \citet{2021MNRAS.508.3877G}. While the {\it Gaia} data are sufficient for identifying white dwarfs with high confidence, follow-up spectroscopy is required (Fig.~\ref{fig:wdspectra}) to determine their physical properties (effective temperature, surface gravity, atmospheric composition, magnetic field strength, and multiplicity) and derive fundamental properties (mass, cooling age, and progenitor mass) that are necessary to address the science areas outlined above.\looseness-1

WEAVE will target $\simeq$100\,000 white dwarfs with $G\le20$ (Fig.~\ref{fig:wdhrd}) for LR spectroscopy, and will roughly \textit{triple} the number of white dwarfs with high-quality spectroscopy in the Northern hemisphere \citep{kleinmanetal13-1, gentile-fusilloetal15-1}. The WEAVE White Dwarfs survey will establish the first large and homogeneous spectroscopic sample that is not subject to complex selection effects, and is therefore ideally suited for detailed statistical analyses of white dwarfs in the context of galactic, stellar, and planetary structure and evolution. The combination of accurate {\it Gaia} parallaxes and photometry with spectroscopic mass determinations will result in an extremely stringent test of the mass-radius relation of white dwarfs \citep[see e.g.][]{2017MNRAS.470.4473P,tremblayetal17-1,2019MNRAS.482.5222T,2019ApJ...871..169G}. The blue coverage (down to 3700\,\AA) of WEAVE is essential for covering the higher Balmer lines, which are important diagnostics for the surface gravity \citep{kepleretal06-1}, and to probe for atmospheric pollution by planetary debris via  Ca\,H/K absorption. Our large sample size will result in the identification of rare white dwarf species, tracing the extremes of parameters space of short-lived phases in their evolution, such as stellar remnants from stars near the core-collapse boundary \citep{gaensickeetal10-1}, products of thermonuclear supernovae \citep{vennesetal17-1, raddietal18-1, raddietal18-2, shenetal18-1}, of as-yet not fully understood binary interactions \citep{dufouretal07-1, kepleretal16-2}, and possibly examples of entirely new evolutionary channels \citep{marshetal16-1}. WEAVE spectroscopy will also provide radial velocity measurements that are an important complement to the {\it Gaia} proper motions -- most white dwarfs have no features in the {\it Gaia} RVS spectra, or are too faint for {\it Gaia} radial velocities altogether. This first large sample of white dwarfs with full 3D kinematics will allow us to distinguish them into their thin and thick disc, and halo populations \citep{paulietal03-1, paulietal06-1, anguianoetal17-1}, with an expected $\simeq$1 per cent of halo white dwarfs. Furthermore, the kinematics will provide constraints on the age-velocity dispersion relation, and insight into the mass distribution of white dwarfs that formed via binary mergers \citep{wegg+phinney12-1}.
\looseness-1

Given their low luminosity, the surface density of white dwarfs varies only mildly, with on average $\simeq5$ stars per square degree, or $\simeq15$ per WEAVE pointing. The WEAVE White Dwarfs survey is therefore multiplexed into the entire WEAVE Survey footprint and the white dwarfs will also be used for the spectro-photometric flux calibration of the entire WEAVE Survey (Section\,\ref{sec:fluxcal}).\looseness-1

\subsection{The WEAVE-Apertif survey}
\label{subsec:WA}

\begin{figure*}
\includegraphics[width=\linewidth]{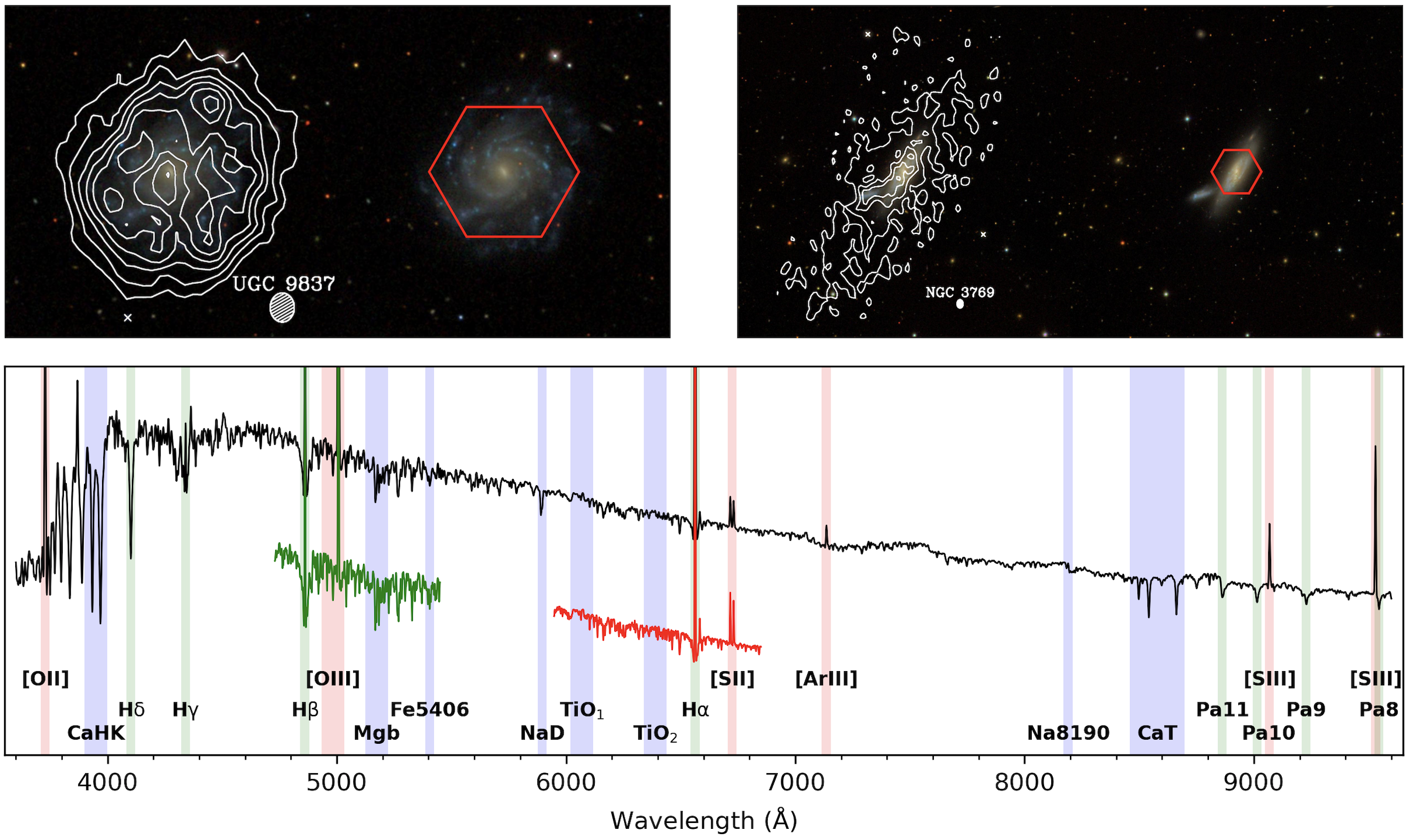}
\caption{Examples of typical galaxies to be observed in the WEAVE-Apertif survey. Top panels: \HI\ contours plotted over SDSS colour images. The WEAVE LIFU footprint is indicated in red over the SDSS image. The galaxy on the left, UGC 9837, is the proto-typical example of a {\it regular} galaxy, whereas the galaxy on the right, NGC 3769, represents a {\it perturbed} case.  Bottom panel: example of spectral instrumental setups used for the survey. The black line shows the wavelength range of our LR observations ($R\sim2500$), while the red and green lines indicate the wavelength coverage of the two HR arms ($R\sim10\,000$). Shaded areas mark the location of some of the main spectral features in our wavelength range, which can be found in absorption (blue), emission (red), or both (green).}
\label{fig:WA_example}
\end{figure*}

The detailed analysis of galaxies in the local Universe is crucial to understanding galaxy evolution. However, recent local IFU surveys of galaxies (e.g.\ CALIFA, \citealt{Sanchez_etal_2012}; SAMI, \citealt{sami}; MANGA, \citealt{manga}) suffer from an optical selection bias that makes it difficult to address the effects of environment on the evolution of galaxies. The gas content of galaxies is less strongly bound to the gravitational potential well than its stellar content. Thus, one possible approach to study the effect of environment on galaxy evolution is to characterize the morphology of galaxies seen in the atomic gas, which indicates interactions that have taken place, and to select galaxies with a range of features. To this end, a wide-area survey of resolved neutral atomic hydrogen (\HI) is needed.\looseness-1

\subsubsection{Apertif and synergy with WEAVE LIFU observations}

In the last decade, the Westerbork Synthesis Radio Telescope (WSRT) was upgraded with a new receiver: an innovative focal-plane array system named Apertif \citep{2008AIPC.1035..265V, 2022A&A...658A.146V} that allowed for the execution of wide-field radio surveys, and \HI\ surveys out to cosmological distances ($z\sim0.2$).  Apertif survey operations were conducted between 2019 July 1 and 2022 February 28, surveying a large portion of the Northern hemisphere in 1.4\,GHz radio continuum, polarization, and the 21-cm spectral line of neutral atomic hydrogen.  The Apertif instrument was the first working focal-plane array capable of full Westerbork resolution ($15\,\mathrm{arcsec}\times15\sin\mathrm{(Dec)}\,\mathrm{arcsec}$ beam\footnote{The factor of sin(Dec) is distinctive to WSRT as an East-West array.}) over a field of view of 6.5\,deg$^2$ per pointing, after tiling, and with \HI\ spectral resolution of down to 2.6\,km\,s$^{-1}$.
The 300\,MHz bandwidth could be tuned to operate between roughly 1130--1750\,MHz, although observations were heavily impacted by radio frequency interference below $\sim1300$ MHz.  With about half of the sensitivity of the original single-pixel WSRT front-end, the wide field of view of Apertif nonetheless drastically enhanced the survey speed of WSRT by imaging an area on the sky about 25 times the size of the full moon in a single pointing.\looseness-1

Apertif conducted a two-tiered \HI\ imaging survey\footnote{\url{https://www.astron.nl/telescopes/wsrt-apertif}} that provides a strong synergy with WEAVE: a medium-deep survey of 150\,deg$^2$ and a wide-area survey of approximately 2200\,deg$^2$ between $\sim+27$ to $65$\degr declination (Hess et al., in preparation). The wide-area survey targeted areas with multi-wavelength coverage from Northern sky surveys including LOFAR, SDSS, and the Hobby-Eberly Telescope Dark Energy Experiment \citep[HETDEX,][]{hill2008} in the Spring sky, and PanSTARRS \citep{2016arXiv161205560C} in both the Fall and Spring skies. The medium-deep survey targeted areas with substantial ancillary data in the Herschel-ATLAS \citep{h-atlas} North Galactic Pole field \citep{Smith17}, and a volume of particular environmental interest in the Perseus-Pisces supercluster. Apertif released its first year of imaging data to the community in late 2020\footnote{\url{https://www.astron.nl/telescopes/wsrt-apertif/apertif-dr1-documentation}} \citep{2020AAS...23513607A}, and the data are described in further detail by \cite{2022A&A...667A..38A}.  A companion paper describes the radio continuum source catalogue for DR1 \citep{2022A&A...667A..39K}. Processing of the full survey data is currently ongoing. Together, the two Apertif surveys are expected to yield thousands of \HI-detected galaxies, of which about 10--15 per cent have optical dimensions and a surface brightness distribution that fit the field of view of WEAVE's LIFU for our required signal-to-noise ratio. The Apertif \HI\ surveys will provide redshifts, neutral gas content, environment densities, morphologies, kinematics, and dynamical masses at the aforementioned resolutions. Furthermore, Apertif will also detect and measure \HI\ absorption against radio-loud AGNs, yielding information on gas accretion and outflows, provide spatially resolved, extinction-free star-formation maps from the radio continuum emission, and identify OH megamasers as locations of intense star formation \citep{Hess21}. \looseness-1

The natural goal when exploiting the synergy between a spatially resolved \HI\ survey and the data coming from the LIFU mode of WEAVE is to compare in detail the cold gas properties of the \HI-selected sample with the spatially resolved properties (stellar and ionized gas) of those galaxies. WEAVE-Apertif will study the transformation of gas into stars within galaxies by comparing the optical and ISM properties of galaxies belonging to the so-called main star-forming sequence, the red sequence, and the intermediate region, distinguishing whether the last represents a transition population triggered by a quenching episode in the fairly recent past \citep[e.g.][]{schawinski14} or smoothly evolving on timescales comparable to the age of the Universe \citep[e.g.][]{2015MNRAS.451..888C}.

With this goal in mind, we will select a sample of around 400 galaxies from the Apertif survey covering the following parameters with at least five galaxies per bin: \HI\ mass, \HI\ morphology (perturbed or not), stellar mass, and star formation. The unique aspect of this survey (compared to other IFU surveys) resides in the \HI\ selection of targets, which enables the detailed analysis of galaxies in different stages of the transformation process. We will use the LIFU in both low and high spectral resolution modes to study nearby galaxies. A large-scale integral-field survey with WEAVE creates a strong synergy with the Apertif imaging surveys, significantly increasing the power of both instruments as compared to each on its own.\looseness-1

Fig.~\ref{fig:WA_example} illustrates the kind of regular and perturbed galaxies that we expect to probe during the WEAVE-Apertif survey and the typical spatial coverage of the LIFU. In addition, the bottom panel shows examples of the spectral coverage and instrumental modes that will be used for the different science cases planned for the survey. While the LR mode will be most useful for emission-line morphologies/diagnostics and stellar-population studies, the HR mode will allow us to measure the stellar kinematics of both the stars and ionized gas with unprecedented accuracy, especially for low-mass systems.\looseness-1

\subsubsection{Scientific goals of the WEAVE-Apertif Survey}
 
In the optical, galaxies are distributed according to their colour and luminosity (based on the colour-magnitude diagram (CMD), e.g.\ \citealt{bell04}) in two main groups -- the red sequence and blue cloud, with a transitioning class named the green valley. The physical processes governing the movement of a galaxy from one group to another are still unclear, with the following questions still open: Why is there a bimodality in galaxy properties today, such that there are both star-forming blue galaxies and red-and-dead galaxies? Where does the gas fuelling the star formation in these galaxies come from? And how is star formation in galaxies regulated?\looseness-1
 
WEAVE-Apertif will analyse the spatially resolved stellar and ionized gas properties of galaxies, sampling the CMD with respect to their \HI\ properties (e.g.\ mass and morphology) and general environment to address the possible mechanisms that regulate the conversion of gas into stars within galaxies, including potential rejuvenation or quenching episodes. We will establish what processes -- revealed by the Apertif \HI\ images, such as major mergers, accretion of small gas-rich satellites, tidal- or ram-pressure gas stripping -- have a direct impact on i) the star-formation history of galaxies, ii) the spatial distribution of star formation in their discs, iii) the physical state of the ionized ISM, and iv) the foreseeable evolution of galaxies. To this aim, we will observe the full sample of $\sim400$ galaxies in the LR LIFU mode of WEAVE ($R\sim2500$) with a signal-to-noise ratio per \AA\ per spatial bin sufficient to derive stellar population parameters. These data will allow us to derive spatially resolved mean stellar ages and metallicities, as well as $[\alpha/\mathrm{Fe}]$ ratios. The data will also make possible the detection of  kinematic twists, warps, and decoupled components. The WEAVE-Apertif survey will also deliver spatially resolved stellar age distributions, ionized-gas metallicities, and star-formation rates that, together with the \HI\ data, will provide a true legacy survey.\looseness-1

As the effects of the environment on the general properties of galaxies will have been revealed by the LR dataset, a detailed study of a subsample of galaxies using the high-resolution capabilities of the LIFU will help to establish the effects of \HI\ accretion history in the secular (i.e.\ internal) evolution of spiral-galaxy discs \cite[for a review, see][]{vgorkom13}. The impact of the \HI\ accretion history on the chemo-dynamical properties of galaxies can be quantified by analysing the stellar kinematics and stellar properties of a sample of galaxies with perturbed \HI\ morphologies, and comparing them to those of galaxies with no signs of perturbed \HI\ morphologies. Our study will allow us to provide answers to crucial questions such as the following: What is the process driving radial migration of stars? What is the level of scattering and radial motions induced by spiral arms and bars in stars? What is the impact of the accretion of satellites in the chemo-dynamical properties of galaxies? And how does this all connect with observations in the Milky Way, as revealed, for example, through the Galactic Archaeology survey described in Section~\ref{subsec:GA}?\looseness-1

To accomplish these aims, we will analyse in detail the line-of-sight velocity distribution (velocity, velocity dispersion, and higher-order moments) of a sample of low-inclination nearby objects (including both high- and low-surface-brightness (H/LSB) galaxies), exploiting WEAVE's HR LIFU mode ($R\sim10\,000$). The sample will include 30--50 galaxies with perturbed \HI\ morphologies and approximately the same number of galaxies with non-perturbed \HI\ morphologies, all taken from the parent low-resolution sample of 400 galaxies. The success of the project and its legacy impact rely on obtaining quality data reaching $\mu_g=24\,\mathrm{mag\,arcsec^{-2}}$ at $\sim3\sigma$.\looseness-1

Finally, these datasets will enable us to constrain the disc mass of the full sample of 400 Apertif galaxies, allowing us to relate dynamical mass to the environment. The dark matter content in both HSB and LSB galaxies can be constrained by combining kinematic data from both \HI\ and optical data and applying the traditional technique of rotation-curve decomposition as well as other dynamical modelling techniques. However, the actual total mass residing within a galaxy's disc remains a controversial matter. This is an important calculation to get right, as the mass-to-light ratio of a disc determines the disc's contribution to the galaxy's rotation curve, which in turn constrains the density and scale-length of the halo. The WEAVE-Apertif survey will be able to expand on previous efforts on this topic \citep[e.g.\ DiskMass survey;][]{diskmassI} and constrain the surface mass density of discs using stellar velocity dispersions, with the HR mode being of particular value to the challenges presented by low-mass systems. The mass modelling will be possible for the entire sample of galaxies included in the two projects mentioned above, using different modelling techniques.\looseness-1

In line with many of the other WEAVE surveys, an extra programme has also been designed within the WEAVE-Apertif survey to help maximise the use of sub-optimal weather conditions and/or under-subscribed observing conditions, while simultaneously extending the scientific output of the survey by relieving some of our most stringent constraints on sample selection. Within this programme, there will be no restrictions to stay within the Apertif footprint, and the removal of this condition will allow for a much more flexible scheduling that will enable us to venture into the ALFALFA survey \citep{gio2007} to obtain upper limit \HI\ mass estimates. In addition, we will allow for the inclusion of highly inclined galaxies and compact  ($\leq$0.2~arcmin) high-surface-brightness systems. This will enable us to obtain observations for certain types of galaxies that would otherwise be missed in the main survey, for example Markarian galaxies, which usually display peculiar \HI\ properties.

\subsection{The WEAVE Galaxy Clusters survey}
\label{subsec:WC}

\begin{figure}
	\includegraphics[width=\columnwidth]{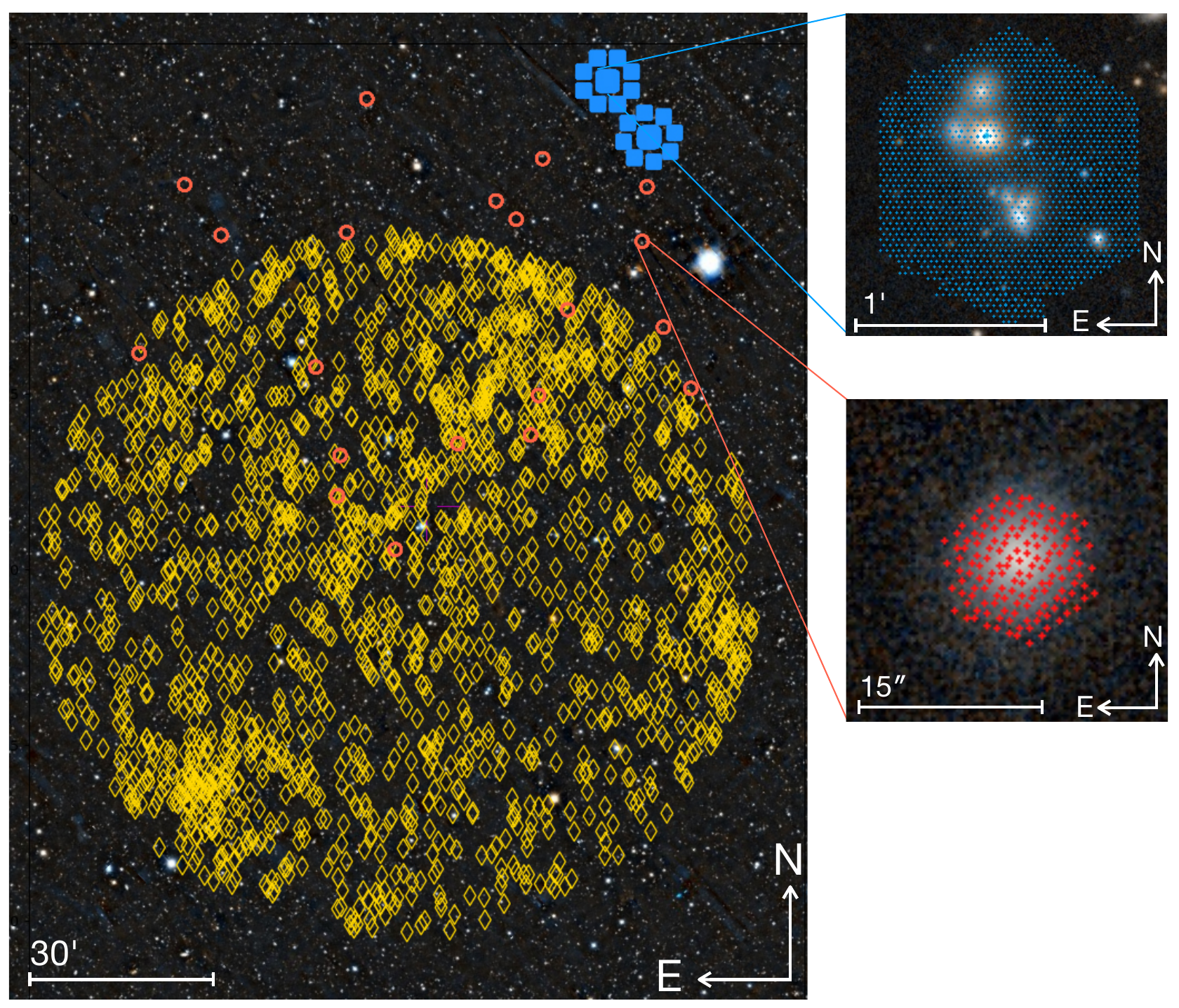}
    \caption{The WEAVE Galaxy Clusters survey encompasses three layers that observe in the three instrument modes, as illustrated in this image of cluster Abell 2142 (background image from PanSTARRS1). Layer 1 focuses on cluster-dwarf-galaxy evolution and uses the mIFU mode centred on selected dwarf galaxies (red circles), which will be selected from previous MOS spectra. The zoomed-in image shows the mIFU arrangement of 37 fibres placed on a single dwarf galaxy. Layer 2 focuses on the filamentary large scale structures that feed galaxy clusters, and uses WEAVE's $\sim$1000-fibre MOS mode (yellow diamonds). Each cluster will be covered by up to 20 pointings. Layer 3 focuses on the evolution of central cluster galaxies, and cosmological constraints that can be drawn from cluster scaling relations. It uses WEAVE's LIFU mode (blue squares), highlighted in the zoomed-in image that shows a LIFU pattern placed on a group falling into Abell 2142.}
    \label{fig:WC}
\end{figure}

On the largest scales, matter is distributed as a vast network of filaments and sheets that connect dense galaxy clusters and groups. This cosmic web \citep{Bond1996} is the environment in which galaxies form and evolve. Relative to the small volume they occupy (much less than one per cent of the volume of the Universe), clusters and groups contain the largest fraction of the mass of the Universe \citep{Cautun2012, Tempel2014, Veena2019}. Observations of galaxies in clusters are therefore of great interest, both for a test of galaxy formation in dense environments and as a measure of environmental influences on galaxy evolution. Consequently, in order to make meaningful progress in our understanding of galaxy evolution, we must consider galaxies evolving in relation to their environment, defined both by their local density and the global large-scale structure.  \looseness-1

Observations of cluster galaxies have firmly established the importance of the environment in addition to galaxy mass for shaping galaxy properties: morphology, colour, star-formation rate, stellar age, and AGN fraction correlate with both local galaxy density and location inside or outside clusters \citep[e.g.][]{1980ApJ...236..351D, Blanton2005, Postman2005, Smith2005, Bamford2009}. In clusters, the fraction of early-type galaxies, defined either morphologically or by the amount of current star formation, is much higher than in the average, lower-density Universe, often referred to as the `field'. Related studies show that cluster galaxies possess much less cold gas than field galaxies \citep[e.g.][]{1990AJ....100..604C, 2011MNRAS.412..800S} and that galaxy luminosity functions are steeper for cluster galaxies, owing to the change in the relative contribution of massive quiescent galaxies \citep[e.g.][]{2003MNRAS.342..725D}. 
These well-established findings may be summarized as manifestations of the morphology-density relation \citep{1980ApJ...236..351D} and are related to the result of a series of processes involving the mass assembly, star formation, and morphological evolution of these galaxies \citep[see e.g.][]{Boselli2006}. A variety of interactions with the hot intracluster medium \citep[e.g.][]{Gunn1972, Bekki2001,Kawata2008, Zinger2018}, as well as interactions and fly-bys of galaxies in this medium \citep[e.g.][]{Gnedin2003, Aguerri2009, Park2009, Sinha2012, Mendez-Abreu2014}, affect the gas -- and often the stellar -- content of galaxies. Importantly, the accumulated effect of various galaxy encounters and the loss of gas, ultimately resulting in the suppression of star formation, may stimulate the transformation of late-type into early-type galaxies in dense environments \citep{Moore1996, Boselli2006, Cappellari2013,Kuchner2017,Joshi2020}.\looseness-1

Despite active research on this topic, it is still a matter of debate as to which of the proposed physical mechanisms drive the evolution of galaxies in high-density environments and their relative importance \citep[for a review, see][]{Boselli2014}. The difficulty lies in disentangling subtle and competing processes that act on different timescales. In addition, quantification of the observed changes has mainly been focused on the end-point in the virialized regions of clusters. However, half of a nearby cluster's galaxy population originates from the cosmic web \citep{McGee2009, Dressler2013}. Clusters grow by accreting galaxies from their surroundings, and a significant fraction of cluster galaxies have therefore been environmentally affected long before they fall into the cluster. This effect is called `pre-processing' and is receiving increasing attention by the community \citep[e.g.][]{Wetzel2013}. \looseness-1

The morphology-density relation is particularly strong for dwarf galaxies \citep{Peng2010, Kovac2014, Tal2014, Wetzel2015}: early-type dwarfs (dwarf elliptical and dwarf spheroidals) dominate the galaxy population in clusters, but they are very rarely found in the field \citep{Geha2012, Grossi2016}. Despite their smooth appearances, these systems are more complex than they might appear: dwarf ellipticals can host stellar late-type features such as discs and spiral arms \citep{Jerjen2000,Lisker2006,Lisker2007,Janz2012}, some with clear signs of rotation \citep{Pedraz2002, Toloba2009} and atomic gas or dust components \citep{diSeregoAlighieri2007, diSeregoAlighieri2013, Hallenbeck2017}.
Star-forming dwarf galaxies accreted into a cluster are easily stripped of their gas and dust when interacting with the cluster environment \citep{Grossi2015, Grossi2016}, but the variety of signatures suggests that more than one process may determine dwarf-galaxy evolution. \looseness-1

Evidently, stellar mass is a dominant driver for galaxy evolution \citep[e.g.][]{2004MNRAS.353..713K,Peng2010}. However, intrinsic properties like galaxy mass are highly dependent on their surrounding large-scale environment through their assembly process. Thus, a comprehensive theory of galaxy formation and evolution must account for the influence of the environment on galaxy evolution while carefully controlling for stellar mass. Within this context, several, large, multi-wavelength studies of galaxy clusters have successfully used cluster number counts as a function of mass and redshift to constrain cosmological models \citep{Rozo2009, deHaan2016,2016A&A...586A.139P,2016A&A...594A..24P,2016A&A...594A..27P,Pacaud2018}. The complexity of the physical processes acting in clusters complicates the obtaining of reliable mass determinations, forcing us to fall back on having to rely on combining a number of scaling relations at various wavelengths. This is still a paramount challenge for performing precision cosmology using clusters for next generation surveys such as the LSST, and surveys with Euclid and the Nancy Grace Roman Space Telecope \citep[see e.g.][]{2020A&A...643A..20S}. \looseness-1

The WEAVE Galaxy Clusters survey was designed as a response to these challenges. The main goal of the survey is to provide a better understanding of the impact of different environmentally induced physical mechanisms on galaxy evolution. Three layers focus on three complementary aspects: WEAVE Galaxy Clusters aims to (1) gain a detailed understanding of the formation history of low-mass dwarf galaxies in dense environments; (2) follow mass and stellar assembly of clusters from the cosmic web through filamentary accretion; and (3) obtain an accurate calibration of the scaling relations and perform precision cosmology on a complete sample of Sunyaev-Zeldovich (SZ) effect-selected clusters \citep{2016A&A...594A..27P}, and infer the star-formation histories of the galaxies in the cores of these clusters (see Fig.~\ref{fig:WC}, which summarizes the three observing modes for WEAVE Galaxy Clusters). \looseness-1

\begin{center}
\begin{figure}
\includegraphics[width=0.48\textwidth,trim=0.2cm 0.2cm 16cm 0.3cm, clip=true, angle=0]{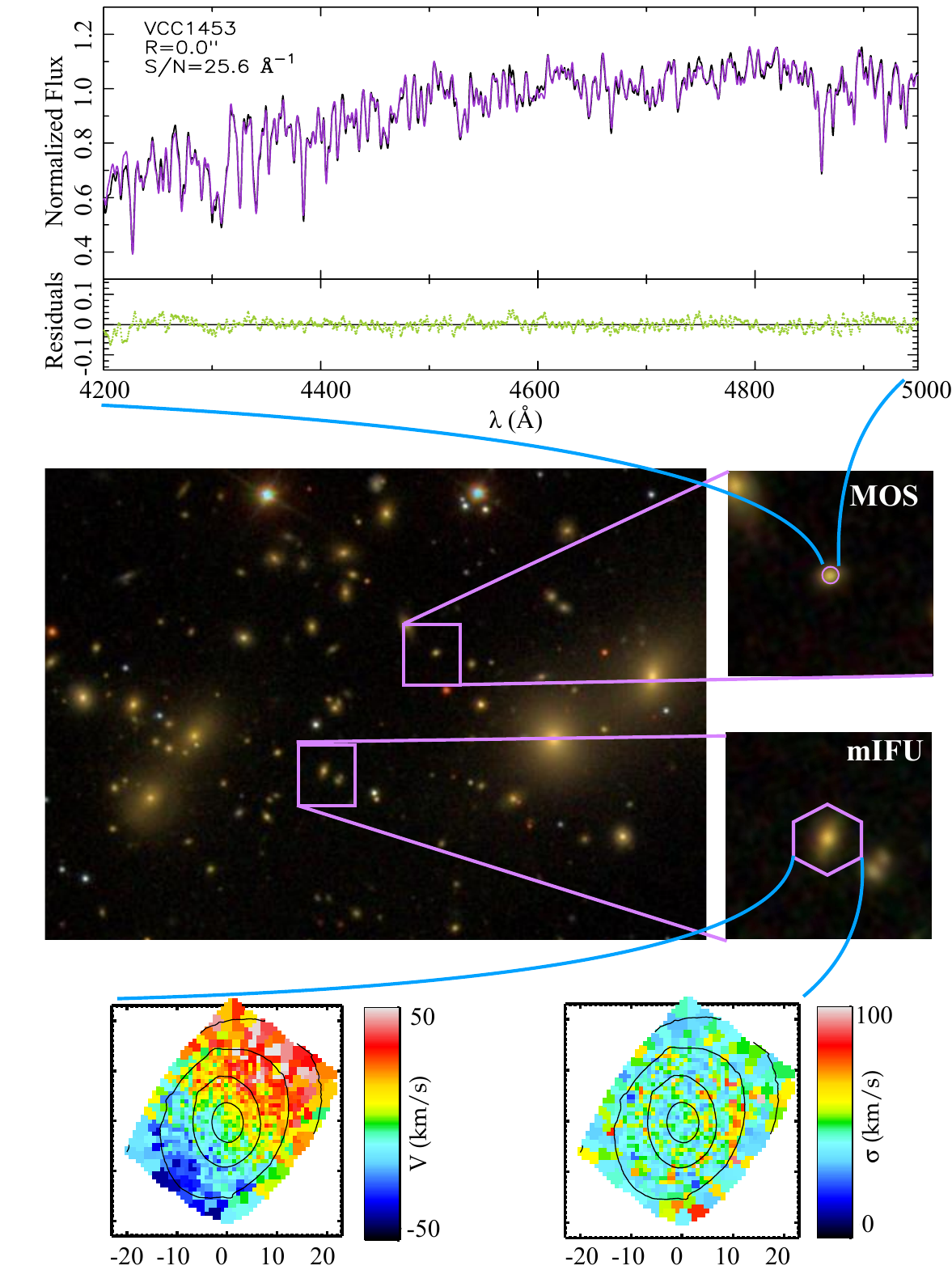}
\caption{The WCN survey aims to obtain high-resolution spectroscopy in both MOS and mIFU observing modes for a significant population of dwarf galaxies in every cluster. The example in the middle panel shows archival images for A2152 from the SDSS with two selected dwarf galaxies with absolute magnitudes $M_r\sim-17$. Top panel: a one-dimensional spectrum of an example dwarf galaxy (not shown in the image in the middle panel) is extracted from \citet{toloba2014}. Bottom panel: two-dimensional velocity and velocity dispersion maps of another example dwarf galaxy from \citet{rys2013}.}
\label{fig:WCN}
\end{figure}
\end{center}

\begin{itemize}

\item {\bf{Layer 1 -- WEAVE Nearby Clusters (WCN) survey:}}
Low-mass systems dominate the galaxy population at all redshifts, and are crucial for the hierarchical build-up of galaxies. However, the processes driving the formation of dwarf galaxies and how the environment affects their evolution are poorly understood \citep[e.g.][]{Silk2012}. The transformation from a star-forming to a passive dwarf galaxy could involve ISM removal by ram-pressure, a series of tidal interactions that kinematically heat the stellar disc and thereby modify its morphology \citep{Gnedin2003,Lisker2009,Toloba2015}. The main goal of the Nearby Clusters survey is to study the properties and formation history of dwarf galaxies in high-density environments.\looseness-1

Specific scientific questions addressed in this layer of the survey include: Are dwarf galaxies primordial or the end-products of galaxy transformations? How do dwarf galaxies contribute to `archaeological downsizing' of galaxies \citep[e.g.][]{Thomas2005, Nelan2005}? What processes drive the transformation of dwarf galaxies in clusters? What are the internal dynamics of dwarf galaxies; are they dark-matter-dominated systems, and what are their angular momenta? How do these processes depend on properties of the galaxies themselves, e.g.\ stellar populations, metallicities, mass, and star-formation history, as well as their local environment?\looseness-1

The WCN survey is a low-spectral-resolution programme combining the MOS and mIFU modes of WEAVE (see Fig.~\ref{fig:WCN}). WCN will observe an X-ray flux-limited sample of 47 nearby galaxy clusters in the redshift range $0.01 < z < 0.04$, covering a large range in cluster mass ($13.2 < \log(M_{\ast}/\Msun) < 14.5$), corresponding to X-ray luminosities of $\log(L_{500}) \in [42,45]$. This survey will provide single-fibre spectroscopic information for several thousands of dwarf-galaxy ($M_r<-16$) members, thus providing the largest spectroscopically confirmed sample of dwarf-galaxy cluster members observed so far. In addition, we will obtain spatially resolved spectroscopic information for $\sim 1000$ dwarf cluster members using the mIFU mode. \looseness-1

The cluster sample is compiled from ROSAT All-Sky Survey data: the ROSAT Brightest Cluster Sample \citep[BCS;][]{Ebeling1998} and its extension \citep[eBCS;][]{Ebeling2000}. The BCS is 90 per cent complete for fluxes higher than $4.4 \times 10^{-12}\,\mathrm{erg\,cm^{-2}\,s^{-1}}$ in the ROSAT 0.1--2.4\,keV band. The eBCS extends the BCS down to $2.8 \times 10^{-12}\,\mathrm{erg\,cm^{-2}\,s^{-1}}$ with 75 per cent completeness. The WEAVE field of view allows us to observe galaxies in the range $0 \leq r/r_{200} < 1$--$2$ in a single pointing for the selected redshift range, where {$r_{200}$} is the cluster radius within which the mean density is 200 times the critical energy density of the universe at that redshift. The general survey strategy of using 1-hour observing blocks alongside dedicated, existing photometry ensures that selected targets will have a signal-to-noise ratio ($S/N$) $\geq 5$ per {\AA} in the galaxy spectra for galaxies with mean surface brightness within a 1.3\,arcsec aperture of $22\,\mathrm{mag\,arcsec^{-2}}$. We thus expect to obtain LR spectra with high enough $S/N$ to obtain reliable radial velocities for all galaxies down to $M_r\sim -16.0$ in our selected clusters.\footnote{At the mean redshift of the cluster survey, the limiting magnitude of the observations will be $M_r \sim -15.0$, and for the closest ones $M_r\sim -14.0$.} This is equivalent to tracing the galaxy luminosity function down to at least $M^* +6$ for all of the clusters, where $M^*$ is magnitude of the knee of the luminosity function. The sample is statistically large enough to study the effects of cluster properties (e.g.\ mass, galaxy density, cluster velocity dispersion) on the evolution of dwarf galaxies. We will therefore be able to explore systematic differences in dwarf galaxy properties related to the main properties of cluster environments, together with cluster-to-cluster variations.\looseness-1

\begin{figure}
\centerline{\includegraphics[width=\columnwidth]{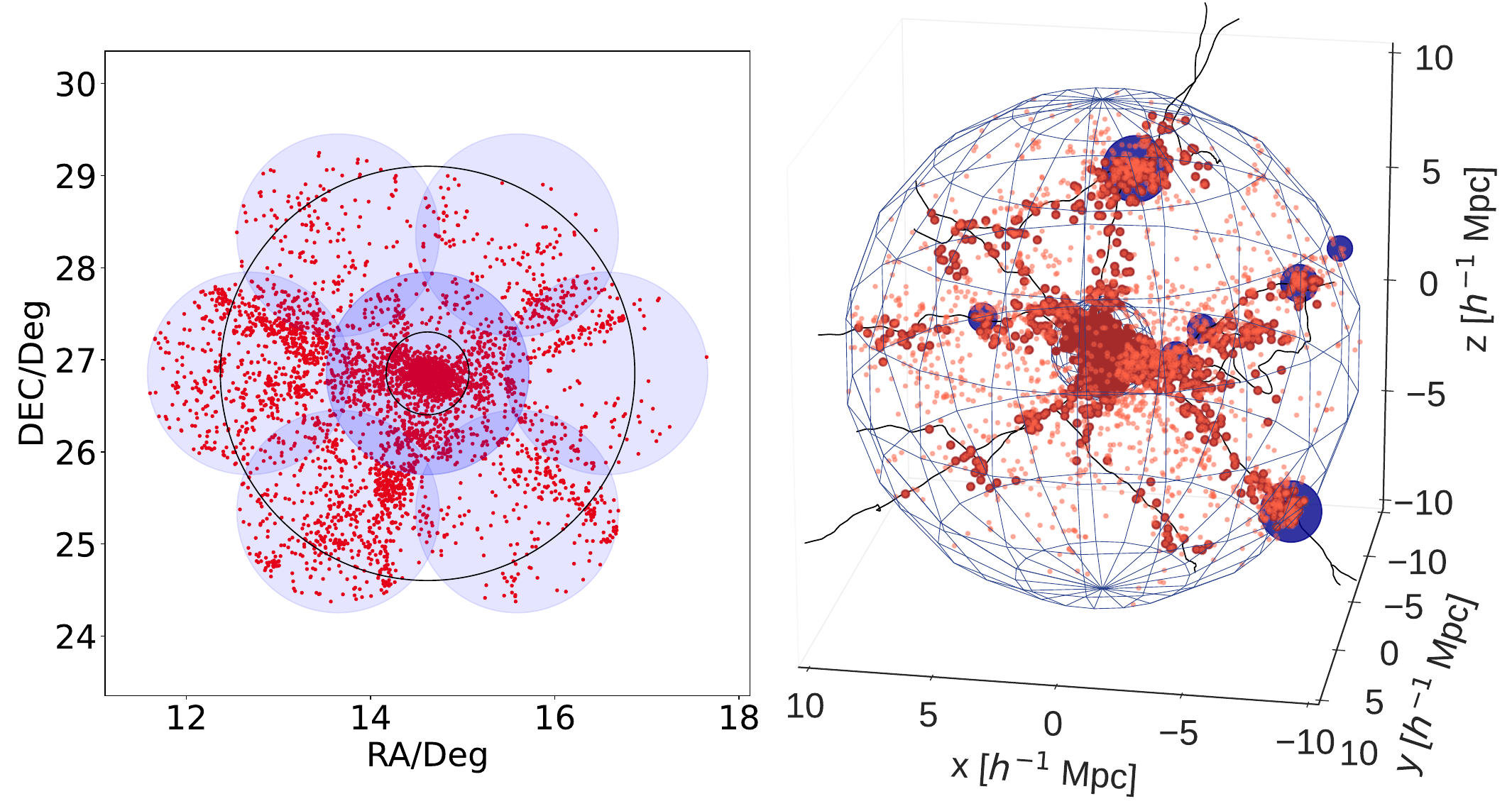}}
    \caption{The WWFCS will observe 20 nearby clusters with several WEAVE pointings to cover at least $5R_{200}$. The left side shows the observing strategy of WWFCS based on simulations from \textsc{The ThreeHundred} simulations tailored to the survey \citep{2022MNRAS.517.1678C}. Each blue disk represents one WEAVE pointing; the black circles are at 1 and $5r_{200}$. This particular cluster is covered with 7 pointings, but others require fewer or more. The right side of the figure shows filaments identified using the software \textsc{Disperse} and groups in the infall region of the same cluster simulation (see \citealt{2020MNRAS.494.5473K} for details).}
    \label{fig:WWFCS}
\end{figure}

\item {\bf{Layer 2 -- WEAVE Wide-Field Cluster survey (WWFCS)}:}
To understand the effect of environment on galaxy evolution, we must understand processes acting during the assembly of clusters. This includes the collapse and feeding of clusters via filaments and groups from the cosmic web. This layer of the WEAVE Galaxy Clusters survey concentrates on the still-to-be-explored accretion physics of cluster infall regions. The WWFCS will observe galaxy clusters out to five times their viral radius to include galaxies in filaments directly feeding the clusters as well as infalling groups. WWFCS will help answer important questions such as: What kind of galaxies feed clusters? Does their fraction vary with the mass and dynamical state of the cluster? Is pre-processing in the form of gas-specific (e.g.\ ram-pressure stripping) or gravitational (e.g.\ galaxy-galaxy mergers) processes significant in groups or filaments outside the cluster cores? How and where do changes in morphology and star formation occur, over what timescales and at which mass? How is AGN activity related to filaments? \looseness-1 
The WWFCS will observe 16--20 clusters at $z\sim0.05$ with $13.8 < \log(M_*/\Msun) < 15.5$ (where $M_*$ is the stellar mass of the cluster) that are drawn from the WINGS sample \citep{Fasano2006}, making use of the availability of extensive ancillary data in the inner region. We will obtain a complete census of galaxies in and around clusters over a broad range of masses ($10^9 < (M_{*}/\Msun) < 10^{11.5}$), corresponding to a magnitude limit of $r=19.75$. To ensure uniform coverage out to at least $\sim5 R_\mathrm{vir}$, each cluster will be covered by between 6 and 18 MOS pointings, each with WEAVE's $2^\circ$ field of view (see left side of Fig.~\ref{fig:WWFCS}). Given the low density contrast of filaments \citep{Martinez2015}, it is important to target the highest possible fraction of cluster and infall region members. This will be achieved by selecting galaxies with magnitudes, colours, and importantly, photometric redshifts. Photometric redshifts are obtained from the 12-band photometry of the Javalambre Photometric Local Universe Survey \citep[J-PLUS;][]{Cenarro2019} fields and open-time observations using the spectral energy distribution fitting code \textsc{LePhare} \citep{arnouts11}.
Tests have shown that rejecting targets with $\sim3$ times the typical photometric redshift error of J-PLUS introduces at most $\sim3$ per cent incompleteness while increasing the number of cluster structure members by a factor of $\sim2$ in comparison to a pure magnitude and colour cut. This leads to an unprecedented total of 4000--6000 structure member galaxies per cluster, which will ensure well-sampled density fields, crucial for mapping and characterizing the filamentary structure around these clusters (see right side of Fig.~\ref{fig:WWFCS}).\looseness-1

The WWFCS programme will use WEAVE's MOS mode in low spectral resolution, covering 370--960\,nm at $R\sim5000$. The spectra will yield emission-line diagnostics, accurate spectral breaks, and absorption-line information for all galaxies. Velocity dispersions and accurate absorption-line information will be available for the brighter half of the galaxies individually, and statistically for the fainter half through spectra stacking. This will shed light on the preferred routes of mass accretion and investigate the effect that this has on galaxy properties. Furthermore, the spectroscopic redshifts will be crucial for identifying cluster volumes for accurate filament extraction in redshift space \citep{2021MNRAS.503.2065K}.
Defining filaments is a non-trivial task due to the multi-scale and diffuse nature of the cosmic web \citep{Rost2020}. The challenge is to map galaxy environments in sufficient detail within -- and to sufficient distance around -- the dominating clusters, where the filaments converge (see right side of Fig.~\ref{fig:WWFCS}). In preparation for this survey, we have evaluated the performance of filament-finding strategies for WWFCS and quantified their robustness on hydrodynamical simulations of \textsc{The ThreeHundred} project \citep[][]{2020MNRAS.494.5473K,2021MNRAS.503.2065K,2022MNRAS.510..581K}. The simulations are tailored to represent WWFCS cluster observations, which will also aid the interpretation of the survey's findings \citep{2022MNRAS.517.1678C}.\looseness-1

\begin{figure}
	\includegraphics[width=0.48\textwidth]{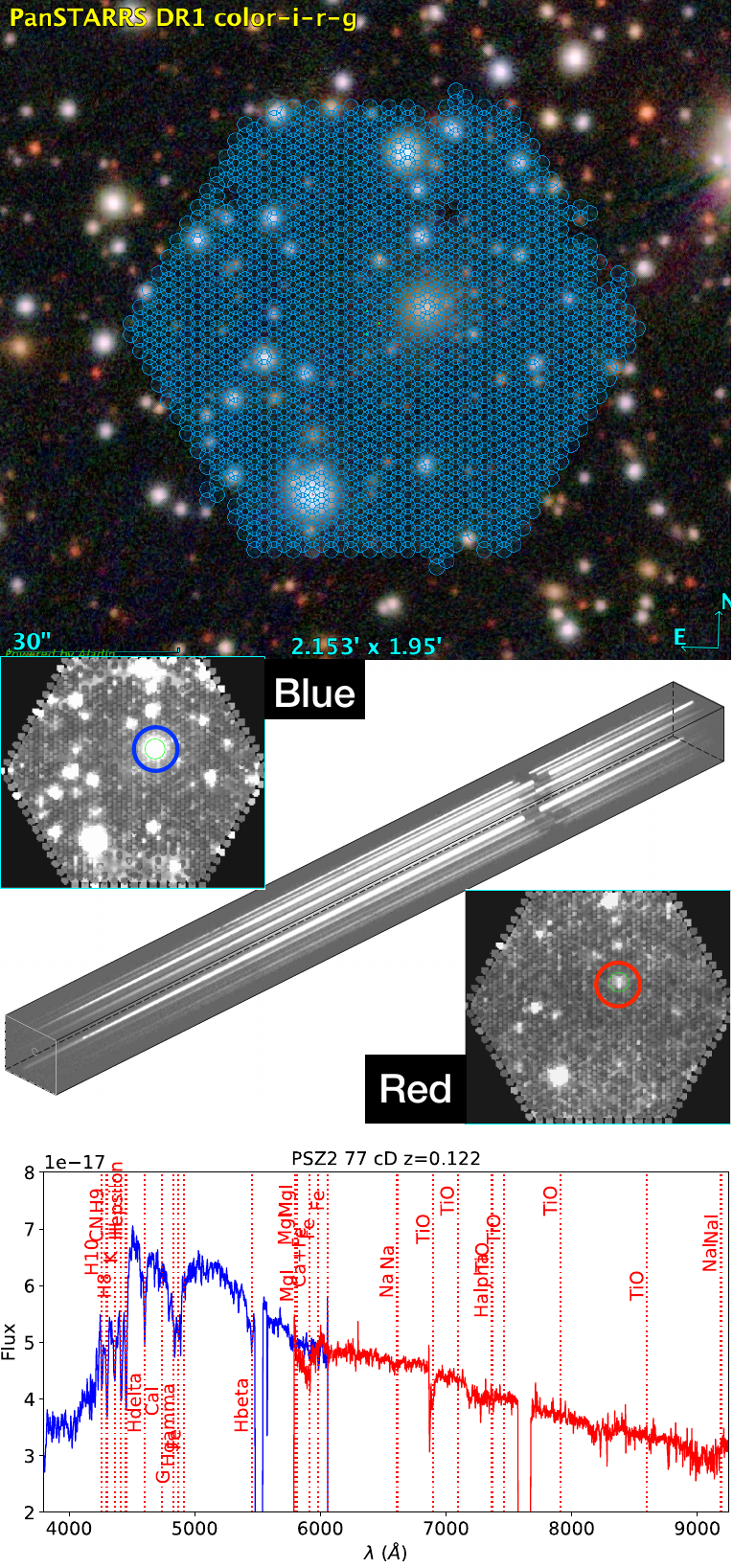}
    \caption{Top panel: LIFU footprint (with three dithers) superimposed on PanSTARRS1 imaging of cluster PSZ2-77. Middle panels: sum of simulated blue-arm (top left) and red-arm (bottom right) data cubes from OpR3b. The circles highlight the brightest cluster galaxy; green circles label the extraction apertures in each data cube. A projection of the blue-arm data cube (short wavelengths at bottom left, long wavelengths at top right) is shown diagonally across the middle panels. Bottom panel: one-dimensional spectral extraction of the (simulated) brightest cluster galaxy with spectral features indicated in the plot. The blue- and red-arm spectral regions are covered by the blue and red spectra, respectively.}
    \label{fig:WCC}
\end{figure}

\item {\bf{Layer 3 -- WEAVE Cosmological Clusters survey:}}

The main goal of the Cosmological Clusters survey is to study the evolution of galaxies in the cores of clusters out to $z\sim0.5$ and to place constraints on cosmological parameters and global scaling relations using a complete sample of SZ-selected clusters. The baryonic component of clusters contains a wealth of information about the processes associated with galaxy formation, including the efficiency with which baryons are converted into stars and the effects of the resulting feedback processes on galaxy formation, which then allows us to estimate the total baryon ($\Omega_\mathrm{b}$: e.g.\ \citealt{2005MNRAS.364..909V,2011ASL.....4..204B}) and dark matter densities. Moreover, cluster abundance studies as a function of mass and redshift, $N(M,z)$, are powerful cosmological tools \citep[e.g.][]{2002ARA&A..40..643C} that allow us to set constraints on cosmological parameters such as the dark matter and dark energy densities, or the equation of state of the dark energy \citep[e.g.][]{2009ApJ...692.1060V,2010MNRAS.406.1759M}. Furthermore, by selecting galaxy clusters at $z\lesssim0.5$, WEAVE can directly observe the evolution of the stellar populations of galaxies in cluster cores over a cosmic epoch over which galaxy transformations are already well documented \citep[e.g.][]{1982ApJ...263..533D,1984ApJ...285..426B,1997ApJ...490..577D} and which can be directly compared with other WEAVE surveys like StePS (see Section~\ref{subsec:StePS}). Specific science questions that will be addressed include: 
obtaining an accurate mass calibration and global scaling relations for a sample of 75 SZ clusters selected from Planck \citep{2016A&A...594A..27P};
using the calibration to provide error bars a factor of two smaller than the existing ones on the key cosmological parameters $\sigma_8$ and $\Omega_\mathrm{m}$;
studying the evolution of the stellar populations for a large sample of massive galaxies in the central regions of a galaxy cluster in the last 6\,Gyr of evolution, using the method pioneered by \citet[][]{2007MNRAS.376..125S};
making comparisons between the $z=0$ cluster galaxies in Layer 1 and 2 surveys and the $z\sim0.5$ field galaxies in StePS to understand how and when early-type galaxies cease and quench their star-formation activity, and transition onto the red sequence.\looseness-1

The Cosmological Clusters survey will use both the LIFU and MOS modes in the low-resolution mode (see Fig.~\ref{fig:WCC} for an example of a dithered LIFU footprint on a cluster). This survey will observe a total of $\sim70$ SZ-selected clusters, with low-projected-density clusters (typically at $z<0.25$) observed with the MOS mode and high-projected-density clusters (typically at $z>0.25$) observed with the LIFU.\looseness-1
\end{itemize}

\subsection{The Stellar Populations at intermediate redshifts Survey (StePS)}
\label{subsec:StePS}

\begin{figure*}
\begin{center}
\includegraphics[width=0.85\textwidth,trim=0cm 0.0cm 1cm 1cm, clip=true, angle=0]{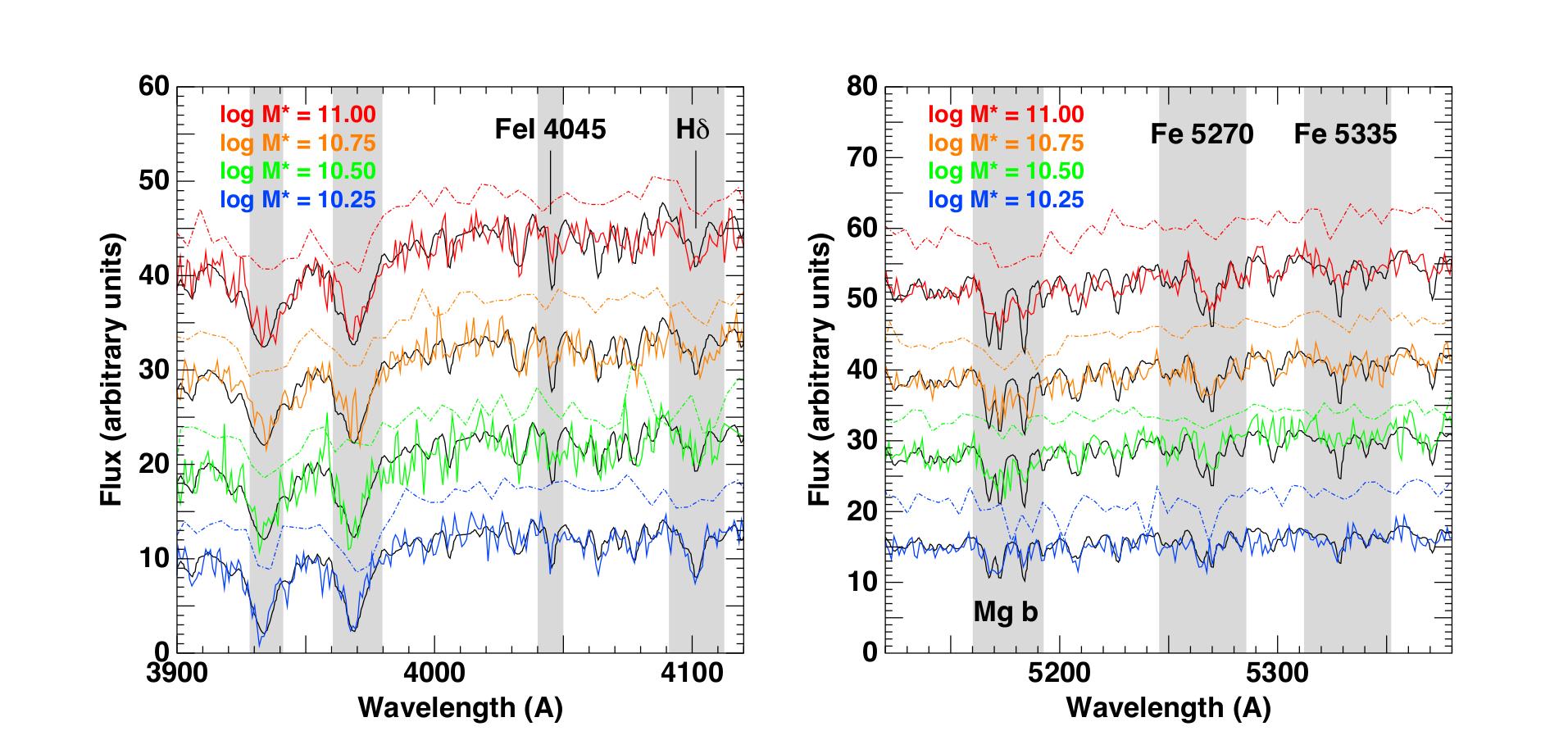} 
\caption{Comparison of typical SDSS \citep[$z\sim 0.1$, solid coloured lines]{2000AJ....120.1579Y} and VIPERS \citep[$z\sim 0.5$, dashed lines]{Guzzo2014} spectra of four early-type galaxies of different stellar masses (see legend). VIPERS spectra are vertically offset by 10 units for clarity. The $S/N$ of all of these spectra are close to the median value for galaxies of that mass in the corresponding survey. The black curves show model spectra of single stellar populations with solar metallicities and ages of 2.5--10\,Gyr (from bottom to top) from the MILES spectral library \citep{Vazdekis2010}. Key spectral indices used for estimating the stellar ages or metallicities are indicated, from left to right (CaH+K, Fe4045, H$\delta$, Mg$b$, Fe5270, Fe5335), the grey bands  indicating the width of the central Lick wavebands. The low resolution ($\sim16$\,{\AA} FWHM) of the VIPERS spectra smooth out (or fill in) the key spectral absorption features, and the low $S/N$ prevent reliable measurements of the Lick indices for individual galaxies. The higher resolution SDSS spectra retain these absorption features, allowing us to reconstruct their finer features.}\label{fig:StePS}
\end{center}
\end{figure*}

One of the major goals of extragalactic astrophysics is to understand the physical processes that control the formation and evolution of luminous structures. While large-scale structure evolution in our Universe is theoretically understood and well reproduced by simulations on scales larger than $\sim 1$\,Mpc, on smaller scales the stellar mass growth of galaxies within dark matter haloes and thus the complex mechanisms sometimes called `gastrophysics' \citep{Bond1993} depend on highly non-linear processes (e.g.\ star formation, energetic feedback mechanisms, and mergers), whose detailed mechanisms are still largely unknown \citep{Kuhlen2012, BullockBoylan-Kolchin2017}.
Advanced theoretical models are needed to shed light on the mechanisms that regulate the connection between galaxies and their dark matter haloes, but the complexity of the problem is such that observations are essential to empirically constrain both theories and simulations of galaxy assembly history \citep{WechslerTinker2018}.\looseness-1

In the last few decades, the study of galaxies in the local Universe has greatly enriched our knowledge and understanding of galaxy evolution. Theoretical and empirical approaches are now anchored at $z\sim0$ by the large, uniform, complete spectroscopic measurements from the Sloan Digital Sky Survey \citep[SDSS;][]{2000AJ....120.1579Y}, providing high $S/N$ and `high resolution' ($R\sim 2000$) spectra of a large, representative sample of local galaxies.
However, despite their high quality and statistical richness, data from the local Universe are not, in general, the ideal tool for retrieving finer details of the star-formation and evolutionary history of galaxies. Archaeological reconstruction -- retrieving star-formation histories at earlier times from spectra observed at $z\sim0$ -- is an extremely difficult task: most of the local galaxies are too old \citep{Gallazzi2005} to resolve differences in the early star-formation history, due to the similarity of stellar spectra at ages $>5$\,Gyr.
An alternative approach is the lookback approach, studying and making a census of galaxies at earlier and earlier cosmic epochs, comparing snapshots of the galaxy population taken at different cosmic epochs. This approach enables us to trace back in time the evolution of galaxies as a population, but linking the galaxies observed in the local Universe directly to their actual progenitors at higher redshifts is a difficult task, due to our fundamental lack of understanding of what the actual progenitors of present-day galaxies of different types are.\looseness-1

None of the recent or current large ($N_\mathrm{gal} \geq 10\,000$) surveys of distant galaxies ($z \geq 0.3$) -- e.g.\ AGES \citep{Kochanek2012}, SHELS \citep{Geller2014}, GAMA \citep{Driver2012} and BOSS \citep{Dawson2013} at the lower redshift end; DEEP2 \citep{Newman2013}, VIPERS \citep{Guzzo2014}, zCOSMOS \citep{Lilly2009}, VVDS/VUDS \citep{LeFevre2005,  LeFevre2015} and GMASS \citep{Kurk2013} at the higher redshift end -- have adequate spectral resolution and/or quality to derive stellar population properties in sufficient detail for archaeological studies. Low resolution and low $S/N$ spectroscopic observations can provide galaxy redshifts, rough measurements of emission lines, and a reliable estimate of the $D_{4000}$ break, but detailed information on the stellar and gas content of galaxies can only be achieved through higher $S/N$ and higher resolution ($R\sim5000$) spectroscopy -- ingredients needed to provide reliable measurements of the absorption features in the stellar continuum and detailed stellar population and gas modelling. The only notable survey to date designed to obtain high-quality, high-resolution spectra in the relatively distant Universe is LEGA-C, whose data are limited to the redshift window $0.6 \leq z \leq 1.0$ and to a sample of $\sim 3200$ galaxies within the COSMOS field. The LEGA-C spectra have a median $S/N \sim 20$, and resolution $R \sim 3500$ \citep[see][]{VanDerWel2016, Straatman2018}.\looseness-1

StePS will use the WEAVE spectrograph to perform a high-$S/N$ survey of galaxies at moderate redshift, targeting a volume-limited sample lying mainly in the redshift interval $0.3\lesssim z \lesssim 0.7$ \citep{2023A&A...672A..87I}. StePS is thus designed to fill the interesting, yet largely unexplored niche of the intermediate redshift slice, fitting in between the SDSS and the LEGA-C survey data. StePS and the SDSS alone already encompass 6\,Gyr of the age of the Universe, i.e.\ nearly one half of its present age, over the period when the cosmic star-formation-rate density continues to drop, while galaxies have almost doubled their stellar mass in dark matter haloes that are still assembling \citep{ConroyWechsler2009}.\looseness-1

To emphasize the need for high spectral resolution, consider that while Lick indices \citep{Worthey1994} tend to have central bandpasses of the order $20$--$30$\,{\AA}, the features themselves may only be $2$--$5$\,{\AA} wide. Moreover, accurately fitting the absorption lines is vital for measuring the Balmer emission lines, which often sit on top of deep absorption features that cannot be modelled and accounted for in typical low-resolution spectra ($R \sim 1000$, see Fig.~\ref{fig:StePS}).\looseness-1

Individual spectra of good $S/N$ are another important ingredient. While spectral stacking \citep[e.g.][]{Gallazzi2008} can enable us to obtain high $S/N$ spectra that provide the correct (light-weighted) average properties in a given bin of galaxy properties, it provides no information about the scatter around this average or the higher moments of the distribution. Pinning down these moments is important for quantifying the amount of stochasticity involved in galaxy evolution processes and the timescales of transitional phases. This clearly adds much more understanding to galaxy evolution than simply tracing the average time evolution of the properties of galaxy populations in given bins. Good-quality individual spectra are crucial for obtaining a description of any statistical distribution of physical properties that goes beyond the most typical. In addition, only a study of individual galaxies would allow for the identification of rare oddball objects, which may be the ones that will teach us the most interesting physics.\looseness-1

StePS will obtain approximately 25\,000 spectra of high $S/N$ and resolution $\sim1$\,{\AA} using WEAVE's LR mode ($R\sim5000$), spanning the wide wavelength range of $3660$--$9590$\,{\AA}.
StePS targets will be selected within four areas on the sky totalling $\sim25\,\mathrm{deg}^2$. The areas are a subset of the CFHTSLS-W1 ($02^{\mathrm{h}}18^{\mathrm{m}}00^{\mathrm{s}} -07^{\circ}00\arcmin00\arcsec$) and W4 ($22^{\mathrm{h}}13^{\mathrm{m}}18^{\mathrm{s}} +01^{\circ}9\arcmin00\arcsec$) regions, and of the ELAIS-N1 region ($16^{\mathrm{h}}12^{\mathrm{m}}10\fs0 +54^{\circ}30\arcmin00\arcsec$). An area centred on the COSMOS field ($10^{\mathrm{h}}00^{\mathrm{m}}28^{\mathrm{s}} +02^{\circ}12\arcmin21\arcsec$; \citealt{Scoville2007}) is also included, where richer ancillary data -- including HST imaging -- are available. This large sample will enable us to characterize the ingredients that shape the history of galaxy assembly with sufficient statistics, exploring galaxy evolution in a sufficient number of bins of different galaxy masses, galaxy morphologies and colours, environments, and cosmic epochs.\looseness-1

We have estimated that $S/N \geq 10$\,{\AA}$^{-1}$ in the observed $I$-band is an essential requirement to estimate the physical parameters of interest with reasonable accuracy, including -- but not limited to -- stellar ages, stellar and ionized gas metallicities and kinematics, and star-formation rates (see e.g. \citealt{2019A&A...632A...9C} and \citealt{2023arXiv230802635D}) for a detailed investigation of the potential of WEAVE-like spectra at such typical $S/N$ values).
Such $S/N$ ratios can be obtained in WEAVE's red arm with an exposure time of $\sim7$ hours in the $R\sim5000$ mode for galaxies down to total galaxy magnitudes $I_{AB} \leq 20.5$.
At a resolution of $R\sim5000$, which in WEAVE is defined as the low-resolution mode, and at the required signal-to-noise level, even narrow features of galaxy spectra will be easily resolved, allowing for an accurate measurement of the intrinsic stellar velocity dispersion.  At these -- what would be termed in this area of study -- `high' resolutions, galaxy spectra reveal a vast number of spectral features, providing an immense resource for fitting models, including the ability to disentangle multiple stellar populations, such as the effect of a frosting of a low-level of young stars overlaid on an underlying old stellar population, or the effects of recent short bursts of star formation followed by rapid quenching. Equally, accurately fitting the absorption lines is vital for measuring the Balmer emission lines, which often sit at the bottom of deep absorption features that cannot be modelled and be accounted for in lower resolution spectra.
The magnitude limit of $I_{AB} \leq 20.5$ corresponds to a logarithmic stellar mass limit (using a Chabrier IMF) of  $\log({M_*}/\Msun)=10.2$ at $z \sim 0.3$, 11.0 at $z \sim 0.5$ and 11.3 at $z \sim 0.7$, thus sampling the high stellar mass tail of the galaxy population. The galaxy surface density of targets, selected using high-quality photometric redshifts to be at $z \geq 0.3$, is $\sim 1500$ per square degree, well-suited for WEAVE's multiplex capabilities with a multi-pass strategy involving several passes of each sky region.\looseness-1

For the majority of observed galaxy spectra, StePS will derive stellar ages, stellar and gas metallicities, dust attenuation, electron densities, and ionization parameters. We will infer the past evolution of galaxies at different masses and redshifts and relate their star-formation histories to their intrinsic (e.g.\ stellar mass, galaxy morphology, and dominant power source) and environmental properties, making use of the option to stack spectra when their $S/N$ are low. Using  available ancillary photometric data will allow us to create well-defined stacks of spectra of even relatively small homogeneous sub-samples.
The observed spectra will also be used to provide information on gas kinematics (including the presence of outflows) and stellar velocity dispersions, allowing us to perform a dynamical classification of the observed galaxies and to explore in detail the link between star-formation history, mass-assembly history, and dynamics.\looseness-1

Last but not least, the StePS redshift range, coupled with the wide spectral range of WEAVE, will enable observations of the near-ultraviolet and $U$-band rest-frame spectral windows, bands that can provide unprecedented constraints on the youngest components in galaxies \citep{Vazdekis2016,SalvadorRusinol2020}.\looseness-1

\subsection{The WEAVE-LOFAR survey}
\label{subsec:WL}

\begin{figure*}
\subfigure{\includegraphics[width=0.495\textwidth,trim=0cm 0cm 1cm 0.7cm, clip=true]{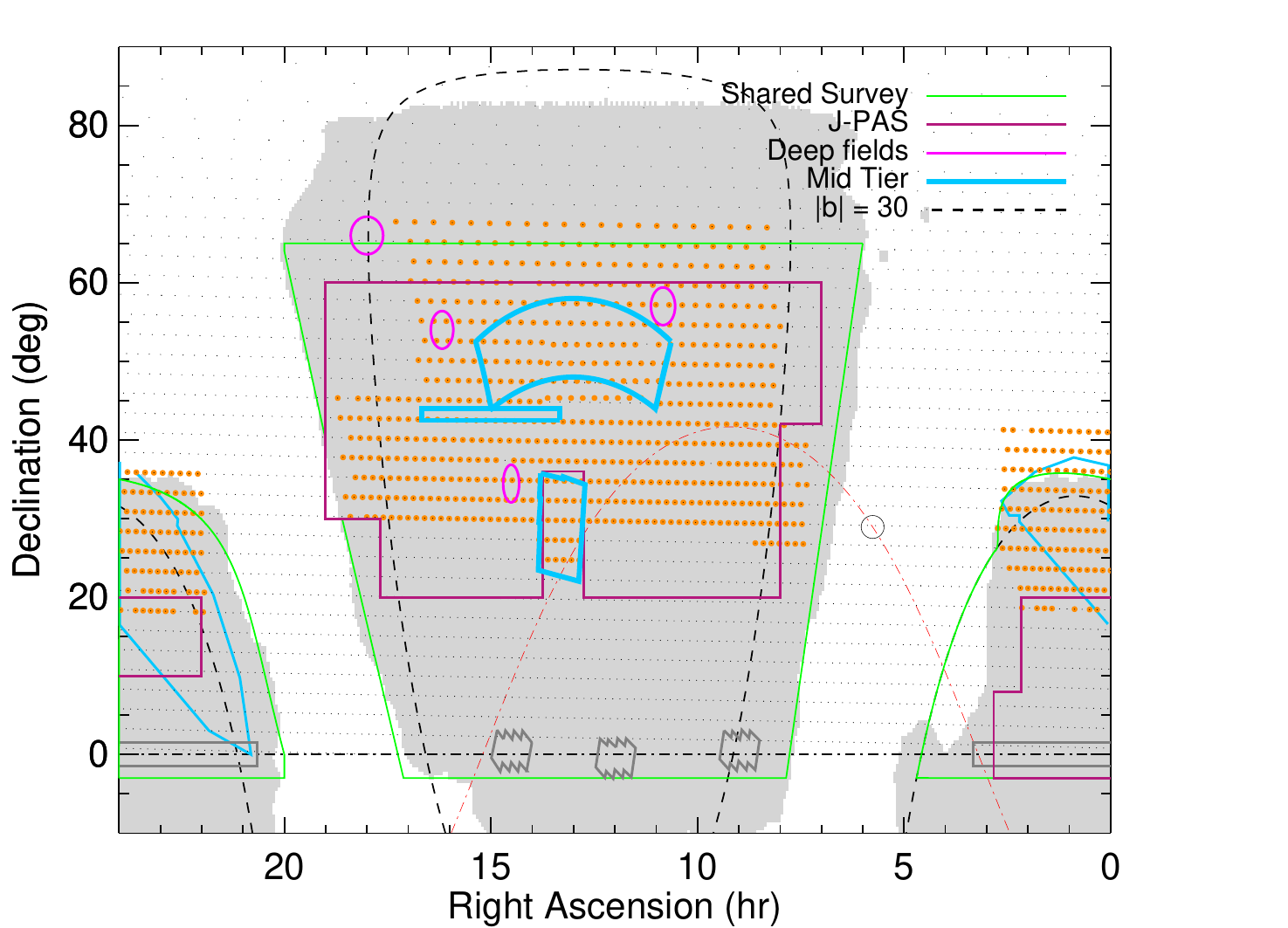}}
\subfigure{\includegraphics[width=0.495\textwidth,trim=0cm 0cm 1cm 0.7cm, clip=true]{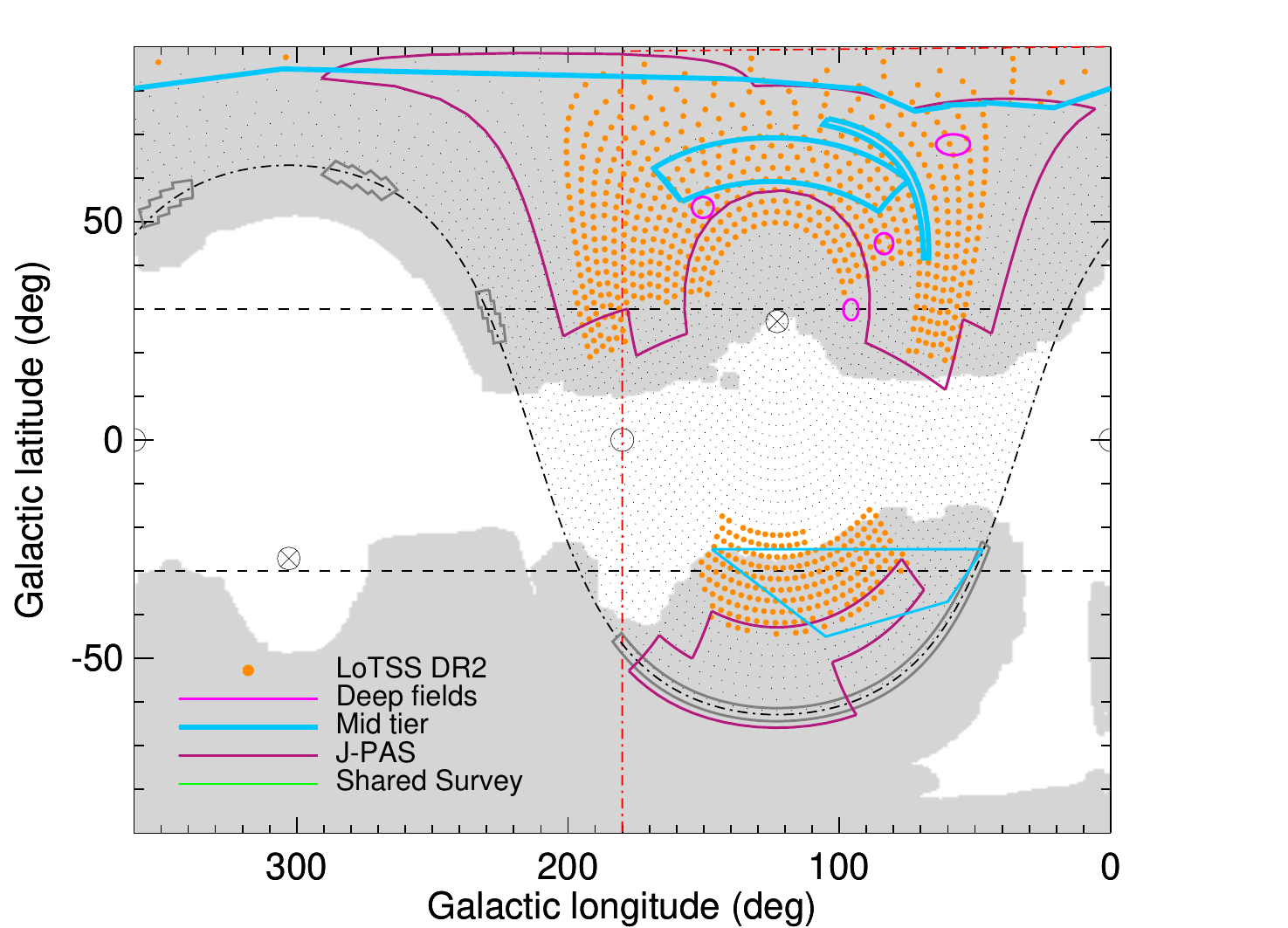}}
\caption{Outlines of the main WEAVE-LOFAR fields, in the context of the LoTSS data, in equatorial (left) and Galactic (right, compare with Fig.~\ref{fig:GAall_lb}) coordinates. In both versions, the LoTSS hemisphere grid is shown as black points, while orange circles indicate the LOFAR pointing centres included in the second data release of LoTSS, covering approximately 5700\,deg$^2$ of the extragalactic sky. Also overlaid are the principal WEAVE-LOFAR Deep Fields (purple circles), and outlines of mid-tier fields (in light blue, including HETDEX, the region covered by LoTSS DR1; see also Fig.~\ref{fig:hetdex}). The provisional area that will be observed by J-PAS is outlined in purple, while the perimeter of the fields shared with the WEAVE Galactic Archaeology and WEAVE-QSO teams is outlined in green. The coverage of the Legacy Surveys \citep{2019AJ....157..168D} that we use for cross-identification in LoTSS is indicated by the grey shaded background. In the right panel, the Galactic Centre and Anticentre are indicated by black open circles, while the poles in the equatorial reference frame are indicated by black circled crosses. In the left panel, the Galactic Anticentre is indicated by a grey open circle.}
\label{fig:lofar_progress}
\end{figure*}

The WEAVE-LOFAR survey \citep[and in preparation]{smith16b} will provide the spectroscopic information to unlock the power of the International Low Frequency Array Telescope\footnote{\url{http://www.lofar.org}} \citep[LOFAR;][]{vanhaarlem13}. Since its inception, an important driver for LOFAR has been to carry out a series of surveys of the sky at low radio frequencies, in particular to advance our understanding of the formation and evolution of galaxies, galaxy clusters, and AGNs. The LOFAR Surveys Key Science Project \citep[LSKSP;][]{rottgering11} is carrying out this aim, using a suite of related surveys.

The observations with LOFAR's High-Band Antennae (HBA) include those used for LOFAR Two-metre Sky Survey \citep[LoTSS: ][]{shimwell17}, and are providing observations across the whole of the Northern sky between 120 and 168\,MHz.
A lower frequency counterpart of LoTSS, using the LOFAR Low Band Antennae (LBA), is the LOFAR LBA Sky Survey \citep[LoLSS:][]{2021A&A...648A.104D}, covering the frequency range 42--66\,MHz.
Because of the influence of ionospheric effects on low-frequency observations, LoTSS required the development of new direction-dependent calibration schemes \citep[e.g.][]{tasse14,williams15,hardcastle16,vanweeren16,2018A&A...611A..87T}. It is now possible to routinely generate science-quality, pipeline-processed images and mosaics with unsurpassed sensitivity, image quality, resolution, and field of view. The most recent second data release of LoTSS \citep[DR2;][]{2022A&A...659A...1S} builds on the huge success of DR1 \citep{shimwell19,williams19,duncan19}, and consists of LOFAR maps and catalogues, plus extensive ancillary information for more than four million sources identified at 150\,MHz at $>5\sigma$ over 5700\,deg$^2$. The median root mean square of the LoTSS data is 83\,$\upmu$Jy\,beam$^{-1}$, while the resolution of the LoTSS mosaics is 6~arcsec, and the astrometric accuracy relative to the {\it Gaia} reference frame is $<0.2$~arcsec, ideal for selecting WEAVE targets. Progress continues to be rapid, and well over half of the Northern sky has so far been observed.

The LSKSP team recently completed the first data release for its Deep Fields \citep[fully described in four works:][]{2021A&A...648A...1T,2021A&A...648A...2S,2021A&A...648A...3K,2021A&A...648A...4D}, along with a slew of new science
\citep[e.g.][]{gloudemans2020,hardcastle2020,2021A&A...648A..12H,mandal2020,morganti2020,osinga2020,smith2020,wang2020,2022MNRAS.511.3250M,2022MNRAS.513.3742K}. The LoTSS Deep Fields first data release is based on hundreds of hours of 150\,MHz observations of the best multi-wavelength `famous fields' visible from Northern Europe. The LoTSS Deep fields DR1 includes tens of deg$^2$ of observations spread over three fields: Bo\"otes and Lockman Hole \citep[described in][]{2021A&A...648A...1T}, as well as the ELAIS-N1 field \citep[which contains the deepest data, reaching 20\,$\upmu$Jy RMS:][]{2021A&A...648A...2S}. The Deep Fields data release includes extensive ancillary information, including host galaxy identifications for over 97 per cent of the sources \citep{2021A&A...648A...3K}, as well as photometric redshifts \citep{2021A&A...648A...4D}. Future Deep Fields releases will be based on even deeper 150\,MHz data, and include the field at the North Ecliptic Pole (NEP). Each field's position is indicated in Fig.~\ref{fig:lofar_progress}.

LOFAR selection of targets for WEAVE is ideal since 150\,MHz observations identify targets on the basis of radio activity (whether that activity is due to star-formation or accretion), and independent of dust obscuration. Whilst the latter effect means that LOFAR-selected samples are more representative of the activity in the Universe than samples identified at other wavelengths (e.g. in the optical), the former ensures that the WEAVE-LOFAR sample is rich with emission-line sources. WEAVE's low-resolution mode having continuous wavelength coverage between 360 and 960\,nm and high throughput means that we expect a redshift success rate approaching 100 per cent out to $z = 1$, and to be able to obtain redshifts for samples of extreme sources out to beyond $z = 6$.

However, spectra are required for much more than simply measuring redshifts; it is only by using these data that we are able to produce the most robust source classifications, and distinguish between sources dominated by star formation, i.e. star-forming galaxies (SFGs), and those dominated by accretion onto an AGN \citep[e.g. using emission line classifications based on the ratios of Balmer and forbidden lines;][]{BPT}. Spectroscopy also allows us to reliably distinguish between different accretion modes among the AGNs, determining whether they are dominated by efficient accretion of cold gas (resulting in high-excitation sources) or by inefficient accretion of hot gas \citep[resulting in low-excitation sources; e.g.][]{best2012,osullivan15}. The ratios of equivalent widths of forbidden and Balmer lines as well as the [\ion{O}{iii}] emission line are widely used for this purpose \citep[e.g.][]{Laing1994,Tadhunter1998,Buttiglione2010,best2012}. These capabilities are essential given that, even in the widest-area LOFAR observations, the 150\,MHz source population is diverse at the faint radio-flux densities sampled, consisting predominantly of SFGs and radio-quiet AGNs. LoTSS data are considerably deeper than all other existing surveys of comparable areas including FIRST \citep{becker95}, NVSS \citep{condon98} and VLASS \citep[e.g.][]{lacy2018,myers2018} and, interestingly, at the faintest flux levels probed by the deepest LOFAR data the SFGs and radio-quiet AGNs comprise more than 90 per cent of the radio source population (Best et al., in preparation); spectroscopic follow-up of large statistical samples of SFGs and radio-quiet AGNs detected by the deep LOFAR surveys will enable detailed characterization of these populations.
Spectroscopy of radio sources also permits us to measure velocity dispersions, estimate metallicities (the high resolution of WEAVE -- even in its `low-resolution' mode -- is essential here), and derive virial black-hole mass estimates.

Some of the topics that WEAVE-LOFAR will address include: the star-formation history of the Universe; the evolution of accretion and AGN-driven feedback; the nature of the Epoch of Reionisation (EoR); cosmology; cluster halos and relic radio sources; and radio galaxies and protoclusters.  We expect, for example, to identify a statistical sample of around 50 radio galaxies at $z>6$ \citep{saxena17,2018MNRAS.475.5041S}; these sources are pivotal for conducting 21-cm absorption experiments, giving a new window on the change of state brought about in the intergalactic medium during the EoR by the first stars\slash galaxies\slash AGNs. EoR 21-cm studies offer the best way to answer key questions about this poorly understood period in cosmic history: how long did it last, and which sources were responsible? The MOS component of WEAVE-LOFAR will obtain more than a million spectra of 150\,MHz-selected sources, almost three orders of magnitude more than the largest existing spectroscopic surveys targeting radio sources. These large numbers of sources will lend immense statistical power to this uniformly selected and homogeneous spectroscopic data bank, offering us the chance to simultaneously model the populations of SFGs and AGNs as a function of stellar mass, environment, and redshift. In this way, we can shed new light on the complicated interplay between the key processes thought to shape galaxy formation and evolution. WEAVE-LOFAR spectroscopic redshifts will also enable much improved cosmological parameter determination from LoTSS, building on the earlier work by \citet{2020A&A...643A.100S}, \citet{2021MNRAS.502..876A} and \citet{2022ApJ...928...38T}.

\begin{figure*}
\includegraphics[width=0.85\textwidth]{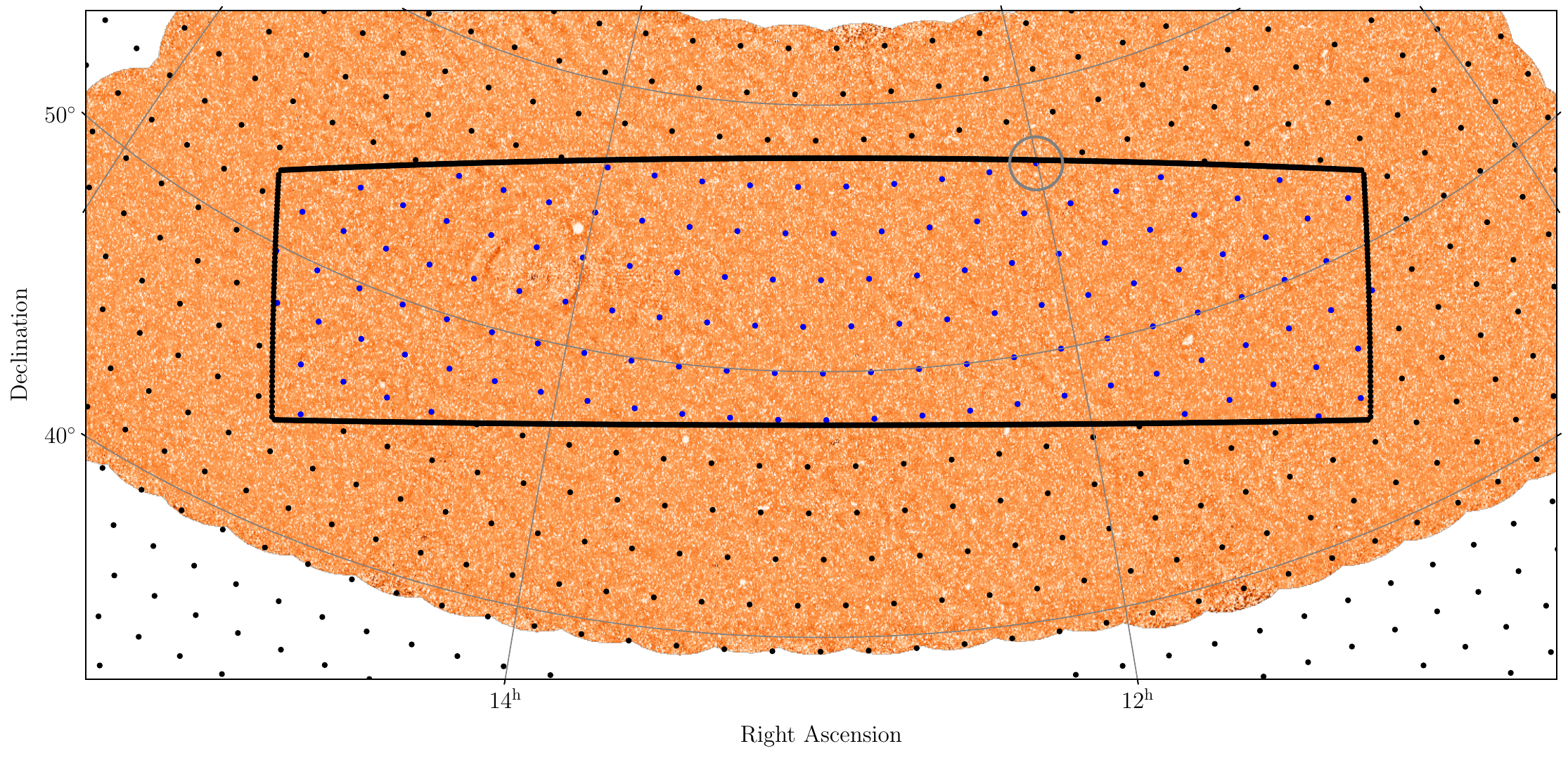}
\caption{150\,MHz mosaic of the region surrounding the HETDEX field using LoTSS data from the second data release (red colour scale, from Martin Hardcastle, private communication). The HETDEX spring field is one of the principal areas of sky in the WEAVE-LOFAR `Mid' tier. Overlaid on the map (the full-resolution version of which reaches a median noise level of 69\,$\upmu$Jy/beam and contains $>500$k sources detected at $>5\sigma$ at 150\,MHz) is the pointing grid for the fields shared with the WEAVE Galactic Archaeology and WEAVE-QSO surveys, covering the whole sky north of Dec = $-10^\circ$ \citep[black dots, derived using the method of][]{saff1997}. The blue circles indicate those WEAVE pointings on the hemisphere grid within the outline of the HETDEX survey \citep[black thick line,][]{hill2008}. Note that this grid differs from the one used for LoTSS on account of LOFAR's even larger field of view. The black thick line also corresponds to the WQ-HighDens footprint boundary for WEAVE-QSO. The grey circle indicates the field of view of WEAVE in its MOS mode, to give a sense of the scale of this field.}
\label{fig:hetdex}
\end{figure*}

To achieve these goals, WEAVE-LOFAR has been designed to predominantly utilize WEAVE's MOS mode, using a tiered layout. The tiers -- chosen to statistically sample the redshift-luminosity plane, the changing demographics of the radio source population, and the processes driving their evolution -- are as follows:
\begin{itemize}
\item {\bf{Wide tier}}, targeting sources brighter than $\sim$10\,mJy at 150\,MHz selected over an area of up to 10\,000\,deg$^2$ from LoTSS;
\item {\bf{Mid tier}}, targeting 150\,MHz sources brighter than $\sim 1\,\mathrm{mJy}$, including over the HETDEX spring field (Fig.~\ref{fig:hetdex}; where more than $400$\,deg$^2$ of the LoTSS first data release are located) and targeting fainter 150\,MHz sources $> 500\,\upmu\mathrm{Jy}$ over the unique Herschel-ATLAS NGP field;
\item {\bf{Deep tier}}, covering an area of up to 50\,deg$^2$ and targeting all radio-detected sources detected in the LoTSS Deep fields data covering the Bo\"otes, Lockman Hole, ELAIS-N1, and NEP fields. 
\end{itemize}

\noindent We emphasize that we do not make use of any selection criteria other than the 150\,MHz flux limits. 
Given the flux limit, sources detected in the Wide tier will have low-frequency radio spectra index information from the combination of LoTSS (144\,MHz) and LoLSS (54\,MHz) data.
In the Wide and Mid tiers, we will use 1-hour integrations for every target. In the Deep tier, we will revisit those `hard' sources (perhaps 30 per cent of the whole sample) that do not yield redshifts within a single 1-hour OB for up to four further hours. By doing this, we expect to obtain redshifts for 15--20 per cent of these hard sources and, in doing so, sample a part of the radio source population that would otherwise have remained completely inaccessible. Since we are relying on the prevalence of bright emission lines in our LOFAR-selected sample for achieving a high redshift success rate, it follows that for many of our targets the WEAVE LR spectra will not detect the continuum. Spectral stacking experiments -- relying on the large samples coupled with the high quality and uniformity of the WEAVE LR spectra -- are therefore a critical means of studying the continuum of 150\,MHz sources (and all of the information it contains) as a function of stellar mass, environment, redshift, and source classification.

We also have plans to use the WEAVE integral-field spectrographs (mIFU and LIFU) in relatively under-subscribed weather conditions to study individual sources in more detail and, in doing so, build on the broad picture that will be produced by the MOS survey. Using the WEAVE mIFUs in LR mode, we will obtain resolved spectroscopy of around 1000 bright background sources identified in the WL-Wide tier, and which have foreground neighbours that appear close in projection. This sample will enable us to identify absorption systems in the line of sight, and allow for the study of the properties of the extended neutral gas reservoirs. These reservoirs can only be detected in absorption against a bright background continuum source, but they play a critical role in fueling star formation \citep[e.g.][]{maddox2015,dutta2017}. We will also use the LIFU in HR mode to study the details of AGN feedback in nearby cluster cores, including the relationship between the radio jets, the intracluster medium and the central cluster galaxy. We will revisit the same targets as often as the conditions and WEAVE \scheduler\ allow, building increasingly sensitive data-cubes. With this data set, we can conduct the best-yet study of the $10^6\,\mathrm{K}$ plasma probed by coronal lines (revealing the cooling in action), as well as the kinematics of the ionized gas, and deconstruct the stellar population in these local analogues of starburst galaxies containing powerful AGNs at $z > 1$.

The scope of WEAVE-LOFAR is unprecedented, and it is therefore all but guaranteed that the WEAVE-LOFAR survey will also discover some wholly unexpected phenomena within the rich legacy data set that it will produce.

\subsection{The WEAVE-QSO survey}
\label{subsec:WQ}

\begin{figure}
\begin{center}
\includegraphics[width=0.75\columnwidth,trim=0.8cm 0.8cm 0.8cm 0.8cm, clip=true]{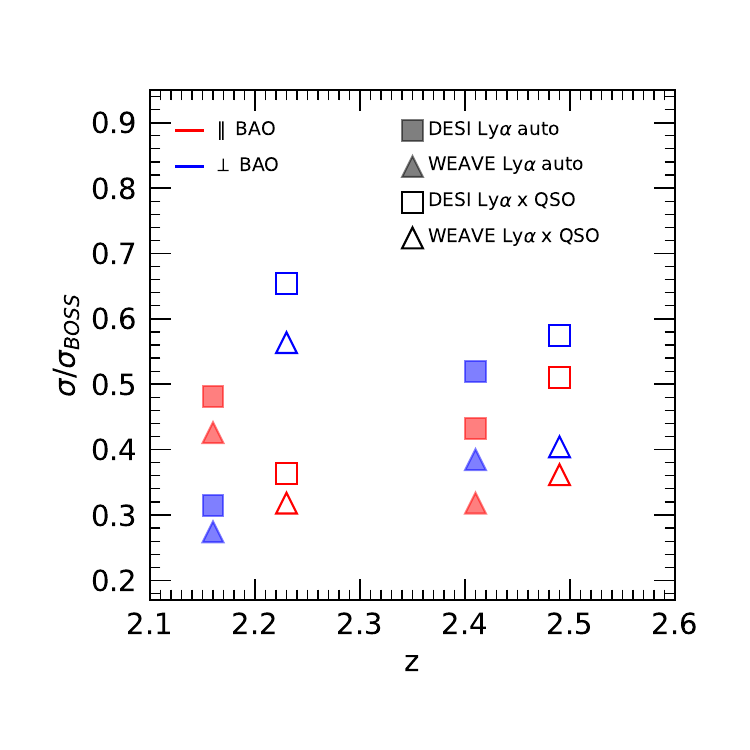}
\end{center}
\caption{Ly$\alpha$-forest BAO precision as a function of analysis redshift assuming a 6000\,deg$^2$ WEAVE-QSO survey with quasars at either $2.1<z_q<2.5$ (triangles at $z \sim 2.15$ for auto-correlation (filled) and $z \sim 2.23$ for cross-correlation (open)) or $2.5<z_q<3$ (triangles at $z \sim 2.4$ for auto-correlation (filled) and $z \sim 2.5$ for cross-correlation (open)), compared with the planned DESI survey \citep[squares;][]{2016arXiv161100036D}, which will cover both redshift ranges. Results are derived from random realizations of the Ly$\alpha$-forest auto-correlation (filled symbols) and the Ly$\alpha$-forest cross-correlations (open symbols) with quasar positions. These are broken down into a Hubble parameter constraint (radial BAO, $\parallel$) and an angular diameter distance constraint (transverse BAO, $\perp$). Both redshift bands offer a significant improvement on BOSS (and eBOSS) BAO precision, but the higher-redshift band is preferred, given WEAVE's performance with respect to DESI. Note that the Ly$\alpha$ forest is always observed in the foreground of the quasars, hence the lower redshift of the measurements compared to the redshift range of the quasars. In addition, the redshift of the predictions is the average of certain pairs -- pairs of Ly$\alpha$ pixels in the case of the auto-correlation, and pairs of Ly$\alpha$ pixels and quasars for the cross-correlation, resulting in a lower redshift for the auto-correlation than for the cross-correlation in each case.}
\label{fig:bao_summary}
\end{figure}

\begin{figure}
\begin{center}
\includegraphics[width=\columnwidth]{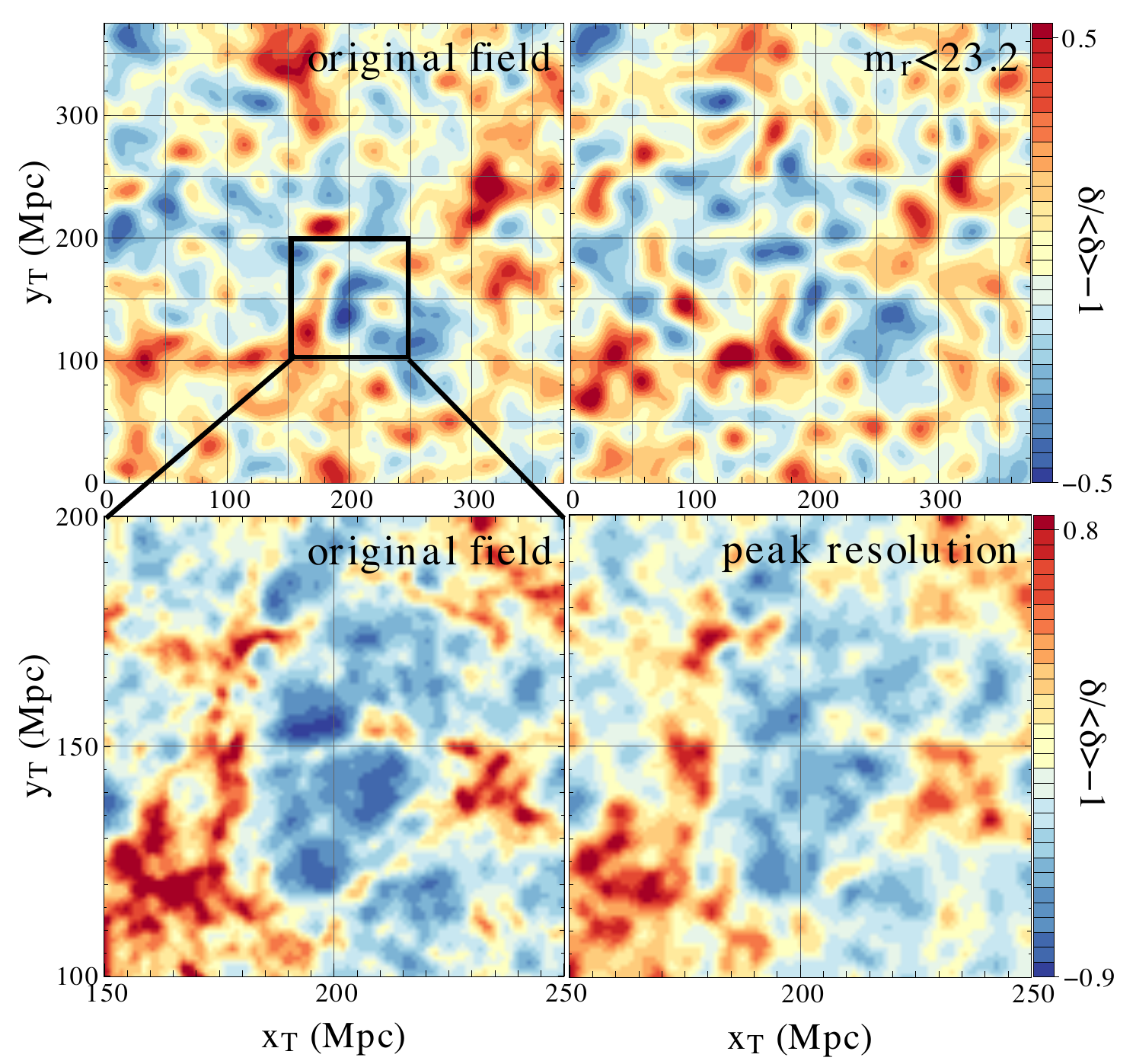}    
\end{center}
\caption{Transverse slice of the original (left) and reconstructed (right) density field in two WEAVE-QSO configurations at $z=2.12$. Top right: a projected $60\,\mathrm{cMpc}\,h^{-1}$-thick slice of the tomographic reconstruction illustrating the potential WEAVE-QSO sample in WQ-HighDens. The transverse correlation length $L_T$ is set to $14\,\mathrm{cMpc}\,h^{-1}$.  Top left: original density field smoothed with a Gaussian kernel at the same scale. 
Bottom right: projected slice ($20\,\mathrm{cMpc}\,h^{-1}$ in thickness) of the peak tomographic reconstruction expected over a few Mpc around 60 small patches where quasars appear closely clustered. This corresponds to a transverse separation of $\sim1.5\,\mathrm{cMpc}\,h^{-1}$. Bottom left: original density field smoothed with a gaussian kernel at the same scale. The colour of the maps encodes the density contrast.}
\label{fig:reconstruct}
\end{figure}

The WEAVE-QSO survey (\citealt{2016sf2a.conf..259P}, and in preparation) will observe around 450\,000 high-redshift quasars over an area of up to 10\,000\,deg$^2$ called WQ-Wide. Nested within this, we intend to survey a denser sampling of quasars targeted with J-PAS (Javalambre Physics of the Accelerating Universe Survey; {\citealt{2014arXiv1403.5237B}}) data over $\sim$6000\,deg$^2$. Further nested within this wide J-PAS area is a special, high-density footprint of $\sim$420\,deg$^2$ in the HETDEX spring field called WQ-HighDens. All objects targeted by WEAVE-QSO will be chosen to recover spectra of quasars with $z_q>2.2$, which is sufficiently high to provide coverage of the Lyman-$\alpha$ (Ly$\alpha$) forest at $z>2$ (hereafter we use the term `Ly$\alpha$-forest quasars' for these objects). The objective is to deepen our understanding of the intergalactic medium (IGM), the circumgalactic medium (CGM), and large-scale-structure cosmology through the study of intervening Ly$\alpha$ and metal-line absorption.

The science objectives of WEAVE-QSO are associated with two observational challenges: the probing of three-dimensional large-scale structures that span multiple quasar sight-lines (both statistically and in the form of maps), and the detailed study of small (extragalactic) scales in one dimension along isolated sight-lines. The former is contingent on WEAVE-QSO's expected observations of an unrivalled number density of Ly$\alpha$-forest quasars, while the latter rests on the WEAVE Survey's unprecedented resolution and signal-to-noise ratio among massive spectroscopic surveys. 
This science is based on a near-complete and pure sampling of Ly$\alpha$ forest quasars.\footnote{Our projections are based on estimates of the quasar population given by \citet{PalanqueDelabrouille2016}.}
WEAVE-QSO fields are predominantly at high Galactic latitudes, where the science return is optimized by wide-area coverage in WQ-Wide. Fields will therefore be shared with the WEAVE Galactic Archaeology and WEAVE-LOFAR surveys in the LR mode, with some limited data also in the HR mode through additional field-sharing with the Galactic Archaeology survey. WEAVE-QSO will also have a smaller, higher-density LR-mode footprint in WQ-HighDens, sharing fields only with Galactic Archaeology targets.
\looseness-1

A major goal of WEAVE-QSO is to help address one of the main cosmological challenges of our time: determining the cause of the acceleration in the expansion of the Universe (termed `dark energy'). A natural approach to the problem is to measure the expansion history of the Universe at various epochs to characterize its emergence. Baryon Acoustic Oscillations \citep[BAO: e.g.][]{2003ApJ...598..720S} are potentially observable over the entire post-recombination Universe and so provide a convenient standard ruler with which to achieve this. In quasar absorption spectroscopy, we measure large-scale structure in `skewers' of density through the Universe, along the lines of sight to these quasars. In this way, we can probe the large-scale 3D distribution of gas using statistical correlations between lines of sight, allowing a measurement of BAO and hence expansion at $z > 2$. This method was first demonstrated with the Baryon Oscillation Spectroscopic Survey \citep[BOSS; e.g.][]{2013A&A...552A..96B}, and reached maturity in final BOSS results \citep{2015JCAP...05..060B,2017A&A...608A.130D} with a sample of $\sim150\,000$ Ly$\alpha$-forest spectra observed over five years, and further developed in eBOSS \citep{2016AJ....151...44D,2019arXiv190403400D,2019A&A...629A..86B,2020ApJ...901..153D} to reach over 200\,000 quasars with $z_q>2$.\looseness-1

WEAVE-QSO will approximately double the number of observed Ly$\alpha$-forest quasars and, more importantly, achieve an unprecedented number density on the sky. Ly$\alpha$-forest measurements of 3D large-scale structure remain far from the cosmic variance limit in next-generation surveys. Furthermore, the value to the survey of selecting any given quasar for these measurements is a weak function of signal-to-noise \citep[e.g.][]{2011MNRAS.415.2257M}. We therefore conclude that the principal driver of correlation-function precision (and hence expected BAO constraints) is the number density of quasars to be observed.
WEAVE-QSO will provide competitive constraints despite observing little more than half the number of Ly$\alpha$-forest quasars expected in DESI, the other next-generation quasar survey \citep{2016arXiv161100036D}. WEAVE-QSO will achieve the necessary boost in BAO precision by concentrating on a higher density sampling. This will be achieved through observing more quasars per unit area over a smaller footprint (up to 6000\,deg$^2$ instead of the 14\,000\,deg$^2$ currently planned for DESI, \citealt{2016arXiv161100036D}), a narrower redshift range, and a magnitude limit fainter by $\Delta r \approx 0.5$.\looseness-1

In the magnitude range $21.5<r<23.5$ and within the wide J-PAS targeted footprint, WEAVE-QSO will observe a redshift subset of the available Ly$\alpha$-forest quasar population. This is in order to obtain the highest sampling density possible for large-scale structure analyses in 3D. Two options presented themselves as broadly bisecting the Ly$\alpha$-forest quasar population: $z<2.45$ and $z>2.45$. A convenient metric for the 3D large-scale structure statistics is provided by BAO measurement precision, and a convenient baseline for comparison is the DESI survey as currently planned \citep{2016arXiv161100036D}. Fisher forecasts were previously used to assess this choice \citep{2016sf2a.conf..259P}, and we return to the question here of using models based on 100 random realizations of correlation functions derived from the scaled BOSS covariance matrices (following the method set out by  \citealt{2017A&A...608A.130D} and \citealt{2018JCAP...05..029B}). The results are summarized in Fig.~\ref{fig:bao_summary}, assuming a 6000\,deg$^2$ footprint, and show that the WEAVE-QSO survey is more optimal when directed towards this higher-redshift window. This projection neglects the fact that DESI is boosted in the low-redshift window through the cross-correlation of lower-redshift quasars, further reinforcing the preference for higher-redshift quasars.

Consequently, we choose to target faint quasars with $2.45< z_q<3$ for statistical measurements of 3D large-scale structure. Note that this requirement is lifted in the special high-density field of WQ-HighDens, where we target all available Ly$\alpha$-forest quasars. \looseness-1

The high density of quasars required for competitive 3D large-scale structure analyses and the maximal brighter sample needed for high-quality isolated sight-line studies both require $\gtrsim $90 per cent Ly$\alpha$-forest quasar completeness. This will be achieved with J-PAS for the majority of the WEAVE-QSO survey footprint. J-PAS is an imaging survey covering more than 8000\,deg$^2$ (see Fig.~\ref{fig:lofar_progress}) with 56 narrow band filters, of which a coverage of $\sim$6000\,deg$^2$ of the WEAVE-QSO footprint is expected on a timescale practical for WQ target selection. Ly$\alpha$-forest quasars will be derived from this sample using a variety of machine learning methods trained and tested on a variety of simulated and observational data (\citealt{2020MNRAS.496.4931P, 2020MNRAS.496.4941P,2021A&A...653A..31B}; \citealt{2023MNRAS.520.3476Q}; \citealt{2022A&A...661A..99M}; Rodrigues et al., in preparation; P\'erez-R\`afols et al., in preparation).

In addition to using J-PAS data for target selection, WEAVE-QSO will join with J-PAS to cross-correlate the Ly$\alpha$ forest with J-PAS quasars even when not targeted spectroscopically, and to explore cross-correlation with faint Ly$\alpha$ sources in J-PAS images treated as an intensity map. The ability to statistically study various IGM/CGM properties using such high-density quasar sampling and high completeness has been demonstrated in BOSS and eBOSS or tested in simulations (e.g.\ \citealt{2018JCAP...05..029B,2018MNRAS.480..610G, 2018MNRAS.480.4702P,2018ApJ...859..125S, 2020MNRAS.499.2760S, 2020arXiv201200772M}).\looseness-1

Our measurement of 3D large-scale structure is not limited to statistical estimators such as correlation functions of the Ly$\alpha$-forest.  Cosmic-web mapping through large-scale structure tomography will be pursued via three distinct programmes: a wide tier (in WQ-Wide) at low density (and so mapping low resolution),  a higher-density tier with higher resolution (in WQ-HighDens), and one focused on rare, close groups of quasars (in WQ-Wide). For the wide tier, WEAVE-QSO occupies a novel niche as the only current or imminent survey with an area larger than 15\,deg$^2$. Other relevant facilities include CLAMATO (\citealt{2018ApJS..237...31L}; 0.8\,deg$^2$), PFS (\citealt{2014PASJ...66R...1T}; $\sim15$\,deg$^2$), MOSAIC/ELT (\citealt{2019A&A...632A..94J}; $<10$\,deg$^2$), MSE (\citealt{2019arXiv190404907T}; $\sim80$\,deg$^2$). These surveys will achieve transverse lengths (the average separation between sight lines) of 2--$3\,\mathrm{cMpc}\,h^{-1}$, and while WEAVE-QSO will only achieve $\sim15$--$20\,\mathrm{cMpc}\,h^{-1}$, it will do so up to 6000\,deg$^2$ and in rare cases will reach a peak resolution as high as  $\sim1\,\mathrm{cMpc}\,h^{-1}$. \looseness-1

Higher average resolution ($\lesssim15\,\mathrm{cMpc}\,h^{-1}$) is expected over the $\sim$420\,deg$^2$ of the WQ-HighDens footprint (Fig.~\ref{fig:hetdex}), where we will survey quasars densely without limits on the Ly$\alpha$-forest redshift range.

The potential IGM tomography we may derive from this sample is illustrated in the upper panels of Fig.~\ref{fig:reconstruct} and further detailed in a forthcoming WEAVE-QSO publication (Pieri et al., in preparation). The upper left panel shows a simulated \HI\ distribution generated from a dark matter simulation and the \textsc{lymas} code\footnote{\url{http://www2.iap.fr/users/peirani/lymas/lymas.htm}} \citep{2014Peirani}, smoothed to reflect the desired resolution. The upper right panel shows a projection of WEAVE-QSO-based reconstruction of this simulation in WQ-HighDens through Wiener filtering (e.g.\ \citealt{pichonetal2001,cauccietal2008,2018ApJS..237...31L}) with a random distribution of Ly$\alpha$ forest quasars. Note that we neglect here the potential boost we may derive by incorporating the HETDEX data of Ly$\alpha$ emitters into the reconstruction.\looseness-1

The WQ-Wide footprint is expected to provide peak cosmic web reconstruction resolution in approximately 60 small patches where three or more quasars with $r<21$  are separated by $\sim1\,\mathrm{cMpc}\,h^{-1}$. This is demonstrated in the lower panels of Fig.~\ref{fig:reconstruct}. The bottom left panel shows a simulated \HI\ distribution smoothed to reflect our desired peak resolution. The corresponding tomographic reconstruction provided by these WEAVE-QSO multiplexes is shown in the bottom right panel (though this is naturally not available for the entire $100\,\mathrm{cMpc}\,h^{-1}$-wide region shown). The IGM tomography potential of WEAVE-QSO is explored further in \cite{Kraljic2022}. Beyond these peak-resolution regions and the WQ-HighDens area, the resolution will vary up to around 15--$20\,\mathrm{cMpc}\,h^{-1}$, making it suitable for a wide void/non-void separation \citep{2016MNRAS.456.3610O}. \looseness-1

All of these maps of varying resolution will provide insights into the impact of environment on IGM and galaxy properties, noting that $z>2$ is the epoch where the star-formation rate reaches its peak \citep{2006ApJ...651..142H, 2014ARA&A..52..415M} where intergalactic gas accretion along filaments is critical for the formation of galaxies (e.g.\ \citealt{2005MNRAS.363....2K}). Furthermore, a complete sample of quasars will allow for the study of their connection to large-scale (re)ionization of the cosmic web (e.g.\ \citealt{2019arXiv190304510M}). \looseness-1

Beyond the J-PAS targeted sub-area of WQ-Wide, WEAVE-QSO will observe bright quasars and, over its entire low-resolution footprint, WEAVE-QSO will obtain spectra of almost all bright ($r\lesssim 21$) Ly$\alpha$ quasars. This constitutes up to 4000\,deg$^2$ targeted initially using a combination of {\it Gaia}+unWISE objects  \citep{2019MNRAS.489.4741S} and SDSS-IV DR16 \citep{2020ApJS..250....8L}, with {\it Gaia} spectroscopy to be incorporated progressively (e.g. \citealt{2023A&A...674A..41G}).

With the addition of this sample, WEAVE-QSO will offer unparalleled spectral resolution (mostly $R=5000$ but also $R=20\,000$) with high signal-to-noise ($>4$--$7$ per \AA). Improved spectral resolution will enable the study of gas temperature and structures on smaller scales (e.g.\ \citealt{2013A&A...559A..85P,2013PhRvD..88d3502V}): the relationship between temperature and density in the IGM is a fundamental quantity describing the physical state of baryons during reionization. Such data will allow the study of smaller-scale power in the matter distribution, measuring neutrino masses and alternative forms of dark matter. The improved resolution provided by WEAVE-QSO will aide in breaking the degeneracy between the small-scale suppression of power caused by IGM temperature and these cosmological effects, which are currently marginalized over using  extensive and expensive hydrodynamic simulations.\looseness-1

Galaxies and circumgalactic medium regions are identified in absorption as damped Ly$\alpha$ systems (e.g.\ \citealt{2017MNRAS.469.2959K}), Lyman-limit systems (e.g.\ \citealt{2013ApJ...775...78F}), and strongly blended Ly$\alpha$ systems (\citealt{2014MNRAS.441.1718P}, and Morrison et al., in preparation). The BOSS and eBOSS surveys demonstrate that a great deal can be learnt about such systems (e.g.\ \citealt{2012A&A...547L...1N,2014MNRAS.441.1718P, 2017ApJ...846....4M}). Improved spectral resolution and signal-to-noise provided by WEAVE-QSO will allow greater fidelity of such systems, and machine learning methods have been developed to recover them (e.g.\ \citealt{2020MNRAS.498.1951F}). In particular, damped systems with lower column density and/or higher redshift will be more readily identified. Furthermore, narrow metal absorption lines associated with the interstellar, circumgalactic, and intergalactic media will be more statistically significant, resolved, and informative in WEAVE-QSO spectra. Studying the properties of such systems will be of increasing importance over the coming years given the quality and size of the WEAVE-QSO sample combined with wide-area $z>2$ galaxy surveys such as HETDEX. This is a key epoch for galaxy assembly, as it includes the early portion of the peak in the star-formation rate. The combination of absorber fidelity, gas temperature, improved galaxy identification, small-scale CGM effects, and large-scale structure context, all at the peak of star-formation rate, will make WEAVE-QSO transformative for our understanding of galaxy formation.\looseness-1

In weather conditions deemed sub-optimal for our primary-science purposes, we intend to use the LIFU to target quasar pairs or groups too close to be targeted in the MOS mode. We will also probe extended emission near quasars and unresolved galaxies in front of quasars using the mIFU mode. These science goals are well adapted to poor seeing, as the LIFU fibres have twice the diameter of their MOS counterparts (see Table~\ref{table:FOC_modes}), while extended or unresolved emission need only be within the mIFU.\looseness-1

The WEAVE-QSO survey aims to achieve an unprecedented diversity of intergalactic-medium science goals spanning scales from sub-kpc structures in the CGM to the BAO scale. This will be achieved chiefly by exploiting WEAVE's high spectral resolution and WEAVE-QSO's high target density with respect to previous and contemporary spectroscopic surveys.\looseness-1

\section{The WEAVE Simulator}
\label{sec:sim}

\begin{figure*}
    \centering
    \includegraphics[width=0.45\textwidth]{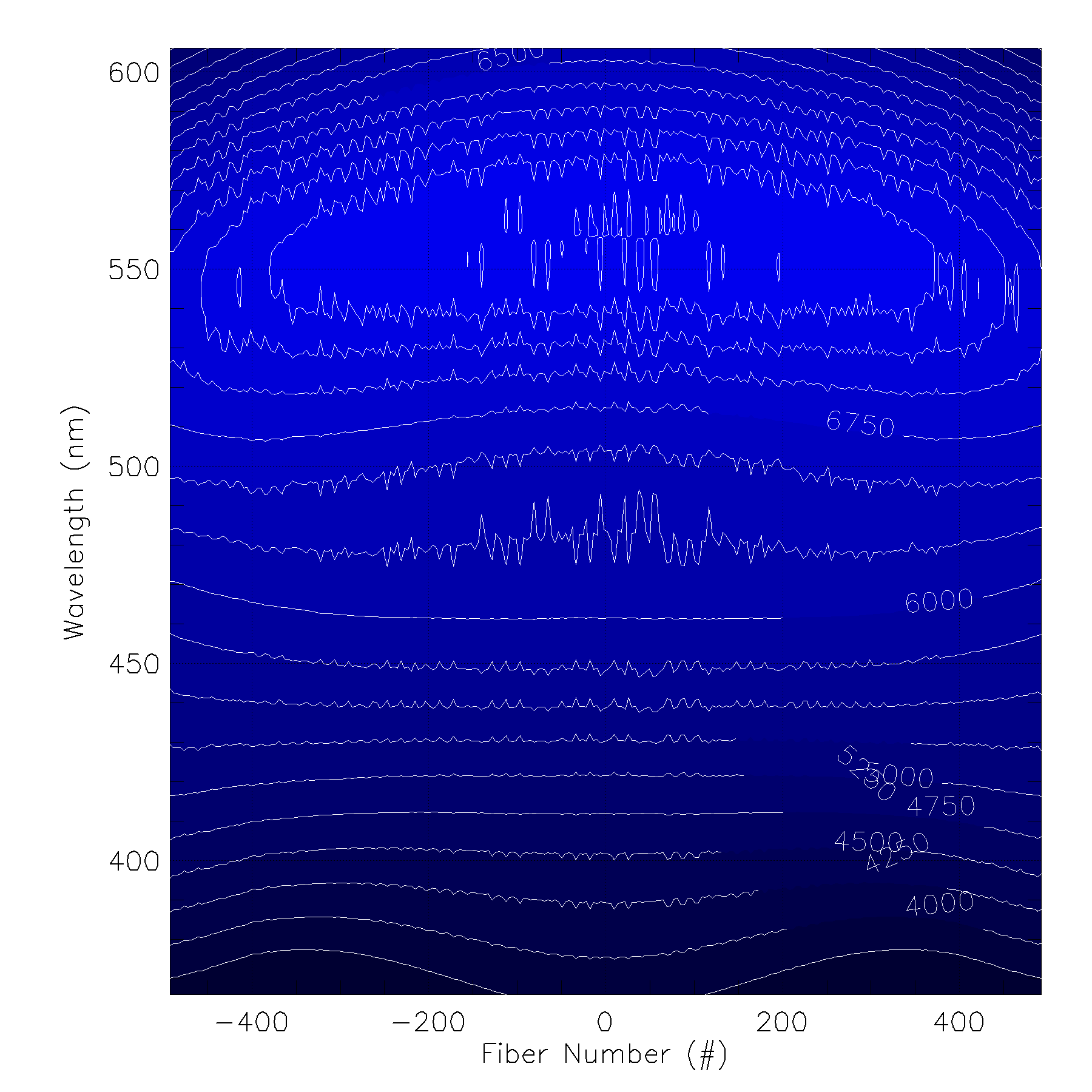}
    \includegraphics[width=0.45\textwidth]{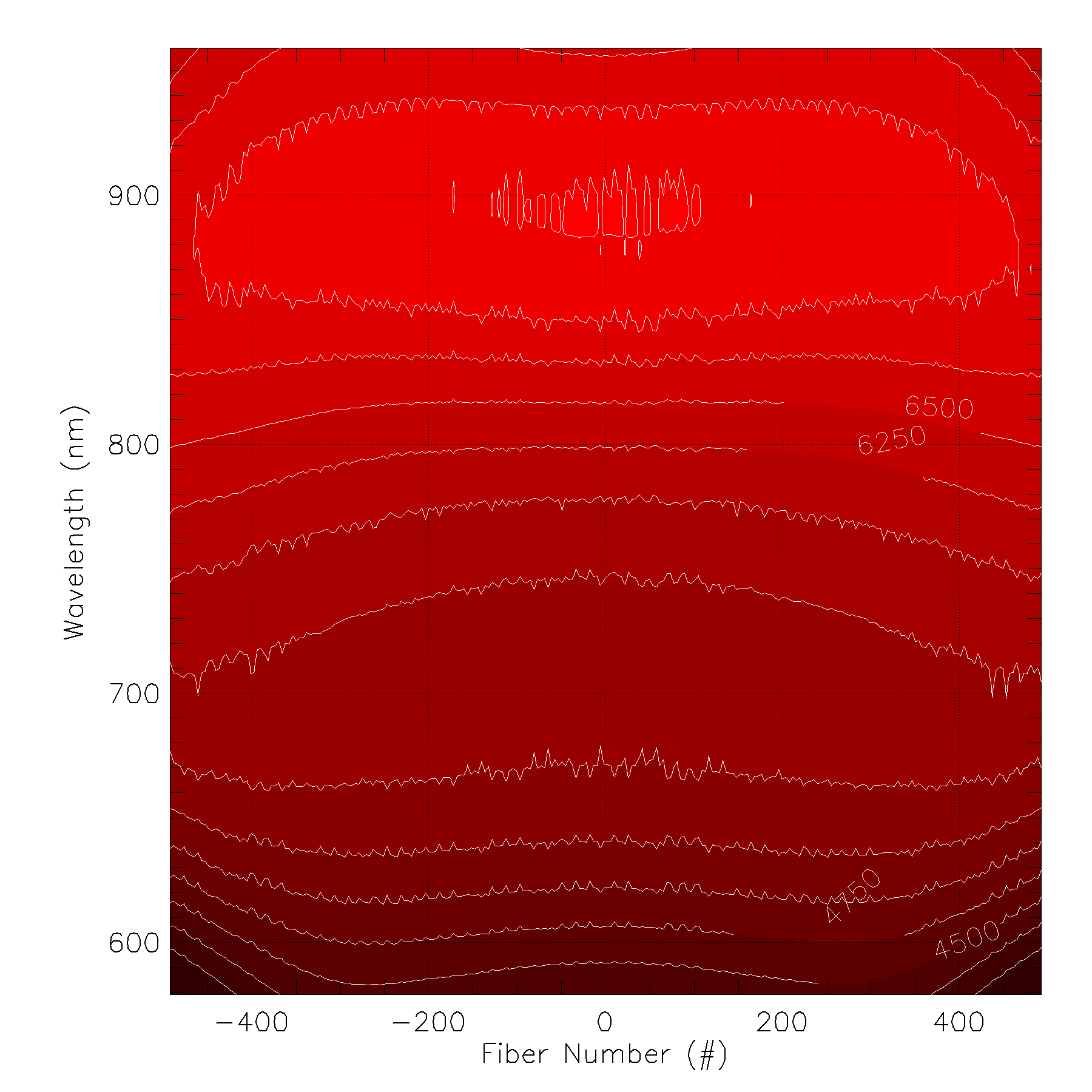}
    \caption{Nominal WEAVE spectral resolution as a function of fibre number ($x$-axis) and wavelength ($y$-axis), computed from the spectrograph optical model. Left: low-resolution blue VPH grating. Right: low-resolution red VPH grating. Contours label regions of equal spectral resolution $R=\lambda/\Delta\lambda$. Note that these resolving powers correspond to MOS and mIFU modes, and are halved in the LIFU mode with its two-times-larger fibres (see Table\,\ref{table:FOC_modes}).
    }
    \label{fig:VPHLR_spectral_resolution}
\end{figure*}

During the preliminary design phase of the project, it was recognised that the spectral resolution delivered by the spectrograph would vary as a function of wavelength \emph{and} as a function of fibre position along the slit (Fig.~\ref{fig:VPHLR_spectral_resolution}). After initial modelling of this effect at a low level for validation of the design, it was decided that a full image simulation would be required to provide representative input for end-to-end testing of the data-processing system. This section provides a description of such a simulator that was created in order to generate representative data for the majority of WEAVE Survey use cases as part of a full Operational Rehearsal \citep[see Section~\ref{sec:OpRs} and][]{2016SPIE.9913E..2XD}, in order to simulate the implementation of the Survey.\looseness-1

\subsection{Point spread functions}
\label{sec:psfgen}

In order to capture the variation in point spread function (PSF) across the field, we used \textsc{zemax}\footnote{\url{https://www.zemax.com}} to generate the PSF at nine equally spaced wavelengths for each fibre in the top half of the slit for each of the five modes given in Table~\ref{table:SPE_modes}. Each PSF is generated as a $25\times25$ pixel image with 15\,\um\ pixels, using $10^7$ rays to ensure that the outer wings are fully sampled. The ray trace assumes a uniform circular output aperture for the fibre, which is not strictly correct but considered to be adequate for the purposes of this simulation. Internal vignetting in the spectrograph is included here. Each PSF generated in this way is centred on the chief ray, and the position of this in focal plane coordinates is recorded. These data are generated once only for each mode using the nominal configuration of the spectrograph optics and stored for use at runtime.\looseness-1

\subsection{Input fields}
\label{sec:siminput}

The simulator is driven by an input in the form of an extended OB XML file, with additional elements for the actual conditions of the observation (time, date, seeing, and transparency), together with an additional descriptive element for each target containing velocity (or redshift), magnitude, bandpass, target FWHM (for extended sources), and the appropriate spectral template to use.
The input field is parsed to determine the location and fibre allocation for each target, and these are used to calculate the vignetting factor at the telescope focal plane (due to the prime-focus corrector optics) and the expected light loss at the spectrograph collimator (due to the non-telecentricity of the input to this fibre at this field position).\looseness-1

\subsection{Sky background}
\label{sec:skycalc}

For each OB, we use the expected date and time of observation to calculate the elevation, moon phase, and moon--target separation for the mid-point of the OB. We then use the ESO Paranal sky simulator \citep[\textsc{skycalc}:][]{2012A&A...543A..92N, 2013A&A...560A..91J} to generate an estimate of the sky emission and absorption spectra for this observation at $R=50\,000$ from $360$--$970$\,nm, which is then degraded to match the resolving power of WEAVE ($R=5000$ for LR and $R=20\,000$ for HR in the MOS/mIFU modes, each further lowered by a factor of two for the LIFU mode; see Table~\ref{table:SPE_modes}). We neglect the zodiacal emission terms, and scale the \textsc{skycalc} output spectrum to the recorded La Palma sky brightness for this date and time. The sky spectra are resampled to 20\,000 pixels across the bandpass of each WEAVE mode to form a basis for each fibre input.\looseness-1

\subsection{Template spectra}
\label{sec:templates}

The template spectra referenced in the OB XML file are provided by the science teams to generate representative samples of spectra to test the data-processing pipelines. Each template is provided in flux units with the assumption that it will be scaled to a given apparent (AB) magnitude for observation. The only practical constraint on the input spectra is that they cover the full WEAVE spectral range. For each fibre allocated to a target, the template is read in, scaled to the appropriate magnitude and velocity or redshift, and converted from flux to incident photons. Finally, each spectrum is scaled to account for vignetting/telecentricity factors described in Section~\ref{sec:siminput}, aperture losses appropriate to the seeing extant at the time of the observations, and the telescope aperture and the exposure time. We apply the instrument throughput for the relevant mode, for which we adopt the as-designed throughput estimates reported in \cite{2014SPIE.9147E..0LD}.\looseness-1

Since the different surveys have targets with very different characteristics, the simulator also allows for targets to have pre-determined aperture losses for extended sources, and pre-determined input fluxes (no magnitude scaling) for emission-line targets. Finally, the input sky spectrum (scaled by the appropriate factors for each fibre) is added, producing the final input spectrum for each fibre.\looseness-1

\subsection{PSF resampling}
\label{sec:psfinterp}

For each fibre in turn, the PSF library described in Section~\ref{sec:psfgen} is then interpolated to give an array of 20\,000 PSFs over the observed wavelength range (each centred on the effective position of the chief ray at the appropriate wavelength). For a subset of these, we directly calculated the same PSF in \textsc{zemax} to check the validity of the interpolation. Each of these 20\,000 PSFs is then resampled using a bivariate spline evaluated over a rectangular mesh to place the photons into the appropriate pixels of the WEAVE CCDs.\looseness-1

For fibres in the lower half of the slit, we use mirror-images of the upper-half PSFs in order to save storage and computation time.\looseness-1

\subsection{Output images}
\label{sec:output}

When the spectra for all fibres have been added to the CCD images, the CCD bias, read-out noise, and Poisson noise are all added. The images are then saved with a complete FITS header, representative of the final WEAVE images. These images can then be sent to the WEAVE SPA pipelines for processing and distribution to the science teams via WEAVE's data archive system.\looseness-1

\subsection{Calibration frames}
\label{sec:calframes}

In addition to science data frames, the simulator also generates calibration data frames in the form of bias, fibre flat-field, detector flat-field, and calibration arc-lamp frames. Fibre flat-fields and arcs include the same input losses at the telescope focal plane as the science data and the correct header information in order to allow the data-processing chains to function. For the flat-field lamp spectra, we assume a black-body spectrum at 3500\,K. For the arc lamp, we use the ThAr lamp spectral data from the National Optical Astronomy Observatory atlas,\footnote{\url{http://iraf.noao.edu/specatlas/thar/thar.html}} sampled at $0.002$\,nm/pixel over the range $360$--$960$\,nm.\looseness-1

\section{Operational Rehearsals}
\label{sec:OpRs}

Using the WEAVE Simulator described in the previous section, a series of exercises were conducted to simulate the implementation of the Survey. These Operational Rehearsals (OpRs) were designed to simulate and test the flow of data in various guises through and between different nodes of the WEAVE data systems (see Section~\ref{subsec:OCS} and \ref{subsec:SPA}), including the OCS, CPS, APS, and WAS. The OpRs have also seen significant involvement of Science Team members, mainly through the work of the SWG (see Section~\ref{subsec:SWG}) and the QAG (see Section~\ref{subsec:QAG}), as well as other key individuals in charge of assembling input target catalogues.\looseness-1

A simplified description of the flow of data and information in the OpR context is as follows:
\begin{enumerate}
    \item input FITS catalogues are submitted to the WASP, hosted at the Cambridge Astronomical Survey Unit (CASU), Cambridge, where all targets are assigned a `CNAME' once a given catalogue has successfully passed verification tests for certain required header keywords, amongst others. The CNAME for a target is based on its right ascension and declination, and is an immutable attribute for identifying that particular target in the WEAVE data-processing and archive systems;
    \item CNAME-containing FITS catalogues are then used by the science teams to create a series of `protofields', which, for each WEAVE MOS and mIFU observations, contains a selection of potential science targets and guide stars:
    \begin{enumerate}
        \item for the MOS mode, protofields containing approximately 2000 potential science targets, as well as lists of guide stars, white dwarf stars (for calibration and the White Dwarfs survey, Section~\ref{subsec:WDs}), and `sky' positions if the user chooses to specify these in advance, form the input to \configure\ (see Section~\ref{subsec:FIB+POS}), the output of which are configured (or `\textsc{configure}-output') XML files;
        \item for the mIFU mode, protofields are similar to those created for the MOS mode, but contain only positions of the objects lying within the boundaries of (and typically centred on) the 20 mIFU bundles, which \configure\ translates into a set of fibre positions for the individual mIFU fibres of each mIFU bundle;
        \item for the LIFU mode, a protofield contains a single pointing that defines the centre of the field (along with a position angle set to optimize the guide star position on the offset LIFU guider), which \configure\ translates into a set of fibre positions for the fixed LIFU fibre array;
    \end{enumerate}
    \item XML files are submitted to the WASP, which performs a series of checks on the files received, ranging from a baseline structure check against template XMLs for the checking of constituent elements, to the verification of instrumental mode and observational requirements against the CNAME-containing FITS catalogues from an earlier stage of the process;
    \item WASP-verified XML files are collated and sent to the OCS, which then ingests the files as OBs into its OB database;
    \item prior to and during each (simulated) observing night, the \scheduler\ (WEAVE observation queue scheduler; see Section~\ref{subsec:OCS}) creates a real-time queue of OBs to be executed, depending on current and predicted weather conditions one hour ahead of observing time;
    \item scheduled fields are `observed' using the WEAVE Simulator (see Section~\ref{sec:sim});
    \item raw data are sent from the `telescope' to CASU, where they undergo initial processing by the CPS;
    \item CPS-output data are transferred to the APS, hosted at the IAC, Tenerife, for further, advanced-level processing;
    \item outputs from the CPS and the APS are ingested by the WAS, hosted at Telescopio Nazionale Galileo, La Palma, ready for inspection and download by the QAG through the archive system.
\end{enumerate}

The first Operational Rehearsal (OpR1) was an exercise centred on testing particular aspects of the data-flow model, ensuring that raw and reduced data could be moved correctly around the data-processing and data-archiving systems. File transfer and data-processing speeds were monitored, allowing weaknesses in the configuration of the systems to be identified and improved.\looseness-1

The second Operational Rehearsal (OpR2) comprised a series of data-flow simulations aimed at pushing data from the telescope all the way to the end user through two stages of data processing and final archiving. Several enactments of OpR2 were performed (the most prominent of these being OpR2.5; see e.g. Fig.~\ref{fig:eVlos}) following the identification and subsequent resolution of a number of issues in the data-processing pipelines, as well as in the inputs provided by the Science Team. The SWG, QAG and other core members of the individual science teams\footnote{\url{https://ingconfluence.ing.iac.es/confluence/display/WEAV/Core+Science+Teams}} were heavily involved in OpR2 through the preparation and submission of FITS catalogues of potential science targets as well as template spectra assigned to these targets, and in using \configure\ to produce configured fields in the form of XML files.\looseness-1

The third Operational Rehearsal (OpR3) -- the latest in the series -- again tested all of the previously listed steps on algorithms and pipelines for the MOS mode (upgraded in the interim) and the LIFU mode, as well as a number of other details related to node interactions. As the last of the OpRs with full involvement of the science teams before an external survey-readiness review (successfully completed in 2020 January) and before WEAVE going on sky, OpR3 has been crucially important for ensuring that node-to-node interfaces work efficiently and seamlessly, and that the workings within each node have been set up to proceed optimally. At the time of writing, the mIFU mode is undergoing a similar, dedicated exercise.\looseness-1

There are naturally some differences between the OpR setup and what is envisaged for real Survey operations. However, the framework of the data flow used for the OpR exercises was established with real Survey operations heavily in mind, mimicking as closely as possible how, for instance, submissions and verification of input FITS catalogues and \textsc{configure}-output XML files would be handled by the WASP. In general, OBs were scheduled (and thus also `observed') on specific `nights' corresponding to a real historical date, whose archived weather conditions at the WHT could be used to set the observing conditions. The WEAVE Simulator would then take, as input, a \textsc{configure}-output XML file, along with the observing conditions for the time of the observation and template spectra corresponding to the targets in the field, and simulate their observation by WEAVE.\looseness-1

Given the importance of the simulated implementation of information and data flow in preparing for the running of the WEAVE Survey, the following subsections provide further details of the key stages of OpR3, highlighting their initial goals as well as some lessons learnt.

\subsection{Operational Rehearsal 3a (OpR3a)}

OpR3a was a scheduling exercise covering a virtual period of three non-contiguous months over the course of just over one year. These three periods, named S1--S3, allowed us to test the ability of the \scheduler\ to handle the creation of real-time queues of OBs on an hourly basis given realistic weather conditions. Each OB specified all required elements for an observation -- field centre, requirements for observing conditions and instrumental setup, etc. -- apart from the location of individual fibres within the field, and the \scheduler\ was used to create nightly queues of OBs to be observed according to archival weather data at the WHT for the corresponding dates being used for the OpR3a simulation.\looseness-1

OpR3a enabled the \scheduler\ to be tested for OB queue creation using a mixture of fields prepared for the LIFU and MOS modes. While science can be performed using the LIFU with seeing of worse than 1.5~arcsec, most science cases using the MOS mode require a seeing of 1.1~arcsec or better. We recall from Section~\ref{sec:instr} that the configuration of a MOS field requires approximately one hour, so the length of a typical observation block is therefore set at one hour. However, a MOS plate cannot be configured during a LIFU observation, as described above, which means that transitions from a LIFU to a MOS observation typically incur significant overhead. This overhead is expected (as specified in the WEAVE Concept of Observations) to be overcome with the aid of mIFU observations, which take less than 15~min to configure. The unavailability of the mIFU mode in OpR3a clearly showed that its presence was highly desirable in order to use intermediate-seeing conditions effectively, and to efficiently transition between the LIFU and MOS modes without compromising on science or losing valuable observing time.\looseness-1

For OpR3a, surveys were also provided the choice of submitting OBs to be observed in the LR mode or the HR mode with either the red+green or red+blue grating combination (see Table~\ref{table:SPE_modes}). Opportunities to request `chained' OBs that specify two or more duplicate OBs (for deep exposures) or `linked' OBs that specify two or more OBs that should be observed within a given time-frame were also provided. In each case, as-yet unobserved OBs belonging to a set that had been started were prioritized over OBs from sets where no OB had yet been observed. In OpR3a, chained and linked OBs were scheduled by manual intervention of the \scheduler\ output. These modes were subsequently implemented into the \scheduler\ and successfully tested in OpR3c (see Section \ref{subsec:OpR3c}).\looseness-1

\subsection{Operational Rehearsal 3b (OpR3b)}
\label{subsec:OpR3b}

\begin{figure}
\includegraphics[width=\columnwidth]{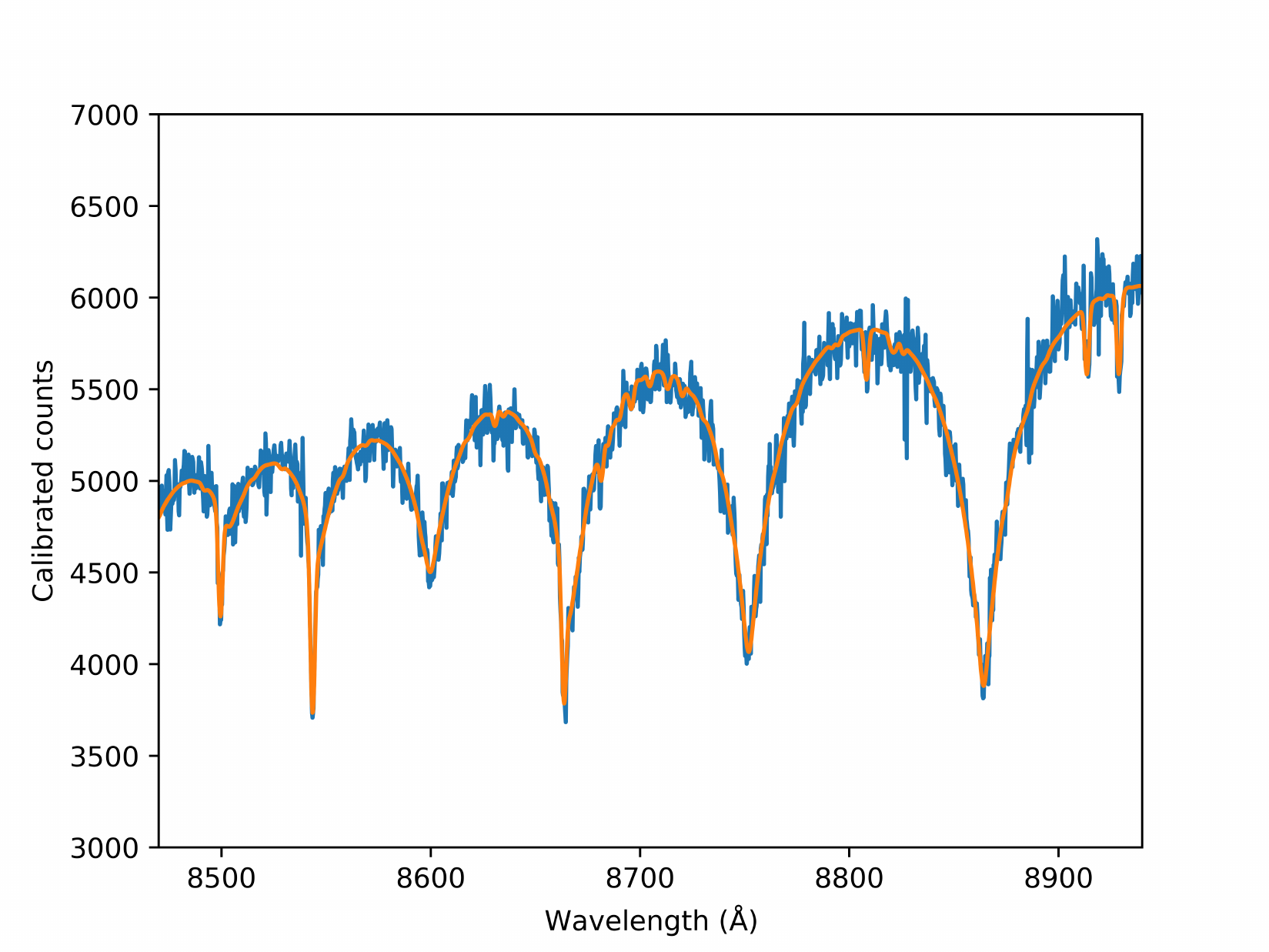}
\caption{Simulated spectrum (blue) of an A star with $T_\mathrm{eff} = 9000\,\mathrm{K}$, $\log g = 4.0$ in the calcium-triplet region, compared with the best fit (orange) obtained using the method of \citet{harris2018}. The biases in the returned parameters (measured minus input) for $T_\mathrm{eff}$, $\log g$, and radial velocity are $-11\,\mathrm{K}$, $-0.003$\,dex, and $+0.7\,\mathrm{km\,s^{-1}}$, respectively.}
\label{fig:SCIP_OpR3b_Astar}
\end{figure}

\begin{figure*}
    \centering
    \includegraphics[width=0.95\textwidth]{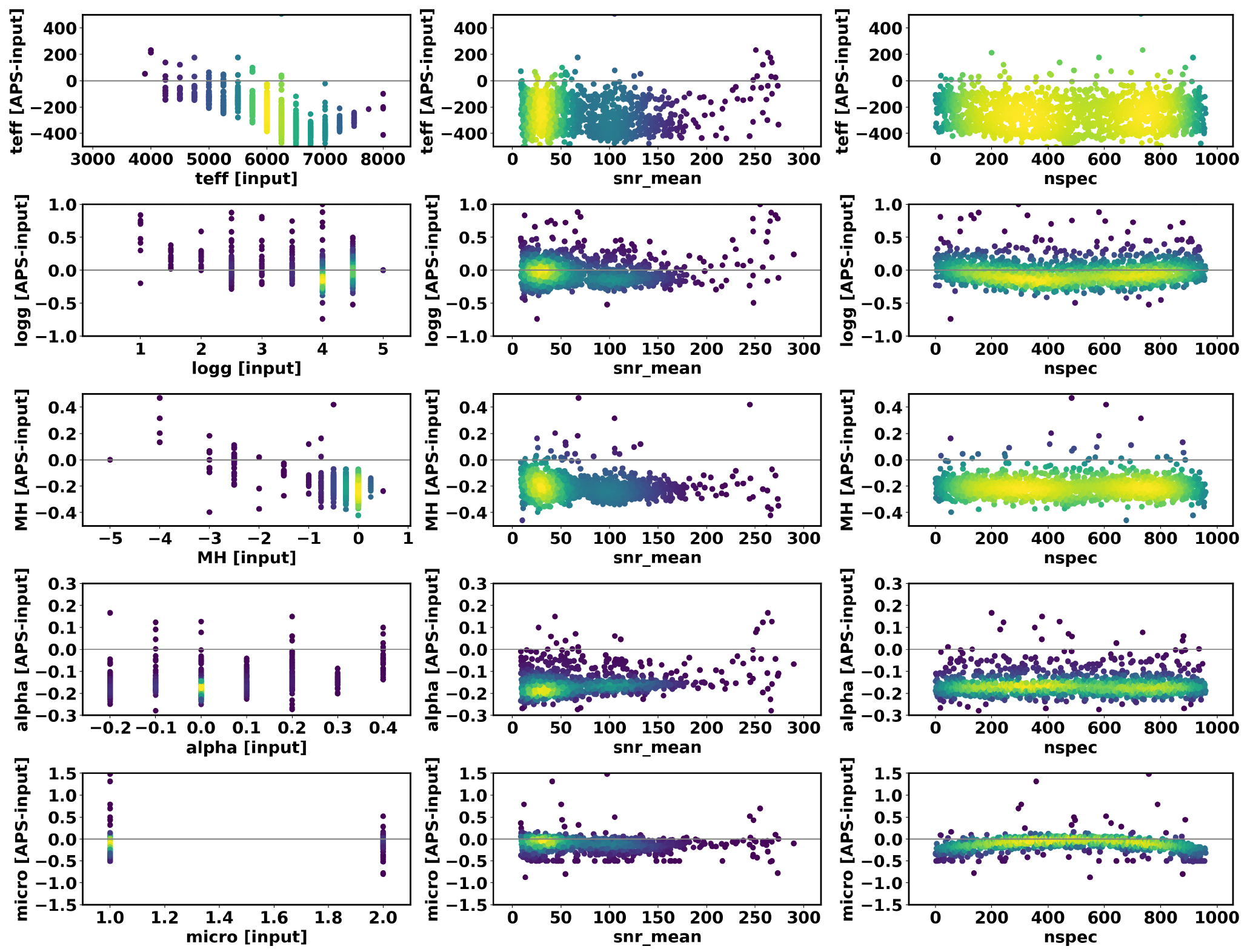}
    \caption{APS output compared with the input to the WEAVE Simulator for high-resolution green+red GA-HR data in OpR3b in the form of density plots, where yellow indicates regions of highest density and purple the lowest-density regions. The panels show the difference ($\mathrm{output}-\mathrm{input}$) for the parameters $T_\mathrm{eff}$, $\log g$, $\mathrm{[M/H]}$, $[\alpha/\mathrm{Fe}]$, and microturbulence (in $\mathrm{km\,s^{-1}}$) from top to bottom, as a function of input value (left), mean signal-to-noise ratio between the blue and red arms (middle), and the fibre location on the detector (right). We note that the curvature seen in the right column (as a function of fibre location) in $\log g$ and microturbulence results from uncorrected spectral resolution variation in the APS pipeline, which has since been addressed in an updated version of the pipeline (see Section\,\ref{subsec:OpR3b}).}
    \label{fig:GA-QAG-HRB}
\end{figure*}

\begin{figure}
    \centering
    \includegraphics[width=0.95\columnwidth]{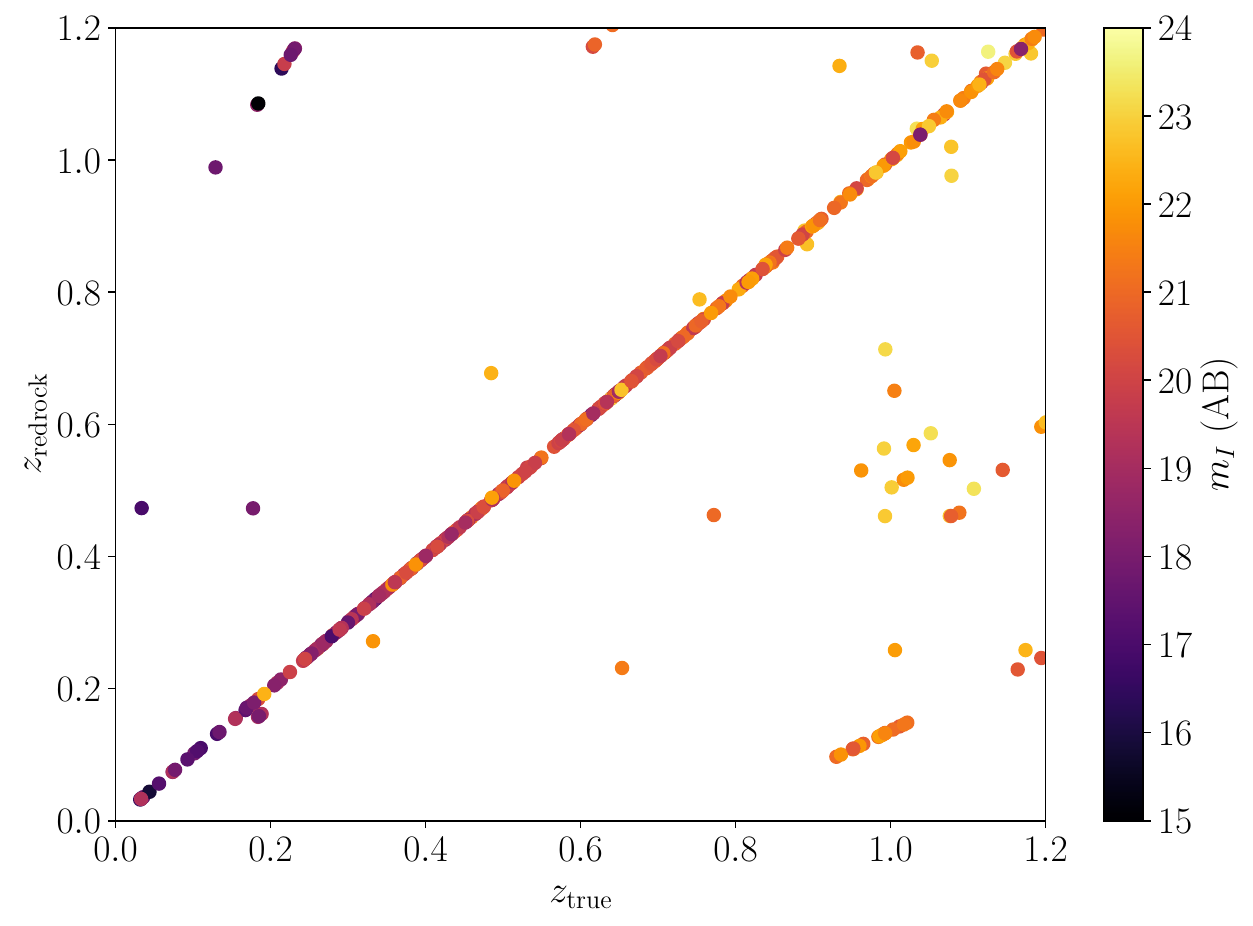}
    \caption{Redshift estimation for WEAVE-LOFAR targets from a model WL-DEEP OB observed in OpR3b, targeting 150\,MHz sources brighter than 100\,$\mu$Jy. The redshift estimate from the APS (derived using the off-the-shelf \texttt{redrock} code is shown on the vertical axis, as a function of the true redshift (taken from the input simulation), and colour-coded by the $I$-band apparent magnitude. While the vast majority of targets are correctly redshifted out to $z = 1$ (93.3 per cent of redshifts are correct to $|z_\mathrm{APS} - z_\mathrm{true}| < 0.001$), there is some evidence of emission-line mis-identification (e.g. the linear features away from the 1:1 line) and bright targets with incorrect redshifts, perhaps highlighting the need to fold in photometric redshift priors and to optimize\slash update the redshifting templates once data begin to arrive in earnest. The redshifting performance for the WEAVE-LOFAR survey will be discussed in more detail in a forthcoming work (Smith et al., in preparation).}
    \label{fig:WL_redshift_comparisons}
\end{figure}

\begin{figure}
    \centering
    \includegraphics[width=0.9\columnwidth]{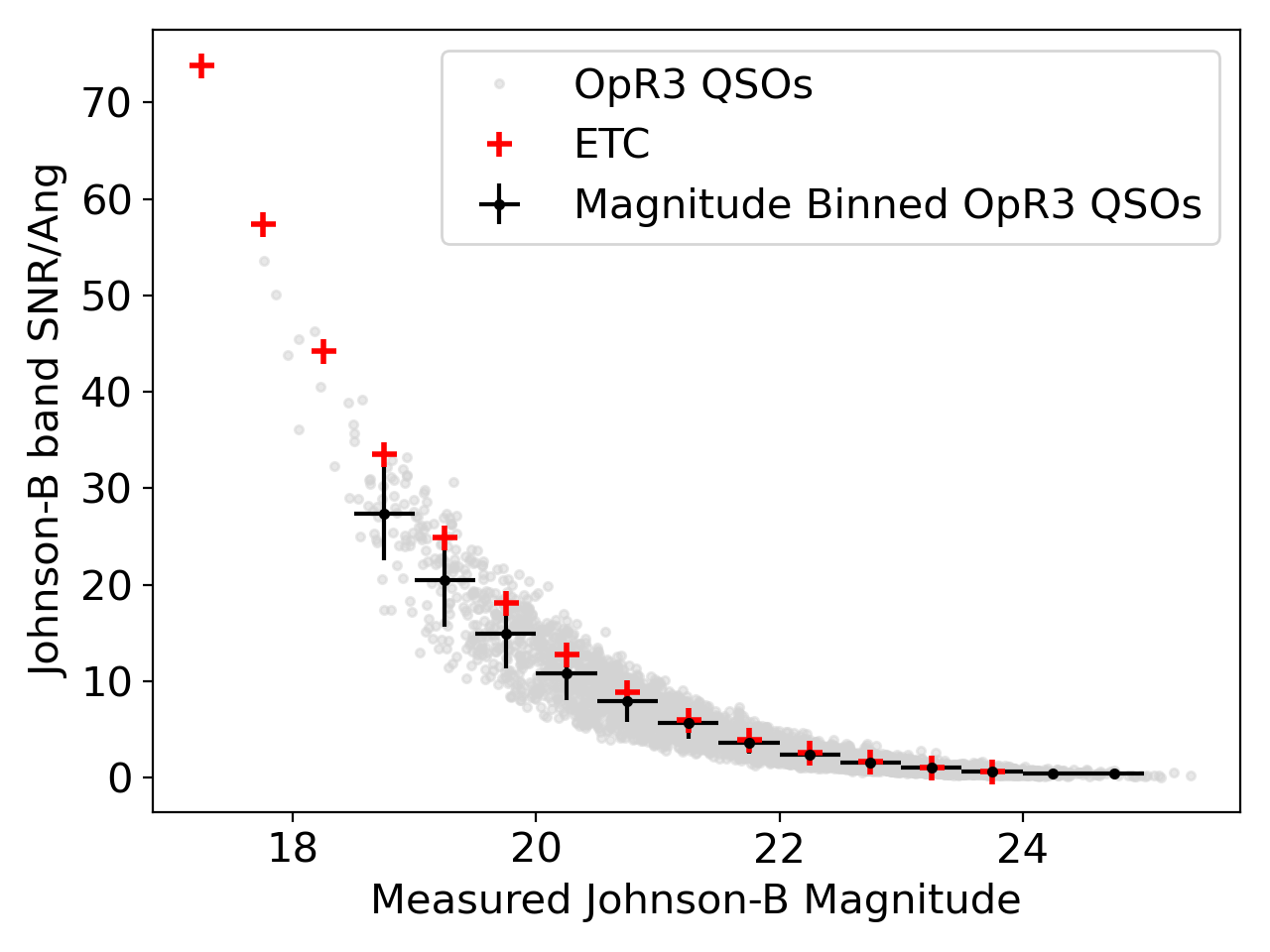}
    \caption{Signal-to-noise per {\AA} as a function of apparent magnitude, estimated from simulated WEAVE-QSO data and the WEAVE exposure time calculator (ETC). Here, magnitude and signal-to-noise ratio are both defined in Johnson filters. The results of the WEAVE ETC (v4.2.4, red) and the measured OpR3 mock data both per spectrum (grey) and binned (black) are shown. As the WEAVE Simulator and ETC are completely independent codes, the excellent agreement at the faint end is reassuring for prospective WEAVE users; the deviation at the bright end is likely due to a combination of the variation of throughput with fibre position in the focal plane due to non-telecentricity and from vignetting, and to issues related to precise spectral shape and redshift of the simulated targets, neither of which the ETC takes into account in the current version.}
    \label{fig:WQ_SNR}
\end{figure}

From each of the semesters modelled in OpR3a, we selected a single week of MOS and LIFU observations (180 OBs in total) for which the target fields were then run through \configure, ingested into the survey system through the WASP, and then passed to the WEAVE Simulator. Additional XML elements were added to each field to specify the actual observing conditions, and to specify the spectral template and magnitude to use for each target. The data generated in this way were then uploaded to the Observatory and used to simulate the full operation of the WEAVE data-flow system through the OCS, quick-look pipeline, CPS, APS and WAS. The `observations' and transfer of data from the `telescope' to CASU (Cambridge) were performed at night (i.e.\ in `real time') to emulate -- and hence also to verify -- the network traffic in a realistic manner. The resulting datasets were then made available to the QAG and SWG for analysis.\looseness-1

A key aspect of this process is that every step in the data flow should occur automatically, driven only by the observation description and calibration templates in the OB, and by information generated internally within the CPS and APS, which have no prior knowledge of the input target parameters beyond the basic photometric data. Flux calibrations were generated from the white dwarf calibration stars within each field, and the sky and object fibres were subject to the variations in throughput arising from the non-telecentricity and vignetting in the optical models.\looseness-1

Figs~\ref{fig:SCIP_OpR3b_Astar}--\ref{fig:WQ_SNR} provide a few illustrative results from OpR3b (see also Fig.~\ref{fig:WCC}). As the first end-to-end test of the complete CPS and APS pipelines, it was not expected that we would meet our science goals in the output of these systems in OpR3b. Rather, the intention was to stress-test the data-flow set-up and the pipelines themselves, which could then be used to make any necessary improvements. Through the rehearsal, a number of undesirable effects were identified and corrected in both the input fields generated by some of the surveys and in the data-processing and analysis chains, illustrating the value of conducting such an exercise in advance of the instrument commissioning phase. An example of such an effect is the noticeable curvature in the residual $\log g$ and microturbulence values as a function of fibre location on the detector arrays for stars observed by the WEAVE Galactic Archaeology survey (shown in Fig.~\ref{fig:GA-QAG-HRB}), symptomatic of an uncorrected spectral resolution variation with wavelength in the high-resolution spectral mode (cf.\ Fig.~\ref{fig:VPHLR_spectral_resolution}, which shows a similar variation as a function of fibre location for the low-resolution spectral mode). On the other hand, Fig.~\ref{fig:WQ_SNR} shows that the WEAVE Simulator and the WEAVE exposure time calculator (ETC), developed independently, are on average in excellent agreement for faint (WEAVE-QSO) targets, suggesting that the ETC can be used with some confidence for planning WEAVE observations.\looseness-1

\subsection{Operational Rehearsal 3c (OpR3c)}
\label{subsec:OpR3c}
OpR3c was a scheduling exercise covering a virtual period of 18 contiguous months. The 18 months were split into trimesters (named T1--T6), each covering three months. These six periods allowed us to test the \scheduler's OB queue creation with WEAVE's full complement of observing modes: the LIFU, the mIFU, and the two MOS plates. In essence, OpR3c was a significantly scaled-up version of OpR3a, taking lessons learnt from both OpR3a and OpR3b and attempting a close-to-reality scheduling simulation for the use of WEAVE on the WHT. As such, we also included a representative sample of OBs belonging to `PI time', which entered the \scheduler\ queue alongside the WEAVE Survey OBs, as will occur during regular Survey operations.\looseness-1

As well as all envisaged observing modes being available for scheduling purposes, the single main difference of OpR3c to OpR3a was in the length of the rehearsal, both in virtual and real time. Two months of `Survey time' were simulated almost every week, with 18 months of virtual observing time scheduled in total. Approximately every two weeks, the OISMT released the outcome of the scheduling exercise of the current trimester to the SWG, offset by one month, e.g.\ for the first trimester (T1), for which the SWG had submitted fields for months 1--3, the OISMT then released scheduling results for months 1--2 to the SWG, which the SWG then used to decide on their submissions for months 4--6 (i.e.\ T2). The SWG members were required, in consultation with their STLs and relevant science team members, to make decisions within three to four working days regarding the next set of fields to submit for the following trimester, based on the knowledge of which fields had been observed. While there is an overall observing strategy for the planned five-year WEAVE Survey for every individual survey, making the decision of which fields to prioritize and which specific targets should be contained within those fields requires knowledge of what has or has not -- despite the OBs having already been submitted -- yet been observed. The actual WEAVE Survey operations will also follow a trimesterly cycle for field submissions by the SWG, making OpR3c a realistic -- if somewhat condensed and highly time-pressured -- simulation of what is to come. 
Note that while the intensive and relatively rigid real-time schedule was set up to enable the SWG and STLs to rehearse decision-making on realistic timescales, the intervals and timing offsets later did require some flexibility and shifting at times, primarily due to the availability of those involved in the exercise.
\looseness-1

One of the principal aims of OpR3c was to identify any missing pieces in the operational flow for WEAVE operations. The following list provides an example of the types of questions raised and issues identified, some of which were resolved in real time during OpR3c, and all of which require resolving prior to the start of WEAVE operations.\looseness-1

\begin{enumerate}
    \item OpR3c allowed science teams to identify which other surveys typically compete for similar observing conditions, modes, and sky positions. This in turn  helped us identify new functionalities to be implemented in the \scheduler, allowing for a better sampling of OBs from the OB database, thereby creating both a better balance of OBs observed across the different surveys as well as better OB selection within each survey according to its science goals. OpR3c further identified some collisions between surveys whose science requirements call for similar observing conditions, which have now been ameliorated through adjustments made to the relative, inter-survey observation-time allocations in the survey planning.
    \item The combined effect of the requests in right-ascension distribution, observing conditions, and instrumental setup by community PI programmes cannot be predicted in advance, and neither, therefore, can the effect that this will have on their interplay with the WEAVE surveys in the OB scheduling process. OpR3c simulated the contribution of PI programmes to the scheduling process by incorporating a wide range of observing requirements (weather conditions and observing modes) that was also varied trimesterly. Although no simulation can truly emulate (or help predict) the effect of combining PI programmes and the WEAVE Survey during WEAVE operations, OpR3c has helped us identify ways in which the balancing of these components could be achieved.   
    \item A small number of empty slots were identified in the \scheduler\ output, indicating sky conditions that could not be used by any WEAVE survey. These tended to correspond to poor-seeing ($>2$~arcsec) during bright time (where there were no OBs available from the Survey or simulated PI programmes) and came to a total of 15 one-hour slots (out of a total of $\sim$ 3500) over the course of 18 months. A PI or WEAVE science programme that can utilize such conditions would therefore be a favourable complement from a scheduling perspective. WEAVE-LOFAR and WEAVE-QSO are amongst the science teams planning to use the WEAVE integral-field spectrographs (mIFU and LIFU) in relatively under-subscribed weather conditions to ameliorate this issue, so that no usable telescope time is left unfilled.   
    \item The weighting of different factors (such as seeing, sky brightness, and moon distance) by the \scheduler\ in choosing which OB should be queued requires re-evaluation and retesting. Giving higher weight to the seeing at the location of the field (which is affected by its airmass, i.e.\ its elevation) relative to other factors tends to disfavour those OBs that never reach low airmasses due to their declination. This biasing has been fixed in the most recent version of the \scheduler.    
    \item OBs should have expiration dates so that the OB database at the OCS (in general) contains only currently observable OBs, and to allow surveys to refresh the content of the OB database in case of changes in their plans. An expiration date of 9 months after ingestion is now set for any OB entering the database. An exception to this rule is if there are OBs in the database belonging to an already started `chain' (i.e.\ duplicate OBs to enable a longer total exposure of the field), in which case the remaining OBs have their expiration dates removed.  
    \item There is a need for the ability to retract OBs (prior to them being observed) or data (at different stages of the data-processing and archiving process). Necessary protocols for this within the SPA are now mostly implemented.
\end{enumerate}

\subsection{Benefits of the Operational Rehearsals}

Interactions between the SPA teams and the Science Team through the work of the SWG and QAG has been -- and continues to be -- crucial for identifying critical areas requiring further development within the data-processing systems (CPS and APS), while feedback from test users of the archive system has been essential in ensuring that the WAS user interface is simultaneously user-friendly and powerful. Within the Science Team, the OpRs have allowed us to identify details of specifics, as well as the order of play, for a multitude of tasks not confined to the following: putting together input target catalogues; how to specify sky positions; how to efficiently manage the configuration of dozens to hundreds of field configurations; the need to create software (codes and scripts) that could, in principle, be run by any individual with sufficient understanding of the process (avoiding single-point failures in manpower); how to deal with fields containing targets from multiple surveys. The OpRs have involved the efforts of more than 70 people, with many processes having been performed under realistic time pressure.\looseness-1

\section{Summary}
\label{sec:summary}

In an era of ever-increasing telescope and survey sizes, following up initial observations of large numbers of objects with medium- and high-resolution spectroscopy with modestly sized telescopes has been widely identified as a vital component of modern astrophysical research. In this context, we present WEAVE, the new wide-field, massively multiplexed spectroscopic survey facility for the William Herschel Telescope.
WEAVE comprises a new, 2-degree-diameter field-of-view, prime-focus corrector system with a nearly 1000-multiplex fibre positioner, 20 individually deployable mIFUs, and a single LIFU. The MOS fibres and mIFU fibre bundles are placed by two robots working in tandem at the WHT prime focus. These fibre systems feed a dual-beam spectrograph that covers 366--959\,nm at $R\sim5000$, or two shorter ranges at $R\sim20\,000$. 

The WEAVE Survey is a spectroscopic survey of the Northern sky, using WEAVE to answer key questions about the evolution of our own Galaxy and other galaxies, dark matter, and dark energy, and is comprised of eight individual surveys.
\begin{enumerate}
    \item The WEAVE Galactic Archaeology survey will complete {\it Gaia}'s phase-space information, providing metallicities for nearly three million main-sequence turnoff and red giant stars, as well as providing detailed abundances for up to 1.5 million brighter stars in four sub-surveys: the high-latitude LR survey, observing more than 1.5 million stars over nearly $9000\,\mathrm{deg^2}$ of the Northern sky; the disc-dynamics LR survey of more than a million stars along 500 sight-lines through the Galactic disc; the HR chemo-dynamical survey of more than 1.5 million stars across $5650\,\mathrm{deg^2}$ (non-contiguous) of the Northern sky; and the open clusters survey of more than one hundred open clusters of a variety of evolutionary states in the HR mode.
    \item The Stellar, Circumstellar and Interstellar Physics survey will observe $\sim400\,000$ evolved stars and the interstellar medium in the Galactic plane, covering $1200\,\mathrm{deg^2}$ in the LR mode and regions in the Cygnus star-forming region and the Galactic Anticentre in the HR mode.
    \item The WEAVE White Dwarfs survey will be one of the largest spectral surveys of white dwarfs ever undertaken, piggybacking on top of the other WEAVE surveys and targeting more than $50\,000$ white dwarfs.
    \item WEAVE-Apertif is an IFU survey of $\approx400$ neutral-hydrogen-selected galaxies in the LR mode, with detailed follow-up of $\sim100$ of these in the HR mode.
    \item WEAVE Galaxy Clusters is a survey of stellar populations, ionized gas properties, and stellar and ionized gas kinematics in cluster galaxies and their outskirts out to $z\sim0.5$ using all of WEAVE's fibre modes (MOS, mIFU, and LIFU), split into the Nearby, Wide-Field, and Cosmological Clusters sub-surveys.
    \item The Stellar Populations at intermediate redshifts Survey is a survey of detailed stellar populations and stellar kinematics in $\sim25\,000$ individual field galaxies, the majority within $0.3\lesssim z \lesssim 0.7$.
    \item WEAVE-LOFAR (WL) will obtain around one million spectra of sources selected on the basis of their activity from the LOFAR Two-metre Sky Survey (LoTSS). The survey will have a wedding-cake structure with three tiers ranging from WL-Wide ($\sim9000\,\mathrm{deg}^{2}$ of the Northern Sky) to WL-Deep (up to $60\,\mathrm{deg}^{2}$ in the LoTSS Deep Fields), together efficiently sampling the luminosity and redshift plane, and observing statistical samples of galaxies and AGNs selected in a manner that is unbiased by dust.
    \item WEAVE-QSO is a survey of the intergalactic medium at $z>2.1$ and a very sensitive probe of baryonic acoustic oscillations at $z>2.4$ using absorption lines along the line of sight to $\sim400\,000$ quasars.
\end{enumerate}

We have developed new, two-phase, dedicated data-processing pipelines for WEAVE, producing both reduced (via the CPS) and science-ready (via the APS) data, with a modular architecture allowing for additional analysis software to be developed as necessary by the science teams to be incorporated into the APS. Outputs from these pipelines populate a dedicated data archive (WAS), which will also store Contributed Data Products (CDPs). WEAVE also requires a new control system at the Observatory (OCS), enabling queue-based observations, in which user-submitted XML files describing the desired telescope and spectrograph settings with MOS or m/LIFU field configurations are converted into OBs and stored in an OB management system that is sampled by the queue \scheduler\ for use by the telescope operator and/or the support astronomer. The entire data system has been exercised in a series of Operational Rehearsals, starting with the end-to-end WEAVE instrument Simulator, which `observes' synthetic data that are then passed through the OCS to the CPS and the APS, and finally into the WAS. Many lessons about the data flow, survey design, and observation scheduling have been learnt by simulating periods between three weeks (full simulated images) and 1.5 years (nightly scheduling) of the WEAVE Survey. As an example, we suggest that PI observations using poor-seeing ($>2$~arcsec) conditions in bright time would favourably complement the WEAVE Survey programme.

The WEAVE Survey will begin observations towards the end of 2023. All processed WEAVE data and CDPs will be released to the worldwide astronomy community through the WAS annually, beginning two years after the start of the WEAVE Survey (i.e. late 2025), and astronomers in the WEAVE Survey Consortium will have access to the data soon after the data have been taken and reduced. A Science Verification (SV) campaign immediately preceding the WEAVE Survey will test various aspects of the surveys and carry out more than a dozen programs submitted by the ING community (astronomers in the UK, the Netherlands, and Spain, plus their collaborators); all SV data is expected to be reduced and released to the ING and WEAVE community within six months of the end of the SV campaign.

WEAVE represents a significant next step forward in our understanding of the Universe around us. The planned surveys will provide data that will help answer the questions: How did our Galaxy form and the stars within it evolve? How were other galaxies assembled? What are dark matter and dark energy?

\subsection*{Acknowledgements}

We would like to thank the reviewers of the WEAVE Survey Strategy Review held at Observatoire de Paris (2015 October 19-20), M.\,Bershady, G.\,Hill, F.\,Najarro, and R.\,Schiavon, for their constructive and critical comments on the science and survey cases presented then, and which formed the basis of the cases presented here as well as the subsequent, extensive WEAVE Science Case and Survey Plan, which were reviewed at the WEAVE Survey Readiness Review held at the University of Oxford (2020 January 15-16) by G.\,Hill, S.\,Martell, K.\,Master, and A.\,McConnachie. We are also very grateful to these reviewers for their constructive and supportive critique. S.\,C.\,Trager and G.\,B.\,Dalton thank the WEAVE Project Board and ING Board for their strong support for the project and many helpful discussions. S.\,Jin and S.\,C.\,Trager thank Francesco de Gasperin for his careful reading of the draft that helped to improve this paper. The authors would also like to thank the anonymous reviewer for their very careful reading of the manuscript.

% Individual acknowledgements of funding/grants
S.\,Jin acknowledges financial support from the European Union's Horizon 2020 research and innovation programme under the Marie Sk{\l}odowska-Curie grant agreement No.\,665593. 
S.\,C.\,Trager acknowledges funding for WEAVE construction through NWO grant 614.061.612 and ongoing support from NOVA, the Netherlands School for Research in Astronomy.
G.\,B.\,Dalton acknowledges ongoing support for the WEAVE project from STFC.
The WEAVE development work carried out at the Cambridge Astronomical Survey Unit (CASU), at the IoA, Cambridge, has been in part funded through UKRI STFC grant ST/X001857/1 and previously ST/M003175/1, ST/N005805/1, ST/P003486/1, ST/M007626/1, and ST/T003081/1.
J.\,A.\,L.\,Aguerri, C.\,Zurita, J.\,M\'endez-Abreu, and A.\,Molaeinezhad acknowledge support from the Spanish Ministries of Economia y Competitividad (MINECO) and Ciencia e Innovaci\'on (MICINN) through the grants AYA2017-83204-P, AYA2013-43188-P, and PID2020-119342GB-I00.
E.\,J.\,Alfaro acknowledges financial support from the State Agency for Research of the Spanish MCIU through the ``Center of Excellence Severo Ochoa'' award to the Instituto de Astrof\'isica de Andaluc\'ia (CEX2021-001131-S) and from the PY20-00753 grant from Junta de Andaluc\'ia, Spain (Autonomic Government of Andalusia, Spain).
T.\,Antoja acknowledges the grant RYC2018-025968-I funded by MCIN/AEI/10.13039/501100011033 and by ``ESF Investing in your future''. 
%T.~Antoja, L.~Balaguer-N\'u\~nez, J.~Carbajo-Hijarrubia, F.~Figueras, C.~Jordi, M.~Romero-G\'omez, M.~Mongui\`o:
This work was partially supported by the Spanish MICIN/AEI/10.13039/501100011033 and by ``ERDF A way of making Europe'' by the European Union through grants PID2021-122842OB-C21 and PID2021-125451NA-I00, and the Institute of Cosmos Sciences University of Barcelona (ICCUB, Unidad de Excelencia `Mar\'{\i}a de Maeztu') through grant CEX2019-000918-M.
G.\,Battaglia acknowledges support from the Agencia Estatal de Investigaci\'on del Ministerio de Ciencia en Innovaci\'on (AEI-MICIN) and the ERDF under grant number AYA2017-89076-P, the AEI under grant number CEX2019-000920-S and the AEI-MICIN under grant number PID2020-118778GB-I00/10.13039/501100011033.
T.\,Bensby acknowledges financial support by grant No.\,2018-04857 from the Swedish Research Council.
S.\,R.\,Berlanas acknowledges funding by the Spanish Ministerio de Ciencia e Innovaci\'on and Agencia Estatal de Investigaci\'on (MCIN/AEI/10.13039/501100011033/FEDER, UE) under the Juan de la Cierva -- Formaci\'on grant (contract FJC 2020-045785-I) and NextGeneration EU/PRTR and MIU (UNI/551/2021) through grant Margarita Salas-Universidad de la Laguna. This research has also received financial support by the MCIN and AEI through grant PID2021-122397NB-C22 and with funding from the European Union NextGenerationEU and Generalitat Valenciana in the call Programa de Planes Complementarios de I+D+i (PRTR 2022, Project HIAMAS, reference ASFAE/2022/017).
P.\,N.\,Best acknowledges support from STFC under grant ST/V000594/1.
S.\,Bonoli acknowledges financial support from the State Research Agency (AEI-MCINN) of the Spanish Ministry of Science and Innovation under the grants PGC2018-097585-B-C22 and PID2021-124243NB-C21 and the Project of excellence Prometeo/2020/085 from the Conselleria d'Innovaci\' o, Universitats, Ci\`encia i Societat Digital de la Generalitat Valenciana.
N.\,Britavskiy acknowledges support by the University of Liege under Special Funds for Research, IPD-STEMA Programme.
L.\,Costantin acknowledges financial support from Comunidad de Madrid under Atracci\'on de Talento grant 2018-T2/TIC-11612 and Spanish Ministerio de Ciencia e Innovaci\'on MCIN/AEI/10.13039/501100011033 through grant PGC2018-093499-B-I00.
J.\,Falc\'on-Barroso and A.\,Ferr\'e-Mateu acknowledge support from the Spanish Ministries of Economia y Competitividad (MINECO) and Ciencia e Innovaci\'on (MICINN) through the grants AYA2016-77237-C3-1-P and PID2019-107427GB-C32. AFM also acknowledges support from RYC2021-031099-I and PID2021-123313NA-I00 of MICIN/AEI/10.13039/501100011033/FEDER,UE,NextGeneration
%\newline %forcing splitting of a long line
EU,PRTR.
M.\,Fumagalli acknowledges support from the ERC under the European Union's Horizon 2020 research and innovation programme (grant agreement No 757535) and from Fondazione Cariplo, grant No.\, 2018-2329.
B.\,T.\,G\"ansicke was supported by grant ST/T000406/1 from the STFC. This project has received funding from the ERC under the European Union's Horizon 2020 research and innovation programme (Grant agreement No.\,101020057).
R.\,Garc\'ia-Benito, A.\,M.\,Conrado, and R.\,Gonz\'alez\,Delgado acknowledge financial support from the grant CEX2021-001131-S funded by MCIN/AEI/10.13039/501100011033 and from PID2019-109067-GB100.
C.\,P.\,Haines acknowledges support from ANID through Fondecyt Regular 2021 project no.\,1211909.
N.\,A.\,Hatch, S.\,C.\,Read, A.\,Aragon-Salamanca, and U.\,Kuchner acknowledge support from the UK STFC consolidated grant ST/T000171/1.
We acknowledge financial support from a Spinoza prize to A. Helmi.
A.\,Herrero, A.\,de Burgos, S.\,Sim{\'o}n-D{\'\i}az acknowledge support from the Spanish Ministry of Science and Innovation (MICINN) through the Spanish State Research Agency through grants PGC-2018-0913741-B-C22, PID2021-122397NB-C21, and the Severo Ochoa Programe 2020-2023 (CEX2019-000920-S). This work has also received financial support from the Canarian Agency for Economy, Knowledge, and Employment and the ERDF/EU, under grant with reference ProID2020010016.
K.\,M.\,Hess acknowledges financial support from the grant CEX2021-001131-S funded by MCIN/AEI/ 10.13039/501100011033, from the coordination of the participation in SKA-SPAIN, funded by the Ministry of Science and Innovation (MCIN); and support from the ERC under the European Union's Seventh Framework Programme (FP/2007--2013)/ERC Grant Agreement no.\,291531 (HIStoryNU).
V.\,Ir\v{s}i\v{c} is supported by the Kavli foundation.
J.\,H.\,Knapen acknowledges financial support from the State Research Agency (AEI-MCINN) of the Spanish Ministry of Science and Innovation under the grant ``The structure and evolution of galaxies and their central regions'' with reference PID2019-105602GB-I00/10.13039/501100011033, and from the ACIISI, Consejer\'{i}a de Econom\'{i}a, Conocimiento y Empleo del Gobierno de Canarias and the ERDF under grant with reference PROID2021010044.
S.\,Lucatello, A.\,Bragaglia, and M.\,Bellazzini acknowledge the support from PRIN INAF 2019 grant ObFu 1.05.01.85.14 (``Building up the halo: chemo-dynamical tagging in the age of large surveys'', PI. S. Lucatello).
J.\,Ma{\'\i}z\,Apell\'aniz acknowledges support from the Spanish Government Ministerio de Ciencia e Innovaci\'on and Agencia Estatal de Investigaci\'on (10.13039/501100011033) through grant PGC2018-095049-B-C22. 
J.\,Mendez\,Abreu acknowledges support from the Spanish Ministries of Economia y Competitividad (MINECO) and Ciencia e Innovaci\'on (MICINN) through the grant PID2021-128131NB-I00, and the support of the Viera y Clavijo Senior program funded by ACIISI and ULL.
I.\,Negueruela and A.\,Marco acknowledge the financial support of the Spanish Ministerio de Ciencia e Innovaci\'on (MCIN) with funding from the European Union NextGenerationEU and Generalitat Valenciana in the call Programa de Planes Complementarios de I+D+i (PRTR 2022, Project HIAMAS, reference ASFAE/2022/017) and through grant PID2021-122397NB-C22. 
I.\,P\'erez-R\`afols acknowledges support from the European Union's Horizon 2020 research and innovation programme under the Marie Sk{\l}odowska-Curie grant agreement No.\,754510.
M.\,M.\,Pieri, I.\,P\'erez-R\`afols, S.\,Morrison, D.\,Som, and M.\,Blomqvist were supported by the Programme National Cosmology et Galaxies (PNCG) of CNRS/INSU with INP and IN2P3, co-funded by CEA and CNES, by A*MIDEX project (ANR-11-IDEX-0001-02) funded by the ``Investissements d'Avenir'' French Government program, managed by the French National Research Agency (ANR), and by ANR under contract ANR-14-ACHN-0021 and ANR-22-CE31-0026.
R.\,Raddi has received funding from the postdoctoral fellowship programme Beatriu de Pin\'os, funded by the Secretary of Universities and Research (Government of Catalonia) and by the Horizon 2020 programme of research and innovation of the European Union under the Maria Sk{\l}odowska-Curie grant agreement No.\,801370.
E.\,M.\,Rossi acknowledges that this project has received funding from the ERC under the European Union's Horizon 2020 research and innovation programme (Grant agreement No.\,101002511 -- VEGA P).
J.\,A.\,Rubino-Martin acknowledges financial support from the Spanish Ministry of Science and Innovation under the project PID2020-120514GB-I00.
D.\,J.\,B.\,Smith, A.\,B.\,Drake, and M.\,J.\,Hardcastle acknowledge support from the UK STFC under grant ST/V000624/1.
R.\,J.\,Smith acknowledges support from the UK STFC consolidated grant ST/T000244/1.
D.\,Sorini was supported by the ERC, under grant no.\,670193, by the STFC consolidated grant no.\,RA5496, and by the Swiss National Science Foundation (SNSF) Professorship grant no.\,202671.
E.\,Starkenburg acknowledges funding through VIDI grant ``Pushing Galactic Archaeology to its limits'' (with project number VI.Vidi.193.093), which is funded by the Dutch Research Council (NWO).
G.\,F.\,Thomas acknowledge support from the Agencia Estatal de Investigaci\'on (AEI) under grant {\it Ayudas a centros de excelencia Severo Ochoa convocatoria 2019} with reference CEX2019-000920-S, and from the Agencia Estatal de Investigaci\'on del Ministerio de Ciencia e Innovaci\'on (AEI-MCINN) under grant {\it En la frontera de la arqueolog\'ia galac\'actica: evoluc\'ion de la materia luminoso y obscura de la v\'ia L\'actea y las galaxias enenas del Grupo Local} with reference PID2020-118778GB-I00.
J.\,S.\,Vink acknowledges support from the UK STFC grant ST/V000233/1.
All INAF researchers are supported by INAF grant 1.05.03.04.05.

%WEAVE Acknowledgement
Funding for the WEAVE facility has been provided by UKRI STFC, the University of Oxford, NOVA, NWO, Instituto de Astrof\'isica de Canarias (IAC), the Isaac Newton Group partners (STFC, NWO, and Spain, led by the IAC), INAF, CNRS-INSU, the Observatoire de Paris, R\'egion \^Ile-de-France, CONCYT through INAOE, Konkoly Observatory (CSFK), Max-Planck-Institut f\"ur Astronomie (MPIA Heidelberg), Lund University, the Leibniz Institute for Astrophysics Potsdam (AIP), the Swedish Research Council, the European Commission, and the University of Pennsylvania. The WEAVE Survey Consortium consists of the ING, its three partners, represented by UKRI STFC, NWO, and the IAC, NOVA, INAF, GEPI, INAOE, and individual WEAVE Participants.
Please see the relevant footnotes for the WEAVE website\footnote{\url{https://ingconfluence.ing.iac.es/confluence//display/WEAV/The+WEAVE+Project}} and for the full list of granting agencies and grants supporting WEAVE\footnote{\url{https://ingconfluence.ing.iac.es/confluence/display/WEAV/WEAVE+Acknowledgements}}.

%Abbreviations for ministries, funding bodies etc. (requested by copy editor)
Abbreviations for ministries, funding agencies, and bodies are as follows:
AEI = Agencia Estatal de Investigaci\'on (Spanish State Research Agency);
ANR = Agence Nationale de la Recherche (French National Research Agency);
CNRS = Centre national de la recherche scientifique (French National Centre for Scientific Research);
CSIC = Consejo Superior de Investigaciones Cient\'ificas (Spanish National Research Council);
DFG = Deutsche Forschungsgemeinshaft (German Research Foundation);
ERC = European Research Council;
ERDF = European Regional Development Fund;
INAF = Istituto Nazionale di Astrofisica (Italian National Institute for Astrophysics;
INSU = Institut National des Sciences de l'Univers (National Institute of Sciences of the Universe, France);
MICINN = Ministerio de Ciencia e Innovaci\'on (Spanish Ministry of Science and Innovation);
MINECO = Ministerio de Econom\'ia y Competitividad (Spanish Ministry of Economy and Competitiveness, Spain);
NOVA = Nederlandse Onderzoekschool Voor Astronomie (Dutch Research School for Astronomy);
NWO = Nederlandse Organisatie voor Wetenschappelijk Onderzoek (Dutch Research Council);
STFC = Science and Technologies Research Council (UK);
UKRI = UK Research and Innovation (U.K.).

%For SCIP figure:
The Digitized Sky Survey (DSS) was produced at the Space Telescope Science Institute under U.S. Government grant NAG W-2166. The images of these surveys are based on photographic data obtained using the Oschin Schmidt Telescope on Palomar Mountain and the UK Schmidt Telescope. The plates were processed into the present compressed digital form with the permission of these institutions.

%For use of Aladin:
This research has made use of `Aladin sky atlas' developed at CDS, Strasbourg Observatory, France.\nocite{2000A&AS..143...33B}

%Open Access:
For the purpose of Open Access, the authors have applied a Creative Commons Attribution (CC BY) public copyright licence to any Author Accepted Manuscript version arising from this submission.\\

%Dedications:
This paper is dedicated to the memories of Jim Lewis and Dave Carter, both of whom made significant contributions to the WEAVE Project, and who passed away in 2019 and 2021, respectively, and to Jordi Torra, whose support on the WEAVE Project Board was invaluable, and who passed away in 2021.

\subsection*{Data Availability Statement}

The simulated WEAVE Survey data can be made available to interested parties through contact with the first three authors of this paper.

\bibliographystyle{mnras}

%%%%%%%%%%%%%%%%% APPENDICES %%%%%%%%%%%%%%%%%%%%%

\appendix
\section{Abbreviations}

\begin{table*}
\centering
\caption{Common WEAVE abbreviations.}
\begin{tabular}{llc}
\hline
& \multicolumn{1}{c}{Commonly used abbreviations in WEAVE} & \multicolumn{1}{c}{Section of paper in which defined}\\
\hline
CPS, APS & Core Processing System, Advanced Processing System & \ref{subsec:SPA} \\
LR, HR & Low Resolution, High Resolution & \ref{subsec:SPE} \\
mIFU, LIFU & mini Integral Field Unit, Large Integral Field Unit & \ref{sec:Introduction} \\
MOS & Multi-Object Spectroscopy & \ref{subsec:FIB+POS} \\
OB & Observing Block & \ref{subsec:OCS} \\
OCS & Observatory Control System & \ref{subsec:OCS} \\
OISMT & On-Island Survey Management Team & \ref{subsec:OISMT} \\
OpR & Operational Rehearsal & \ref{sec:OpRs} \\
QAG & Quality Assessment Group & \ref{subsec:QAG} \\
SPA & Science Processing and Analysis & \ref{subsec:SPA} \\
STL & Science Team Lead & \ref{sec:Consortia} \\
SWG & Survey Working Group & \ref{subsec:SWG} \\
WAS & WEAVE Archive System & \ref{subsec:SPA} \\ 
WASP & WEAVE Automated Submission Platform & \ref{sec:OpRs}\\
WEAVE & WHT Enhanced Area Velocity Explorer & \ref{sec:Introduction} \\
WHT & William Herschel Telescope & \ref{sec:Introduction} \\
\hline
\end{tabular}
\label{table:WEAVE_abbrevs}
\end{table*}

A list of abbreviations used in this paper is given in Table~\ref{table:WEAVE_abbrevs}.

\section{Author affiliations}
\label{sec:affiliations}
% In addresses, "The Netherlands" instead of "the Netherlands". No "NL" preceding postal code for Dutch addresses, in contrast to postal codes for other European countries.
$^{1}$Oxford Astrophysics, University of Oxford, Keble Road, Oxford OX1 3RH, U.K.\\
$^{2}$Kapteyn Astronomical Institute, Rijksuniversiteit Groningen, Landleven 12, 9747\,AD Groningen, The Netherlands\\
$^{3}$RALSpace, STFC, Harwell, Didcot OX11 0QX, U.K.\\
$^{4}$SRON -- Netherlands Institute for Space Research, Landleven 12, 9747\,AD Groningen, The Netherlands\\
$^{5}$Instituto de Astrof\'isica de Canarias, Calle V\'ia L\'actea s/n, E-38205 La Laguna, Santa Cruz de Tenerife, Spain\\
$^{6}$Departamento de Astrof\'isica, Universidad de La Laguna, E-38206, La Laguna, Tenerife, Spain\\
$^{7}$Centre for Astrophysics Research, University of Hertfordshire, Hatfield, Hertfordshire AL10 9AB, U.K.\\
$^{8}$Department of Physics and Astronomy, University College London, Gower Street, London WC1E 6BT, U.K.\\
$^{9}$Department of Physics, University of Warwick, Gibbet Hill Road, Coventry CV4 7AL, U.K.\\
$^{10}$Universit\'e C\^ote d'Azur, Observatoire de la C\^ote d'Azur, CNRS, Laboratoire Lagrange, Bd de l'Observatoire, CS 34229, F-06304 Nice Cedex 4, France\\
$^{11}$INAF -- Osservatorio Astronomico di Brera, Via Brera, 28, I-20121 Milano, Italy\\
$^{12}$Aix Marseille Univ, CNRS, CNES, LAM, Laboratoire d'Astrophysique de Marseille, F-13388 Marseille, France\\
$^{13}$INAF --  Osservatorio Astronomico di Padova, Vicolo Osservatorio 5, I-35122 Padova, Italy\\
$^{14}$Isaac Newton Group of Telescopes, Apartado 321, E-38700 Santa Cruz de la Palma, Tenerife, Spain\\
$^{15}$Dipartimento di Fisica e Astrofisica, Univerisit\`{a} degli Studi di Firenze, via G. Sansone 1, I-50019 Sesto Fiorentino, Italy\\
$^{16}$INAF -- Osservatorio Astrofisico di Arcetri, Largo E. Fermi, 5, I-20125 Firenze, Italy\\
$^{17}$Institute of Astronomy, University of Cambridge, Madingley Road, Cambridge CB3 0HA, U.K.\\
$^{18}$Institut de Ci\`encies del Cosmos (ICCUB), Universitat de Barcelona (UB), Mart\'i Franqu\`es 1, E-08028 Barcelona, Spain\\
$^{19}$Departament de Física Qu\`antica i Astrof\'isica (FQA), Universitat de Barcelona (UB), Mart\'i i Franqu\`es 1, E-08028 Barcelona, Spain\\
$^{20}$Institut d'Estudis Espacials de Catalunya (IEEC), c. Gran Capit\`a, 2-4, E-08034 Barcelona, Spain\\
$^{21}$School of Physics and Astronomy, University of Nottingham, University Park, Nottingham NG7 2RD, U.K.\\
$^{22}$Departamento de F\'isica Te\'orica, Universidad Aut\'onoma de Madrid, E-28049 Madrid, Spain\\
$^{23}$Univ. Grenoble Alpes, CNRS, IPAG, F-38000 Grenoble, France\\
$^{24}$GEPI, Observatoire de Paris, Universit\'e PSL, CNRS: 5, place Jules Janssen, F-92195 Meudon, France\\
$^{25}$Lund Observatory, Department of Astronomy and Theoretical Physics, Box 43, SE-221\,00 Lund, Sweden\\
$^{26}$INAF -- Osservatorio di Astrofisica e Scienza dello Spazio, via P. Gobetti 93/3, I-40129 Bologna, Italy\\
$^{27}$Instituto Nacional de Astrof\'isica, \'Optica y Electr\'onica, Luis Enrique Erro 1, Tonantzintla, Puebla, C.P. 72840, Mexico\\
$^{28}$Institute for Astronomy, Royal Observatory, Blackford Hill, Edinburgh EH9 3HJ, U.K.\\
$^{29}$Universit\'e de Strasbourg, CNRS, Observatoire astronomique de Strasbourg, UMR 7550, F-67000 Strasbourg, France\\
%following was listed as both #29 and #39: 
%$^{29}$Instituto de Astronomia y Ciencias Planetarias, Universidad de Atacama, Copayapu 485, Copiap\'o, Chile\\
$^{30}$UK Astronomy Technology Centre, Royal Observatory Edinburgh, Blackford Hill, Edinburgh EH9 3HJ, U.K.\\
$^{31}$Instituto de Astrof\'isica de Andaluc\'ia, CSIC, Glorieta de la Astronom\'ia s/n, E-18008 Granada, Spain\\
$^{32}$ASTRON, Netherlands Institute for Radio Astronomy, Oude Hoogeveensedijk 4, 7991\,PD Dwingeloo, The Netherlands\\
$^{33}$Astrophysics Research Institute, Liverpool John Moores University, 146 Brownlow Hill, L3 5RF, Merseyside, U.K.\\
$^{34}$Department of Physics, McWilliams Center for Cosmology, Carnegie Mellon University, 5000 Forbes Avenue, Pittsburgh, PA 15213, U.S.A.\\
$^{35}$Institut d'Astrophysique de Paris, 98bis Bd Arago, F-75014 Paris, France\\
$^{36}$Technical University of Munich Institute for Advanced Studies Lichtenbergstr. 2a, D-85748 Garching, Germany\\
$^{37}$INAF -- Osservatorio Astronomico di Capodimonte, Salita Moiariello, 16, I-80131 Napoli, Italy\\
$^{38}$Department of Astronomy, University of Illinois at Urbana-Champaign, Urbana, IL 61801, U.S.A.\\
$^{39}$Instituto de Astronom\'ia y Ciencias Planetarias, Universidad de Atacama, Copayapu 485, Copiap\'o, Chile\\
$^{40}$Instituto Universitario Carlos I de F\'isica Te\'orica y Computacional, Universidad de Granada, E-18071 Granada, Spain\\
$^{41}$Institut de F\'isica d'Altes Energies (IFAE), The Barcelona Institute of Science and Technology, Campus UAB, E-08193 Bellaterra Barcelona, Spain\\
$^{42}$Universitat Polit\`ecnica de Catalunya, Carrer de Jordi Girona, 31, E-08034 Barcelona, Spain\\
$^{43}$Mullard Space Science Laboratory, University College London, Holmbury St Mary, Dorking RH5 6NT, U.K.\\
$^{44}$Space Telescope Science Institute, 3700 San Martin Drive, Baltimore, MD 21218, U.S.A.\\
$^{45}$INAF -- Osservatorio Astrofisico di Catania, Via S. Sofia 78, I-95123 Catania, Italy\\
$^{46}$Department of Physics, Lancaster University, Bailrigg, Lancaster LA1 4YB, U.K.\\
$^{47}$Dpto. de F\'isica Aplicada, Facultad de Ciencias, Universidad de Alicante, Cta. de San Vicente s/n, E-03690 San Vicente del Raspeig, Alicante, Spain\\
$^{48}$Astrophysics Group, Keele University, Keele ST5 5BG, U.K.\\
$^{49}$INAF -- Osservatorio Astronomico di Brera, via Bianchi 46, I-23087 Merate (LC), Italy\\
$^{50}$Donostia International Physics Center, Paseo Manuel de Lardizabal 4, E-20018 San Sebasti\'an, Spain\\
$^{51}$IKERBASQUE, Basque Foundation for Science, Bilbao, Plaza Euskadi 5, E-48009 Bilbao, Spain\\
$^{52}$Leiden Observatory, Leiden University, Niels Bohrweg 2, 2333\,CA Leiden, The Netherlands\\
$^{53}$Leibniz-Institut f\"ur Astrophysik Potsdam, An der Sternwarte 16, D-14482 Potsdam, Germany\\
$^{54}$Laboratoire d'astrophysique de Bordeaux, Universit\'e de Bordeaux, CNRS, B18N All\'ee Geoffroy Saint-Hilaire, F-33615 Pessac, France\\
$^{55}$Departamento de F\'isica Te\'orica, Universidad Aut\'onoma de Madrid (UAM), Campus de Cantoblanco, E-28049 Madrid, Spain\\
$^{56}$Australian Astronomical Optics, Macquarie University, 105 Delhi Rd, North Ryde, NSW 2113, Australia\\
$^{57}$ARC Centre of Excellence for All Sky Astrophysics in 3 Dimensions (ASTRO 3D)\\ %Note that postal code, city and country are not applicable for this affiliation. See https://astro3d.org.au/our-centre/, which states that “ASTRO 3D is comprised of nine Australian collaborating universities (Node Universities) and a number of world-class Australian and International Partner Organisations”.
$^{58}$Centro de Astrobiolog\'{\i}a (CAB/CSIC-INTA), Ctra. de Ajalvir km 4, Torrej\'on de Ardoz, E-28850 Madrid, Spain\\
$^{59}$Institute for Computational Cosmology, Durham University, Durham DH1 3LE, U.K.\\
$^{60}$Centre for Extragalactic Astronomy, Durham University, Durham DH1 3LE, U.K.\\
$^{61}$Departamento de F\'isica Te\'orica y del Cosmos, Universidad de Granada, Campus de Fuentenueva, E-18071 Granada, Spain\\
$^{62}$Departamento de F\'isica, Ingenier\'ia de Sistemas y Teor\'ia de la Sei\~nal, Universidad de Alicante, Cta. de San Vicente s/n, E-03690 San Vicente del Raspeig, Alicante, Spain\\
$^{63}$Dipartimento di Fisica `G. Occhialini', Universit\`a degli Studi di Milano Bicocca, Piazza della Scienza 3, I-20126 Milano, Italy\\
$^{64}$Department of Physics, University of Surrey, Guildford, Surrey GU2 7XH, U.K.\\
$^{65}$INAF -- Osservatorio Astronomico di Trieste, Via Giambattista Tiepolo, 11, I-34131 Trieste TS, Italy\\
$^{66}$Department of Chemistry and Physics, Saint Mary's College, Notre Dame, IN 46556, U.S.A.\\
$^{67}$Research School of Astronomy \& Astrophysics, Mount Stromlo Observatory, Cotter Road, Weston Creek, ACT 2611, Australia\\
$^{68}$Institute of Physics, University of Graz, Universit\"atsplatz 5, 8010 Graz, Austria\\
$^{69}$INAF -- Osservatorio Astronomico di Palermo, Piazza del Parlamento 1, I-90134 Palermo, Italy\\
$^{70}$INAF -- Fundaci\'on Galileo Galilei, Rambla Jos\'e Ana Fern\'andez P\'erez 7, E-38712 Bre{\~n}a Baja, Tenerife, Spain\\
$^{71}$Kavli Institute for Cosmology, University of Cambridge, Madingley Road, Cambridge CB3 0HA, U.K.\\
$^{72}$Department of Physics \& Astronomy, University of the Western Cape, Private Bag X17, Bellville, Cape Town 7535, South Africa\\
$^{73}$Universit\'e Lyon 1, ENS de Lyon, CNRS, Centre de Recherche Astrophysique de Lyon UMR5574, Saint-Genis-Laval, France\\
$^{74}$Department of Astrophysics/IMAPP, Radboud University, P.O. Box 9010, 6500\,GL Nijmegen, The Netherlands\\
$^{75}$Astronomisches Rechen-Institut, Zentrum f\"ur Astronomie der Universit\"at Heidelberg, M\"onchhofstr. 12-14, D-69120 Heidelberg, Germany\\
$^{76}$ESA -- ESTEC, MMO, Keplerlaan 1, PO Box 299, 2200\,AG Noordwijk, The Netherlands\\
$^{77}$Department of Astronomy, Stockholm University, AlbaNova University Centre, SE-106 91 Stockholm, Sweden\\
$^{78}$Max-Planck-Institut f\"ur Astronomie, K\"onigstuhl 17, D-69117 Heidelberg, Germany\\
% all numbers after this line need to shift by +1, and affected authors identified and their affiliations corrected (updated 4 July 2023)
$^{79}$Department of Physics and Astronomy, Uppsala University, Box 516, SE-751 20 Uppsala, Sweden\\
$^{80}$Department of Astronomy and Astrophysics, University of Chicago, 5640 South Ellis Avenue, Chicago, IL 60637, U.S.A.\\
$^{81}$Centro de Astrobiolog\'{\i}a (CAB/CSIC-INTA), ESAC campus, Camino bajo del castillo s/n, E-28692 Villanueva de la Ca{\~n}ada, Madrid, Spain\\
$^{82}$Instituci\'o Catalana de Recerca i Estudis Avan\c{c}ats, Passeig Llu\'is Companys 23, E-08010 Barcelona, Spain\\
$^{83}$INAF -- Osservatorio Astronomico di Cagliari, Via della Scienza 5, I-09047 Selargius CA, Italy\\
$^{84}$School of Physics, University of Exeter, Stocker Road, Exeter EX4 4QL, U.K.\\
$^{85}$Facultad de f\'isica, Universidad de Sevilla, Avda. Reina Mercedes s/n, Campus Reina Mercedes, E-41012 Sevilla, Spain\\
$^{86}$National Astronomical Observatory of Japan, Mitaka-shi, Tokyo 181-8588, Japan\\
$^{87}$NOVA Optical-Infrared Instrumentation Group at ASTRON, PO Box 2, 7990\,AA Dwingeloo, The Netherlands\\
$^{88}$D\'epartement de Physique Th\'eorique, Universit\'e de Gen\`eve, 24 quai Ernest Ansermet, 1211 Gen\`eve 4, Switzerland\\
$^{89}$Dipartimento di Fisica ``E.R. Caianiello'', Universit\`a degli Studi di Salerno, Via Giovanni Paolo II 132, I-84084 Fisciano (SA), Italy\\
$^{90}$Armagh Observatory and Planetarium, College Hill, Armagh BT61 9DG, Northern Ireland, U.K.\\
$^{91}$Joint ALMA Observatory, Alonso de C\'ordova 3107, Vitacura, Santiago 763-0355, Chile\\
$^{92}$National Radio Astronomy Observatory, 520 Edgemont Road, Charlottesville, VA 22903, U.S.A.\\
$^{93}$School of Physics and Astronomy, Sir William Henry Bragg Building, University of Leeds, Leeds LS2 9JT, U.K.\\
$^{94}$School of Physics and Astronomy, University of St Andrews, North Haugh, St Andrews KY16 9SS, U.K.\\
$^{95}$Zentrum f\"ur Astronomie der Universit\"at Heidelberg, Landessternwarte, K\"onigstuhl 12, D-69117 Heidelberg, Germany\\
$^{96}$School of Architecture -- Universidad Europea de Canarias, La Orotava, Tenerife, Spain\\
$^{97}$Universit\'e de Li\`ege, All\'ee du 6 Ao\"ut 19c, B-4000 Sart Tilman, Li\`ege, Belgium\\
\\
$^\star$E-mail: jin@astro.rug.nl\\
$^\diamond$E-mail: sctrager@astro.rug.nl\\
$^\ast$E-mail: gavin.dalton@physics.ox.ac.uk\\
$^\dag$Deceased\\
$^\ddag$Anna Boyksen fellow

% Don't change these lines at end of mnras_template.tex
\bsp	% typesetting comment
\label{lastpage}
\end{document}